\documentclass[aps,twocolumn,showpacs,superscriptaddress,prb,amsmath,amssymb,letterpaper,floatfix]{revtex4-1}
\pdfoutput=1
\usepackage{natbib}
\usepackage{amsmath,bm,amsfonts}
\usepackage{graphicx}
\graphicspath{./}
\usepackage[percent]{overpic}
\usepackage{rotating}
\usepackage{latexsym}
\usepackage{color}
\usepackage{xcolor}
\usepackage[breaklinks,colorlinks=true,linkcolor = red,urlcolor=cyan,citecolor=red]{hyperref}
\usepackage{ulem}

\definecolor{primary-1}{HTML}{2A4B5A}
\definecolor{secondaryA-1}{HTML}{2B624A}
\definecolor{secondaryB-1}{HTML}{8F563F}
\definecolor{complementary-1}{HTML}{8F6A3F}

\renewcommand{\v}[1]{\ensuremath{\mathbf{#1}}} 
\newcommand{\gv}[1]{\ensuremath{\mbox{\boldmath$ #1 $}}}
\newcommand{\avg}[1]{\left< #1 \right>} 

\let\baraccent=\= 
\renewcommand{\=}[1]{\stackrel{#1}{=}} 

\newcommand{\HM}{\mathcal{H}}

\newcommand{\eref}[1]{(\ref{#1})}

\begin{document}

\title{
Quantum spin liquids
in the absence of spin-rotation symmetry:
application to Herbertsmithite
}

\author{Tyler \surname{Dodds}}
\affiliation{Department of Physics, University of Toronto,
Toronto, Ontario M5S 1A7, Canada}

\author{Subhro Bhattacharjee}
\affiliation{Department of Physics, University of Toronto,
Toronto, Ontario M5S 1A7, Canada}
\affiliation{Department of Physics and Astronomy, McMaster University, Hamilton,
Ontario L8S 4M1, Canada}
\author{Yong Baek Kim}
\affiliation{Department of Physics, University of Toronto,
Toronto, Ontario M5S 1A7, Canada}
\affiliation{School of Physics, Korea Institute for Advanced Study, Seoul 130-722, Korea}

\date{\today}

\begin{abstract}
	It has been suggested that the nearest-neighbor antiferromagnetic Heisenberg model on the Kagome lattice may be a good starting point for understanding the spin-liquid behavior discovered in Herbertsmithite. In this work, we investigate possible quantum spin liquid phases in the presence of spin-rotation symmetry-breaking perturbations such as Dzyaloshinskii-Moriya (DM) and Ising interactions, as well as second-neighbor antiferromagnetic Heisenberg interactions. Experiments suggest that such perturbations are likely to be present in Herbertsmithite. We use the projective symmetry group (PSG) analysis within the framework of the slave-fermion construction of quantum spin liquid phases and systematically classify possible spin liquid phases in the presence of perturbations mentioned above. The dynamical spin-structure factor for relevant spin liquid phases is computed and the effect of those perturbations are studied. Our calculations reveal dispersive features in the spin structure factor embedded in a generally diffuse background due to the existence of fractionalized spin-$1/2$ excitations called spinons. For {\it two} of the previously proposed $Z_2$ states, the dispersive features are almost absent, and diffuse scattering dominates over a large energy window throughout the Brillouin zone. This resembles the structure factor observed in recent inelastic neutron scattering experiments on singlet crystals of Herbertsmithite. Furthermore, one of the $Z_2$ states with the spin structure factor with mostly diffuse scattering is gapped, and it may be adiabatically connected to the gapped spin liquid state observed in recent Density Matrix Renormalization Group calculations for the nearest neighbor antiferromagnetic Heisenberg model. The perturbations mentioned above are found to enhance the diffuse nature of the spin structure factor and reduce the momentum dependencies of the spin gap. We also calculate the electron spin resonance (ESR) absorption spectra that further characterize the role of spin-rotation symmetry breaking perturbations in these states. We suggest that the measurement of ESR spectra can shed more light into the nature of the ground state in Herbertsmithite.

\end{abstract}

\maketitle

\section{Introduction}
\label{sec:introduction}

Two-dimensional spin-1/2 Mott Insulators in corner sharing lattice geometries, like the Kagome lattice, have been studied extensively due to their natural tendency to suppress magnetic ordering potentially realizing a quantum spin liquid (QSL) ground state. \cite{Nature.464.199,Lee05092008,0143-0807-21-6-302,PhysRevB.87.060405,PhysRevLett.109.067201,NewJPhys.14.115031,NatPhys.8.902,PhysRevLett.108.157202,Nature.492.406,PhysRevB.83.224413,PhysRevB.81.064428,PhysRevB.78.140405,PhysRevLett.101.026405,PhysRevLett.98.107204,PhysRevLett.98.077204,PhysRevB.66.014422,PhysRevB.68.214415,PhysRevB.77.144415,JApplPhys.69.5962,PhysRevLett.95.087203,PhysRevLett.103.077207,CanJPhys.79.1283,diep2004frustrated,JPSJ.78.033701,PhysRevB.45.12377,JAmChemSoc.127.13462,Yan03062011} While the theoretical understanding of QSLs suggests several interesting features of these highly entangled quantum states of matter, such as effective quantum number fractionalization and/or topological order,\cite{Nature.464.199,Lee05092008,wen2004quantum,PRB.65.165113} its conclusive experimental realization has remained elusive so far.

Of the several candidate materials investigated, the compound, Herbertsmithite
[ZnCu$_3$(OH)$_6$Cl$_2$] seems particularly promising.\cite{JAmChemSoc.127.13462,PhysRevLett.108.157202,Nature.492.406,PhysRevB.81.064428,PhysRevB.78.140405,PhysRevLett.101.026405,PhysRevLett.98.107204,PhysRevLett.98.107204,PhysRevLett.98.077204,PhysRevLett.103.237201,PhysRevLett.100.077203,PhysRevB.84.020411,PhysRevLett.107.237201} Here, the Cu$^{2+}$ ions sit on an isotropic Kagome lattice and carry spin-$1/2$ moments. While the Curie-Weiss temperature ($\theta_{CW}$$\sim$ $-300$ K) suggests a strong and predominantly antiferromagnetic exchange interaction, the spins do not show any sign of long-range order or freezing down to 50 mK.\cite{PhysRevLett.98.107204, PhysRevLett.98.077204} Inelastic neutron scattering on powder samples shows magnetic scattering for a broad range of wavenumbers;\cite{PhysRevLett.98.107204,PhysRevLett.103.237201} furthermore, recent inelastic neutron scattering on single crystals show that even at temperatures as low as $1.6$ K, a diffuse magnetic scattering continuum exists over a large energy window ($1.5-11$ meV) in significant portions of the Brillouin zone and no spin gap down to at least $1.5$ meV.\cite{Nature.492.406} This has raised hopes that a low-temperature QSL phase may be realized in this material. In a QSL, such a continuum of scattering is naturally expected due to the presence of fractionalized spin-1/2 excitations, called spinons, as opposed to the conventional spin-1 magnon excitations in a magnetically ordered systems.

While Herbertsmithite suffers from inter-site disorder, it seems to mostly come in the form of 14\% occupation of inter-layer non-magnetic Zn$^{2+}$ ions by excess magnetic Cu$^{2+}$ ions, with less than $5\%$ Zn$^{2+}$ impurities in the Kagome planes.\cite{PhysRevLett.108.157202} Hence, it has been suggested that this material is, to a very good approximation, a realization of spin-1/2 nearest neighbor (NN) Heisenberg antiferromagnet on the Kagome lattice .\cite{PhysRevB.84.020411,PhysRevB.82.144412,PhysRevLett.100.157205,PhysRevLett.98.107204} We term this the  NN Heisenberg Kagome antiferromagnet (HKAF).

The above suggestion has brought in focus the theoretical studies on the NN-HKAF. These have a long and varied history, and there exist several proposed candidate ground states. Such states include valence-bond solids \cite{PhysRevB.68.214415,PhysRevB.77.144415,PhysRevLett.104.187203,PhysRevB.81.180402,PhysRevLett.93.187205} and gapless\cite{PhysRevB.77.224413,PhysRevLett.101.117203,PhysRevB.63.014413} and gapped QSLs. \cite{PhysRevB.83.224413,JApplPhys.69.5962,PhysRevLett.81.2356,PhysRevB.45.12377,PhysRevB.74.174423} Of these, a gapped QSL was found recently by Density Matrix Renormalization Group (DMRG) calculations.\cite{Yan03062011,PhysRevLett.109.067201,NatPhys.8.902} However, variational Monte Carlo (VMC) studies have found a gapless spin-disordered ground state with energy very similar to that of the DMRG.\cite{PhysRevB.87.060405} These and other studies also seem to suggest that the strictly NN-HKAF may be close to a quantum phase transition. Indeed, recent studies find that the nature of the ground state can be changed by tuning the small next-nearest-neighbor (NNN) exchange interaction.\cite{BAPS.2012.MAR.L19.1,epl.2009.27009,NewJPhys.14.115031}

For Herbertsmithite, this means that, due to the potential proximity to a quantum phase transition, small perturbing interactions may play an important role, particularly at low energies.\cite{PhysRevB.81.144432} Hence, it may be important to account for these smaller perturbing energy scales in order to understand the experiments on Herbertsmithite. Electron spin resonance (ESR) and magnetic susceptibility measurements suggest that perturbations break spin-rotation symmetry, with both a Dzyaloshinsky-Moriya (DM) interaction and an Ising-like easy-axis exchange, each with strengths of around 10\% of the NN antiferromagnetic exchange interaction. \cite{PhysRevLett.101.026405,PhysRevLett.108.157202,JPhysCondMatt.23.164207,PhysRevB.79.134424} 

In this paper, we study the effect of these further perturbations on various QSL states that potentially offer an explanation of the unusual phenomenology of the non-magnetic ground state in Herbertsmithite. We contrast these states with respect to their signature in spin-spin correlations, which is measured in inelastic neutron scattering measurements, by calculating the dynamical spin-structure factor for a host of candidate $Z_2$ and associated $U(1)$ spin liquid states that are allowed by the projective symmetry group  analysis on the Kagome lattice.\cite{PhysRevB.83.224413} In addition, we calculate the ESR absorption spectra that provide useful information in systems without spin-rotation invariance, and hence can help us to identify the nature of the QSL.\cite{PhysRevB.67.224424,JPSJ.75.104711} ESR absorption resulting from DM interactions has been probed in, for instance, the quasi-two-dimensional system Cs$_2$CuCl$_4$.\cite{PhysRevLett.107.037204} In this compound, the ESR absorption can be interpreted in terms of deconfined spin-$1/2$ spinons, in contrast with the spin chain Cu benzoate, where bound states dominate the ESR spectra.\cite{PhysRevB.65.134410,PhysRevLett.84.5880}

Due to the technical nature of the results presented here, before discussing further details, we briefly summarize first the current situation of the spin liquid physics on Kagome antiferromagnets in view of Herbertsmithite, and, second, the main results obtained in this work.
\subsection{Summary of results: Spin liquid physics in Herbertsmithite}
\label{subsec:summary}

The current theoretical framework of constructing QSL states is largely based on slave-particle mean-field theories.\cite{PhysRevB.37.580,Baskaran1987973,PhysRevB.37.3774,PhysRevB.38.5142,wen2004quantum,PRB.65.165113,RevModPhys.78.17} In particular, we shall focus on the slave-fermion mean-field theory where each spin-1/2 operator is written in terms of a fermion bilinear: $\v{S}_i = \frac{1}{2} f^{\dag}_{i\alpha}\gv{\sigma}_{\alpha\beta}f_{i\beta}$ [also see Eq. \ref{eqn:spinoperator}] with the constraint $\sum_{\alpha}f_{i\alpha}^{\dag}f_{i\alpha} = 1$.  These fermions, called spinons, carry spin-1/2 and form the low energy quasi-particles whose excitation spectrum, as we shall see below, mainly characterizes the low energy response of a QSL state. The slave-particle representation introduces gauge redundancy (see Eq. \eref{eq:gaugerotate}), and hence the spinons are minimally coupled to an emergent gauge field. Depending on the structure of the gauge group, we can broadly classify quantum spin liquids as $SU(2)$, $U(1)$ or $Z_2$ spin liquids.

 The spin-spin interactions give rise to quartic spinon interactions, which, within mean-field theory, are decoupled in the spinon hopping (particle-hole) and pairing (particle-particle) channels (see Sec. \ref{sec:meanfielddecoupling}). These decoupling channels together constitute a QSL ansatz, and different mean field QSL states are characterized by the structure of these ans\"atze. However, the above gauge redundancy suggests that the mean-field states are only projective representations of physical states. Hence, the same physical state may be represented by a set of {\it gauge equivalent} mean-field ans\"atze. \cite{wen2004quantum,PRB.65.165113} Different mean-field spin liquid ans\"atze form different projective representations of the lattice (and time-reversal) symmetries of the Hamiltonian, and the spinons also transform accordingly. Thus, a systematic understanding of distinct QSLs requires a careful analysis of the projective symmetry group (PSG) \cite{wen2004quantum,PRB.65.165113} In the presence of both spinon hopping and pairing terms, the mean field calculations for the spinon ground state can be cast in the form of a BCS type Hamiltonian for the spinons [see Sec. \ref{sec:meanfielddecoupling}, Eq. \eref{eq:bcsform}]. If the Hamitonian is spin-rotation symmetric, the hopping and the pairing channels can be completely separated into a spin independent ({\it{singlet}}) part and a spin dependent ({\it{triplet}}) part [Section \ref{sec:meanfielddecoupling}]. In this work we refer to them as {\it{singlet}} and {\it{triplet}} channels (or terms) respectively. In usual studies of models with frustrated antiferromagnetic spin-rotation invariant exchanges, the {\it{singlet}} ans\"atze give rise to stable QSL mean field solutions and hence the {\it{triplet}} channels are absent. Thus, we obtain symmetric QSLs that do not spontaneously break spin-rotation symmetry. However, it is known that in the presence of competing ferromagnetic and antiferromagnetic interactions, both singlet and triplet channels can exist (and may give rise to exotic three dimensional gapped $U(1)$ spin liquid states with topologically protected gapless surface spinon modes\cite{PhysRevB.80.064410,PhysRevB.85.224428}).

A PSG analysis\cite{PhysRevB.83.224413} of the {\it{singlet}} ans\"atze on Kagome lattice shows that there are {\it twenty} symmetric $Z_2$ QSLs (for Kramers doublets) possible on the kagome lattice that do not break any lattice symmetry. Out of these, {\it only} eight $Z_2$ QSL states with gapped or gapless fermionic spinons that can be stabilized within a spin liquid ansatz that contain both NN and NNN hopping and pairing terms for the spinons. These eight $Z_2$ QSLs are found in the neighborhood of {\it only} four different parent $U(1)$ QSLs of the many $U(1)$ states possible on the kagome lattice. Also one of these parent $U(1)$ QSL, one with Dirac-like spectrum, has been suggested to be the ground state of the NN-HKAF model in the recent VMC calculations.\cite{PhysRevB.87.060405}  Hence we start by looking at these parent $U(1)$ QSLs. Since the above spin liquids are time reversal symmetric, there exists a gauge where the singlet $U(1)$ QSL mean-field ans\"atze can be written employing NN real, singlet hopping  channel. In this gauge, the $U(1)$ QSLs are conveniently labeled using the following notation: $U(1)[a,b]$,\cite{PhysRevB.83.224413} where $a$ ($b$) denotes the magnetic flux of the emergent gauge field through the triangular (hexagonal) plaquette of the Kagome lattice. Time reversal symmetry constrains these fluxes to be either $0$ or $\pi$. Accordingly, we have $U(1)[0,0]$, $U(1)[0,\pi]$, $U(1)[\pi,\pi]$, and $U(1)[\pi,0]$ QSLs.\cite{PhysRevB.83.224413,PhysRevB.77.224413} From these four $U(1)$ QSLs, the eight $Z_2$ QSL are obtained by tuning in appropriate spinon-pairing terms (up to second nearest neighbor) allowed by PSG. These pairing terms necessarily break the gauge structure from $U(1)$ down to $Z_2$, hence the name. The nomenclature for the $Z_2$ spin liquids is derived straightforwardly from their parent $U(1)$ states.\cite{PhysRevB.83.224413} The added Latin/Greek letter distinguishes between two or more $Z_2$ QSL obtained from the same $U(1)$ parent.\cite{PhysRevB.83.224413}  Both the $U(1)$ and $Z_2$ QSLs have characteristic spinon band structure. We note that, while this spinon dispersion is not a gauge invariant quantity, it is still useful to understand its general features in order to capture various properties of these states like the nature of the experimentally observable spin-structure factor.\cite{PhysRevLett.100.227201} Hence, we summarize the general characteristic features of the spinon spectra (discussed in detail in Section \ref{sec:singletmfansatze}) in the second column of Table \ref{tab:ansatzelowenergyspinons}. 

More recent work, however, indicates that the NN-HKAF model likely sits close to a quantum phase transition between a $Z_2$ spin liquid and a valence-bond-solid state on tuning the second-neighbor interaction or a magnetically ordered state on tuning the Dzyaloshinsky-Moriya interaction.\cite{PhysRevB.81.144432,NewJPhys.14.115031,PhysRevLett.109.067201,NatPhys.8.902,PhysRevB.66.014422,PhysRevB.81.064428,PhysRevB.78.140405} It is this proximity to the quantum phase transition that may make the system very sensitive to small perturbations which then can have a sizable effect.\cite{PhysRevB.81.144432} The perturbations to the HKAF seen in Herbertsmithite are then likely to be relevant to a proper description of the material, and we endeavour to gain an understanding of how the aforementioned {\it{symmetric}} spin liquid states are affected under these perturbations, particularly the breaking of spin-rotation symmetry induced by DM interaction and Ising anisotropies. 

\begin{table}
\centering
\begin{tabular}{|c||c|c|}
\hline
QSL Label & Singlet Ansatz\cite{PhysRevB.83.224413} & ``Singlet + Triplet'' Ansatz
\\
\hline
\hline
$U(1)[0,0]$ & Fermi surface (F.S.) & F.S. is altered \\
            &                      & but not gapped out\\ \hline

$U(1)[0,\pi]$ & Dirac point (D.Pt.)  & D. Pt. is gapped out ($*$)\\ \hline

$U(1)[\pi,\pi]$ &  Flat bands & Bands acquire dispersion, \\
				& 		\& D.Pt.			  & D.Pt. remains intact \\
				\hline
$U(1)[\pi,0]$ &  Flat bands & Bands acquire dispersion \\\hline
\hline
$Z_2[0,0]A$ & F.S. gapped out to & Band touching points\\ 
& Band touching points & remain\\ \hline
$Z_2[0,\pi]\beta$ & F. S. becomes & Gap is\\
&fully gapped & altered\\ \hline
$Z_2[0,0]B$ & F.S. shrinks compared & F.S. gapped out to\\
&  to parent $U(1)$& band touching points\\ \hline
$Z_2[0,\pi]\alpha$ & D.Pt. changes to & Band touching points\\
& band touching points & remain\\ \hline 
$Z_2[0,0]D$ & F.S. shifted compared & F.S. gapped out to\\
& to parent $U(1)$ & band touching points\\ \hline
$Z_2[0,\pi]\gamma$ & D.Pt. changes to & Band touching points \\
& band touching points & remain \\ \hline
$Z_2[\pi,\pi]B$ & Negligible change & Bands gapped to D.Pt.\\
& compared to  & and band touching points\\
& parent $U(1)$&\\ \hline
$Z_2[\pi,0]B$ & Flat Bands are gapped & F.S. gapped to \\
& to form a F. S. & band touching point\\
\hline
\end{tabular}
\caption{(* {\it This state is unstable to confinement}).
Structure of low-energy spinon excitations in the $U(1)$ and $Z_2$ under
consideration, in the mean-field parameter regime described in Sec. \ref{sec:meanfielddecoupling}.
These behaviors persist within the neighborhood of the respective parameter sets; it gives a general understanding of these states.
The singlet case shows the evolution from $U(1)$ to $Z_2$ spin
liquids upon addition of the bolded terms in Table \ref{tab:ansatzparameters}.
The ``singlet + triplet'' case shows the evolution of all these states upon
addition of nearest-neighbor triplet terms. For $Z_2$ spin liquids, F.S.
indicates a Fermi surface of Fermionic quasi-particles, obtained upon
diagonalization of the mean-field Hamiltonian $\HM_Q$. The low energy
excitations lie about the spinon Fermi-level which is set by the half filling
constraint on the spinons [Eq. \ref{eq:constspinon}].
}
\label{tab:ansatzelowenergyspinons}
\end{table}

Since the spin-rotation symmetry is broken, the singlet and the triplet spinon decoupling channels (both hopping and pairing) can no longer be separated, and a more general QSL ans\"atze incorporating both singlet and triplet decomposition, and allowing their intermixing, needs to be introduced.\cite{PhysRevB.86.224417} While it is possible to use such a general four-component spinor representation, as in the case of a triplet BCS Hamiltonian, here, we find that since the DM vectors and the Ising anisotropies both are perpendicular to the Kagome plane, we can use a two-component spinor representation. The effect of such spin-rotation symmetry breaking terms on the low-energy excitations of these spin liquids is also summarized in the Table \ref{tab:ansatzelowenergyspinons} (third column). Generally, the triplet terms decrease the density of states at the spinon Fermi-level in most of the $U(1)$ and $Z_2$ QSLs. 

Beyond mean field theory, apart from the gapless/gapped spinon excitations, there exist additional low-energy excitations, depending on the gauge structure. For example, in gapless $U(1)$ QSL in two spatial dimensions, there is an emergent gapless photon. For $Z_2$ spin liquids, the ``magnetic flux" associated with the $Z_2$ gauge field leads to gapped topological excitations called {\it{visons}}.\cite{PhysRevB.62.7850} In this work, we shall be mostly concerned with the question of the nature of the ground state and spinful excitations, such as spinons and spin-1 excitations.

Having identified the spin liquids, we turn to their dynamical spin-structure factor, which is just the Fourier transform of spin-spin correlation function. The dynamical spin-structure factor has been recently measured in inelastic neutron scattering experiments on a single crystal sample of Herbertsmithite. It shows diffuse scattering with no signature of the spin gap down to at least $1.5$ meV.\cite{Nature.492.406}  

In our computations, almost all of the above spin-liquid ans\"atze show
patterns of diffuse intensity in the  dynamical structure factor throughout the
Brillouin zone, the overall shape of which is largely influenced by the  parent
$U(1)$ spin liquid.  The structure factor of the $U(1)$ states have
characteristic flat and dispersive features within the diffuse continuum. In
the $Z_2$ spin liquids, the dispersive features are broadened and become
significantly more complex. A central result is that, for some $Z_2$ spin
liquids, these dispersive features are only slightly stronger than the diffuse
background, so the dynamical structure factor looks mostly diffuse. This is particularly true for the states labeled
$Z_2[0,\pi]\beta$ and $Z_2[0,\pi]\alpha$ [the dynamical structure factor is
shown later in Figs. \ref{fig:dsf-z2-singlet-part-2} (e) and \ref{fig:dsf-z2-singlet-part-2} (f),
respectively]. This may be consistent with present neutron scattering results
where similar diffuse scattering is seen over a large energy window 
in large parts of the Brillouin zone.\cite{Nature.492.406} It
is found that the spin-rotation symmetry breaking terms primarily introduce
non-zero intensity at the center of the Brillouin zone. They also split the spinon bands, so that diffuse
scattering is found more evenly {\it throughout all of the Brillouin zone} in
the presence of DM and Ising interactions. The inelastic neutron scattering
indeed shows some intensity at the center of the Brillouin zone, with a broad
maximum near
$0.4J$. 

To better understand the effect of triplet terms on the ans\"atze we consider
the ESR absorption spectra that show a non-trivial response {\it solely} from
spin-rotation breaking terms. The ESR signal shows the largest absorption
intensity at a $\delta$-function peak at the Zeeman field energy scale. Triplet
terms create additional ``satellite'' absorption peaks, offset by their energy
scale. The number, position, and broadening of these extra absorption peaks
differentiate among the different ans\"atze. Particularly for the
aforementioned $Z_2[0,\pi]\beta$ and $Z_2[0,\pi]\alpha$ states, the absorption
spectrum provides a distinct qualitative features [as in 
Figs. \ref{fig:esr-z2-triplet-part-2} (e) and \ref{fig:esr-z2-triplet-part-2} (f)] that can be verified experimentally.

The rest of this paper is organized as follows. We begin in Sec.
\ref{sec:model} by discussing the spin model used to capture the behavior of
Herbertsmithite. Following the introduction of the spin Hamiltonian, in Sec.
\ref{sec:slavefermion} we describe the fermionic slave-particle construction of
mean-field states. In Sec. \ref{sec:psg} we describe the projective symmetry group
analysis of the $Z_2$ ans\"atze, describing their respective $U(1)$
parent QSL states as well as the effect of the spin-rotation symmetry breaking
terms. The spin correlations of these states are characterized in in
Sec. \ref{sec:spincorrelations}. Finally, in Sec.
\ref{sec:inelasticneutronscattering} we discuss the implications of these
correlations in light of the recent neutron scattering results. We summarize
our results in Sec. \ref{sec:discussion}. The details of different calculations
are given in various appendices.

\section{Spin-$1/2$ Model on the Kagome Lattice}
\label{sec:model}

\begin{figure}
	\centering
\includegraphics[width = 0.9 \linewidth]{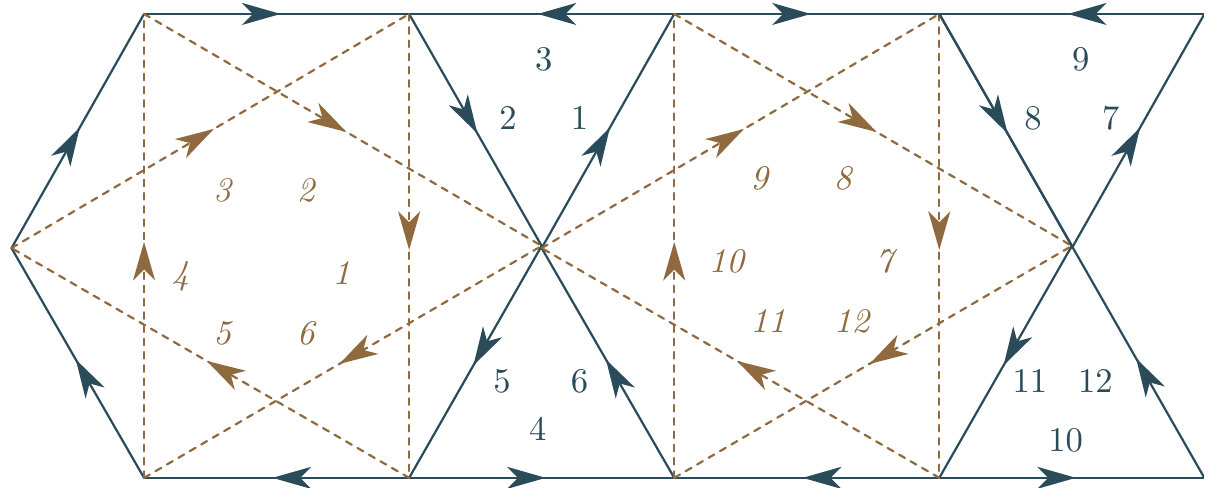}
	\caption{(Color online)
	Bond labels and directions for nearest-neighbor
	(dark blue, solid) and next-nearest-neighbor (light brown, dashed)
	bonds on the Kagome lattice. 
	The nearest-neighbor directions also indicate the
	directions of the Dzyaloshinsky-Moriya terms.
	The bonds labeled 1 are the ones on which mean-field
	parameters are defined as in Table \ref{tab:ansatzparameters}; parameters
	on other bonds are determined from these through the application of
	symmetry operations.
	}
	\label{fig:Kagome-bonds}
\end{figure}

The starting point for our analysis is the spin-$1/2$ model with antiferromagnetic nearest-neighbor Heisenberg interactions on the Kagome lattice,
\begin{align}
		\HM_{\rm{NN-KHAF}} = J\sum_{\avg{ij}} \v{S}_i\cdot\v{S}_j.
		\label{eqn:modelheisenbergpart}
\end{align}
where $J \sim
17$ meV $\sim 200$ K denotes strong antiferromagnetic interactions, and forms the largest energy scale in Herbertsmithite.\cite{PhysRevLett.98.107204,PhysRevLett.98.077204} Experiments also indicate that there may be three kinds of interactions
acting as perturbations to the above Hamiltonian: 
\paragraph{Dzyaloshinsky-Moriya (DM) interaction:} 
\begin{align}
		\HM_{\rm{DM}} = D\sum_{\avg{ij}} \hat{\v{D}}_{ij} \cdot \v{S}_i\times\v{S}_j.
		\label{eqn:modeldmpart}
\end{align}

This has been estimated to have energy scale of $|\v D| \sim 15$ K
(perpendicular to the Kagome planes) from ESR
experiments\cite{PhysRevLett.101.026405} and fitting of the
anisotropy of the magnetic susceptibility data.\cite{PhysRevLett.108.157202} In
Eq. \eref{eqn:modeldmpart}, the orientation of $\v D_{ij}=\hat z$ is out of the
plane when the bonds are counted in a counter-clock wise way around each
triangle (Fig. \ref{fig:Kagome-bonds}).  The in-plane components of the DM
interactions ($\sim 2$ K) appear to be negligible in comparison with the
out-of-plane component, and we will neglect it in our calculations. 

\begin{figure}
	\centering
\includegraphics[width = 0.9 \linewidth]{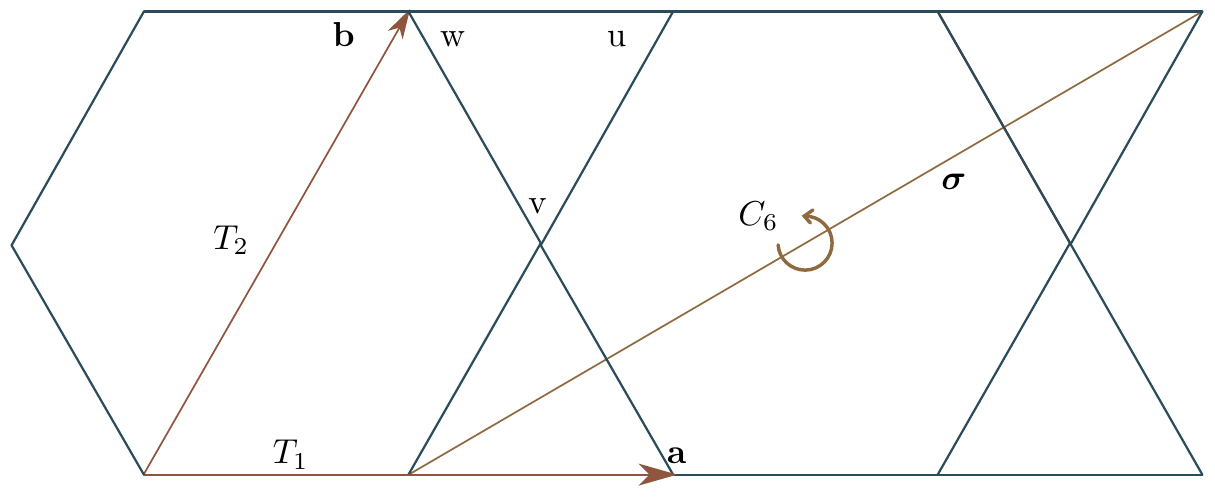}
		\caption{(Color online. Symmetry transformations of the Kagome lattice, including
				translations $T_1$ and $T_2$, a rotation $C_6$ and
				reflection $\gv{\sigma}$.}
	\label{fig:Kagome-transforms}
\end{figure}

\paragraph{XXZ anisotropy:} The susceptibility anisotropy also indicates a
sizable  easy-axis spin-spin interaction along an axis perpendicular to the
Kagome plane, 
\begin{align}
		\HM_{\rm{Ising}} = \Delta \sum_{\avg{ij}} S_i^z S_j^z,
		\label{eqn:modelisingpart}
\end{align}
where $\Delta \sim J/10 \sim 20$ K.
\paragraph{Next-nearest-neighbor coupling:} 
As discussed in the introduction, even small $J_2$ can drastically affect the stability of spin liquids of the HKAF, as found in recent numerical calculations. Accordingly, we consider adding an isotropic next-nearest-neighbor antiferromagnetic Heisenberg interaction $J_2$.
\begin{align}
		\HM_{\rm{NNN}} = J_2\sum_{\avg{\avg{ij}}} \v{S}_i\cdot\v{S}_j.
		\label{eqn:modelnnnpart}
\end{align}

The complete Hamiltonian is then obtained by adding Eqs. \eref{eqn:modelheisenbergpart} -- \eref{eqn:modelnnnpart}. All these terms have full symmetries of the Kagome lattice, shown in Fig. \ref{fig:Kagome-transforms}, and enumerated below. These are as follows:
\begin{itemize}
\item Translations $T_1$ and $T_2$ in the Kagome plane.
\item A sixfold rotation $C_6$ around the center of a hexagon.
\item A reflection $\gv{\sigma}$ through a hexagon.
\end{itemize}

With this Hamiltonian, we now construct possible $Z_2$ spin liquid states with fermionic spinons, both gapped and gapless, and then contrast their properties in context of experiments on Herbertsmithite. We shall limit ourselves to $Z_2$ spin liquid ans\"atze that can be realized with just first- and second-neighbor terms (both hopping and pairing). Since mean-field theory only gives an order of magnitude estimate of the microscopic parameters, we shall investigate the existence of the above spin liquids in somewhat a broad range of the parameters, especially $J_2$, whose effect is known to be quite sensitive in stabilizing $Z_2$ spin liquids. \cite{NatPhys.8.902}
\section{Slave-Fermion Construction of Quantum Spin Liquid States}
\label{sec:slavefermion}

\subsection{Spinon Representation}

In the slave-fermion formalism, as noted earlier, we represent a spin-$1/2$ operator, at a site $i$, in terms of a fermion (spinon) blinear:
\begin{align}
		\v{S}_i = \frac{1}{2} f^{\dag}_{i\alpha}\gv{\sigma}_{\alpha\beta}f_{i\beta},
		\label{eqn:spinoperator}
\end{align}
where $\alpha,\beta \in \{\uparrow,\downarrow\}$ are the two flavours of spinons.
\cite{Baskaran1987973,PhysRevB.37.3774,PhysRevB.37.580,PhysRevB.38.5142}
Here, $\sigma^{a}~~(a=1,2,3)$ are the Pauli matrices. The Hilbert space of the spinons is twice as large as the original $S=1/2$ Hilbert space. The physical spin wave-functions are obtained by projecting the spinon wave-function in the subspace of single spinon per site. This is the equivalent of having the following constraints for every lattice site: 
\begin{align}
		f^{\dag}_{i\alpha}f_{i\alpha} = 1 \nonumber \\
		f_{i\alpha}f_{i\beta}\epsilon_{\alpha\beta}=  
		f^{\dag}_{i\alpha}f^{\dag}_{i\beta}\epsilon_{\alpha\beta}=0,
		\label{eq:constspinon}
\end{align}
where $\epsilon_{12}=-\epsilon_{21}=1$ is the $2\times2$ completely antisymmetric tensor.

\subsection{Mean-Field Decoupling}
\label{sec:meanfielddecoupling}

With the above spinon representation, we can re-write the bilinear spin Hamiltonian of Sec. \ref{sec:model}  as a quartic fermion Hamiltonian in terms of the spinon operators. We write the Heisenberg, DM and Ising anisotropic exchange interactions on the nearest neighbor bonds as in the following:
\begin{align}
&\v{S}_i\cdot \v{S}_j = 
\frac{1}{16}\left(-3\hat{\chi}_{ij}^{\dag} \hat{\chi}_{ij} + \hat{\v{E}}_{ij}^{\dag} \cdot \hat{\v{E}}_{ij}\right)\nonumber\\
&~~~~~~~~~~~+\frac{1}{16}\left(-3 \hat{\eta}_{ij}^{\dag} \hat{\eta}_{ij} + \hat{\v{Y}}_{ij}^{\dag}\cdot \hat{\v{Y}_{ij}}\right),
\\
&\hat{z} \cdot \v{S}_i \times \v{S}_j =\frac{i}{8} (\hat{\chi}^{\dag}_{ij}\hat{E}^z_{ij} -  \hat{E}^{z\dagger}_{ij}\hat{\chi}_{ij} )+ \frac{i}{8} (\hat{\eta}^{\dag}_{ij}\hat{Y}^{z}_{ij} -  \hat{Y}^{z\dagger}_{ij}\hat{\eta}_{ij} ),
\\
&S^z_i S^z_j =\frac{1}{16} (- \hat{\chi}_{ij}^{\dag} \hat{\chi}_{ij}- \hat{E}^{z\dag}_{ij} \hat{E}^z_{ij}+ \hat{E}^{x\dag}_{ij} \hat{E}^x_{ij}+ \hat{E}^{y\dag}_{ij} \hat{E}^y_{ij})\nonumber\\
&~~~~~~~~~+ \frac{1}{16} (- \hat{\eta}_{ij}^{\dag} \hat{\eta}_{ij}- \hat{Y}^{z\dag}_{ij} \hat{Y}^z_{ij}+ \hat{Y}^{x\dag}_{ij} \hat{Y}^x_{ij}+ \hat{Y}^{y\dag}_{ij} \hat{Y}^{y}_{ij}).
\label{eqn:fourfermionterms-ising}
\end{align}
We have defined
\begin{align}
\hat{\chi}_{ij} = f^{\dag}_{i\alpha} \delta_{\alpha\beta} f_{j\beta},\ \ \ \ \hat{E}_{ij}^{a} = f^{\dag}_{i\alpha} \tau_{\alpha\beta}^a f_{j\beta}\nonumber\\
\hat{\eta}_{ij} = f_{i\alpha} (i\tau^2)_{\alpha\beta} f_{j\beta},\ \ \ \ \hat{Y}^a_{ij} = f_{i\alpha} (i\tau^2\tau^a)_{\alpha\beta} f_{j\beta}m,
\end{align}
$(a=x,y,z)$ as the singlet and triplet hopping (particle-hole) channels and
singlet and triplet pairing (particle-particle) channels, respectively.  Note
that unlike the usual slave fermion decomposition, where only the singlet
decoupling channels are used, we have introduced both singlet and triplet
decoupling channels. As discussed earlier, since spin-rotation symmetry is
already broken in the Hamiltonian, there is no {\it a priori} reason to choose only
the singlet or triplet channel. Indeed, in the following we shall see that the
triplet channel plays an important role in our calculations. 

These quartic operators are then decomposed by introducing eight auxiliary fields corresponding to $\chi,\eta,\v E$ and $\v Y$ for all the particle-hole and particle-particle channels. The saddle point approximation in terms of these auxiliary fields then gives the mean-field quadratic fermion Hamiltonian, $\HM_Q$, characterized by the non-zero decoupling channels.

In the singlet case, $\HM_Q$ is conveniently written using a two-component Nambu basis  $\nu_i$, which most clearly reveals the underlying gauge redundancy of $\HM_Q$. The Hamiltonian is given by\cite{PhysRevB.37.3774,PRB.65.165113,RevModPhys.78.17,wen2004quantum}
\begin{align}
	\HM_Q^{\rm sing} = \sum_{ij} \nu_i^{\dag} U_{ij}^{\rm sing} \nu_j,
\label{eq:bcsform}
\end{align}
where the $U_{ij}$ matrix satisfies $U_{ji} = U_{ij}^{\dag}$, and
\begin{align}
	\nu_i = 
	\begin{pmatrix}
	f_{i\uparrow} \\ f_{i\downarrow}^{\dag}
	\end{pmatrix},
	~
	U_{ij}^{\rm sing} = \begin{pmatrix}
				\chi^*_{ij}  & -\eta_{ij}   \\
				-\eta^*_{ij}  & -\chi_{ij} 
		\end{pmatrix}.
		\label{eqn:singletuij}
\end{align}

The exact constraint of single spinon per site [Eq \eref{eq:constspinon}] is then relaxed to an average constraint and imposed by Lagrange multipliers, given by\cite{PhysRevB.37.3774,PRB.65.165113,RevModPhys.78.17,wen2004quantum}
\begin{align}
 \mu^a\sum_{a}  \nu^{\dag}_i \tau^a \nu_i.
\end{align}
where $\mu^a$ ($a=0,1,2,3$) are the Lagrange multipliers. We note that the Nambu ($\nu_i$) basis components involve operators of the same spin type. This form clearly demonstrates the explicit $SU(2)$ gauge redundancy,\cite{PhysRevB.37.3774,PRB.65.165113,RevModPhys.78.17,wen2004quantum}
\begin{align}
 &\nu_i \to W_i \nu_i\nonumber\\
 &U_{ij} \to W_iU_{ij}W_j^{\dag}~~~~ \forall W_i \in SU(2).
 \label{eq:gaugerotate}
\end{align} 

The introduction of triplet terms necessitates, in principle, a four-component basis.\cite{PhysRevB.86.224417,PhysRevB.80.064410} However, since in the following we will use only the $z$-components of the triplet terms, we may use the same basis $\nu_i$ as the singlet case,  writing terms with the matrix form $U_{ij}^{\rm trip}$, given as
\begin{align}
	U_{ij}^{\rm trip} = \begin{pmatrix}
		E^{z^*}_{ij} & Y^z_{ij} \\
		- Y^{z^*}_{ij} & E^z_{ij}
		\end{pmatrix}.
		\label{eqn:tripletuij}
\end{align}
The quadratic form for the triplet part is now given by $\nu_i^{\dag} U_{ij}^{\rm trip} \nu_j$. Importantly, this form has the same gauge redundancy as the singlet case, so the
projective symmetry group classification will be similar, as to be discussed in
Sec. \ref{sec:psg}.
These terms combine in the quadratic Hamiltonian to give
\begin{align}
\HM_Q = \sum_{ij} \nu_i^{\dag}(U_{ij}^{\rm sing} + U_{ij}^{\rm trip}) \nu_j + h.c.
\label{eqn:basistotaluij}
\end{align}

As pointed out earlier, with the introduction of the DM and Ising anisotropy terms on the NN bonds,  the singlet and triplet channels mix and the saddle point is characterized by a combination of the  two kinds of channels. For example, in the hopping sector, we can rediagonalize the interactions  to write (for NN bonds). 
\begin{align}
	J \v{S}_i \cdot \v{S}_j + \Delta S_i^zS_j^z + D \hat{z} \cdot \v{S}_i \times \v{S}_j
	\Rightarrow \nonumber \\
	\omega_- \hat{\chi}^{\dag}_-\hat{\chi}_-
	+ \omega_+ \hat{\chi}^{\dag}_+\hat{\chi}_+
	+ \frac{J+\Delta}{8} \hat{E^x}^{\dag}\hat{E}^x
	+ \frac{J+\Delta}{8} \hat{E^y}^{\dag}\hat{E}^y,
	\label{eqn:omegaplusminushamiltonian}
\end{align}
where we withhold the $i,j$ subscripts on all operators for the rest of this section, and
\begin{align}
	\omega_{\pm} = -\frac{J+\Delta}{8} \pm \frac{1}{4}\sqrt{J^2+D^2},
	\nonumber \\
	\hat{\chi}_{\pm} = \frac{ \sqrt{J^2+D^2} \mp J}{\xi_{\pm}} \hat{\chi}
	\mp i \frac{D}{\xi_{\pm}} \hat{E}^z,
		\nonumber \\
	|\xi_{\pm}|^2 = 2\sqrt{J^2+D^2}\left(\sqrt{J^2+D^2} \mp J\right).
	\label{eqn:singlettripletcombinedchannels}
\end{align}
For the pairing channels, the same is true with $\eta$ in place of $\chi$, and
$\v{Y}$ in place of $\v{E}$.

Both $x$ and $y$ components of the triplet terms lead to unstable mean-field states, so we will only consider ans\"atze with singlet ($\chi$, $\eta$) and $z$-component triplet ($E^z$, $Y^z$) terms. We can write the mean-field Hamiltonian in the basis described by \eref{eqn:singletuij} and \eref{eqn:tripletuij}. 
On a NN bond, we have
\begin{align}
	\HM_{ij}^{\rm NN} = 
		{\omega_-} \left( \chi^*_-\hat{\chi}_- 
		+ \eta^*_- \hat{\eta}_-\right)  
		+
		{\omega_+} \left( \chi^*_+\hat{\chi}_+ 
		+ \eta^*_+ \hat{\eta}_+\right) 		\nonumber \\
		+ h.c. 
		- \omega_-{|\chi_-|^2+|\eta_-|^2}
		- \omega_+{|\chi_+|^2+|\eta_+|^2}.
\end{align}
On a NNN bond (since there is only antiferromagnetic Heisenberg term, and no $D$ or $\Delta$) we decouple only in the singlet channels:
\begin{align}
	\HM_{ij}^{\rm NNN} = 
		-\frac{3J}{8} \left( \chi^*\hat{\chi} 
		+ \eta^* \hat{\eta}\right)  
		+ h.c. 
		+ \frac{3J}{8} \left({|\chi|^2+|\eta|^2}\right).
\end{align}
The quadratic Hamiltonian 
$\HM_{ij} = \HM_{ij}^{\rm NN} + \HM_{ij}^{\rm NNN}$
can now be solved to obtain the spinon wave-function, $|\psi\rangle_{\rm spinon}$.

To gain insight on the actual values of different mean-field parameters that characterize the different QSL ansatz, we perform self-consistent
mean-field calculations, the details of which are outlined in Appendix
\ref{sec:mfresults}. We note that in the self-consistent calculation, we find
that the next-nearest-neighbor exchange, $J_2$, necessary to stabilize {\it six} of the {\it eight} $Z_2$ spin liquids in Table \ref{tab:ansatzparameters}  is: $J_2 > 0.4 J$.
This large value of $J_2$, required to stabilize non-zero next nearest
neighbor hopping and pairing interactions for the spinons, is certainly an artifact of our mean field treatment.

These mean-field results leading to stable solutions act as a guide for choosing a representative mean-field parameter set for calculation of the dynamical spin-structure factor. The general qualitative features of the structure factor so obtained are expected to be less sensitive to the details of the exact parameter values. In any case it is  likely that the parameter values obtained from a mean-field theory are renormalized by quantum fluctuations. Hence, we take mean-field parameters of singlet $Z_2$ terms wherein pairing and hopping terms are of similar magnitudes, to demonstrate the qualitative features of these $Z_2$ states in comparison to the $U(1)$ ones. We will also take next-nearest-neighbor terms on the order of 0.4 times the magnitude of the nearest-neighbor terms and a triplet-singlet ratio of $0.1 \sim D/J$ when considering a spin-rotation symmetry-breaking ansatz.

\section{Projective Symmetry Group and Mean-Field Ans\"atze}
\label{sec:psg}

The spinon wave-function $|\psi\rangle_{spinon}$ so obtained is not a valid
spin wave-function, since it only satisfies constraint Eq. \eref{eq:constspinon}
on average. Hence, it must be projected into a single-spinon-per-site
subspace to obtain the spin wave-function through numerical Gutzwiller projection.
\cite{wen2004quantum}

Alternatively, it is possible to consider the low energy fluctuations about the
mean-field solution. The gauge redundancy of the spinon description (Eq.
\eref{eq:gaugerotate}) immediately suggests that the low energy effective theory
contains spinons minimally coupled to a lattice gauge field. The structure of
the gauge group can be determined from the structure of the mean field QSL
ansatz for $\HM_Q$\cite{wen2004quantum,PhysRevB.86.224417} in the following way. Each mean-field
ansatz is characterized by $U_{ij}\in SU(2)$ on the bonds [Eqs.
\eref{eqn:singletuij} and \eref{eqn:tripletuij}]. Taking the product of such
$U_{ij}$s over various closed loops ($C$) starting from a specific base site
($i$), we obtain the ``flux'' of the $U_{ij}$ link fields which has the following
form:\cite{PRB.65.165113}
\begin{align}
\mathcal{W}_C(i)=\prod_{(jk)\in C} U_{jk}=\sum_{a=0,x,y,z}A_C^a(i)\tau^a
\end{align}
where $A_C^a(i)$ are numbers specific to the loop and $\tau^a$ are the identity ($a=0$) and Pauli matrices ($a=1,2,3$) in the gauge space. For $U(1)$ QSLs, $A^a(i)$ for different loops are proportional to each other for the same $a$ and at least one of $A^a(i)$ is non-zero for $a=1,2,3$ (say $3$). Then it is possible to write\cite{PRB.65.165113}
\begin{align}
U(1) \text{QSL}:~~~~~ \mathcal{W}_C(i)\propto \tau^3 ~~~~\forall C
\end{align}
This means that there exists a gauge in which all $U_{ij}\propto \tau^3$ and hence $\HM_Q$ of the form in Eq. \eref{eq:bcsform} is invariant under all gauge transformation: $\nu_i\rightarrow e^{i\theta_i^z\tau^3}\nu_i$ where, $\theta_i\in [0,2\pi]$. Thus the gauge group is $U(1)$ and accordingly we have a $U(1)$ QSL.\cite{PRB.65.165113}

For $Z_2$ spin liquids, two or more loops based at the same site have
$A_C^a(i)$ that are not proportional to each other and a similar analysis as
above shows that the gauge transformation of the following form is only
possible: $\nu_i\rightarrow\zeta_i\nu_i$, where $\zeta_i=\pm
1$.\cite{PRB.65.165113} This is a $Z_2$ gauge group describing $Z_2$ spin
liquids. The above argument can be extended to the case of the most general
triplet ansatz with a four-component spinon representation.\cite{PhysRevB.86.224417}

The gauge group so obtained is called the invariant gauge group, (IGG) because the mean
field ansatz remains invariant under such gauge transformations. Due to the
above gauge redundancy, the lattice symmetries (and time reversal) under which
the Hamiltonian is invariant act projectively on the mean-field ans\"atze. This
means that usual physical symmetry transformations on the ansatz transforms it
to its gauge equivalent form; hence an added set of gauge transformation is
required following the actual symmetry transformation to bring the ansatz back to
itself. The PSG characterizes, for a given set of
symmetry transformations
SG, the possible distinct sets of gauge transformations $\{G_S\}$ for each operation $S \in \rm{SG}$, where 
\begin{align}
(G_S S) U_{ij}(G_S S)^\dagger = U_{ij}.
\end{align}
A particular set $\{G_S\}$ characterizes a corresponding $\HM_Q$ and hence the
mean field ground state. Therefore, elements of the IGG are the gauge
transformations $G_I$ associated with the identity operation $I$ of the symmetry
group SG. In other words, the symmetry group $\rm{SG} = \rm{PSG}/\rm{IGG}$. One
can choose a gauge where the elements of the IGG are site-independent,
namely $G_I(i) = \pm I$ for $Z_2$  transformations, $G_I(i) = e^{i \theta
\tau^3}$ ($\theta \in [0,2\pi)$) for $U(1)$ transformations, and $G_I(i) = e^{i
	\theta \hat{n} \cdot \vec{\tau}}$ ($\hat{n} \in S^2$, $\theta \in
	[0,2\pi)$) for $SU(2)$ transformations, characterizing $Z_2$, $U(1)$ or
		$SU(2)$ spin liquids, respectively.\cite{PRB.65.165113}

Determination of the allowed sets $\{G_S\}$, and thereafter the ans\"atze
$\HM_Q$, comes from the group multiplication table of SG. For every group
multiplication rule written in the form $ABC^{-1} = I$, a similar constraint
is put on the PSG, namely $(G_AA)(G_BB)(G_CC)^{-1} \in IGG$, a pure gauge
transformation. These constraints determine the allowed PSG $\{G_SS\}$ for the
symmetry group of the Kagome lattice. 
The singlet $Z_2$ spin liquids have been
determined by Lu {\it et. al.},\cite{PhysRevB.83.224413} and the same PSGs
apply in the triplet case, since we use the same basis \eref{eqn:singletuij}.
However, the resulting ans\"atze 
differ between these  cases, primarily in the existence and structure of the triplet terms. We
will leave these details to Appendix \ref{sec:psgdetails}, 
and summarize the results here.

\subsection{Singlet Mean-Field Ans\"atze}
\label{sec:singletmfansatze}

In this section, we characterize the singlet spin liquid ans\"atze on the Kagome lattice that are of interest to us.\cite{PhysRevB.83.224413} These include the eight $Z_2$ spin liquids that occur in
the vicinity of one of four different parent $U(1)$ spin liquids and can be stabilised by tuning spinon hopping and pairing terms up to the second-nearest-neighbor sites. We shall begin by briefly describing the spinon excitation spectra of the four parent $U(1)$ spin liquids at the nearest-neighbor level. The $Z_2$ states are obtained upon appropriate introduction of additional hopping and pairing terms that break the IGG to $Z_2$, as are shown in Table \ref{tab:ansatzparameters}.

\begin{figure}
\centering
\includegraphics[scale=0.35]{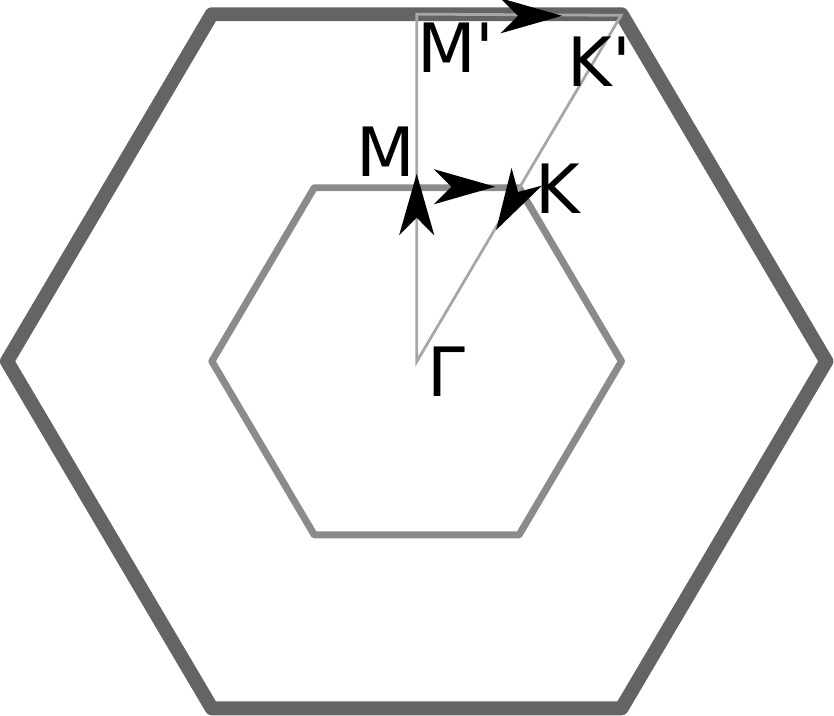}
\caption{The cut along the high symmetry direction of the primitive and extended first Brillouin zone is shown respectively. The boundary of the primitive (extended) Brillouin zone is shown in light (dark) gray. While the spinon dispersion is periodic within the primitive Brillouin zone, the spin-structure factor is periodic in the extended Brillouin zone (see text for details). 
}
\label{fig:bzcut}
\end{figure}

The nomenclature of the four $U(1)$ QSLs have already been introduced in Section \ref{subsec:summary}. These are identified as $U(1)[0,0]$, $U(1)[0,\pi]$, $U(1)[\pi,\pi]$, and $U(1)[\pi,0]$ phases. In Appendix \ref{sec:u1fluxansatze} we show the signs of the hopping parameters used (in the gauge that we have chosen). Here we summarize the features of their spinon spectrum in the above gauge. We note once again that though gauge dependent, the spinon spectra provides valuable clues to the nature of the experimentally measurable and gauge invariant spin-spin correlation functions. 
We plot the dispersion along the high symmetry direction $\Gamma\rightarrow M\rightarrow K\rightarrow \Gamma$ of the original Brillouin zone (Fig. \ref{fig:bzcut}).

\begin{figure}
	\centering
\includegraphics[width = 0.9 \linewidth]{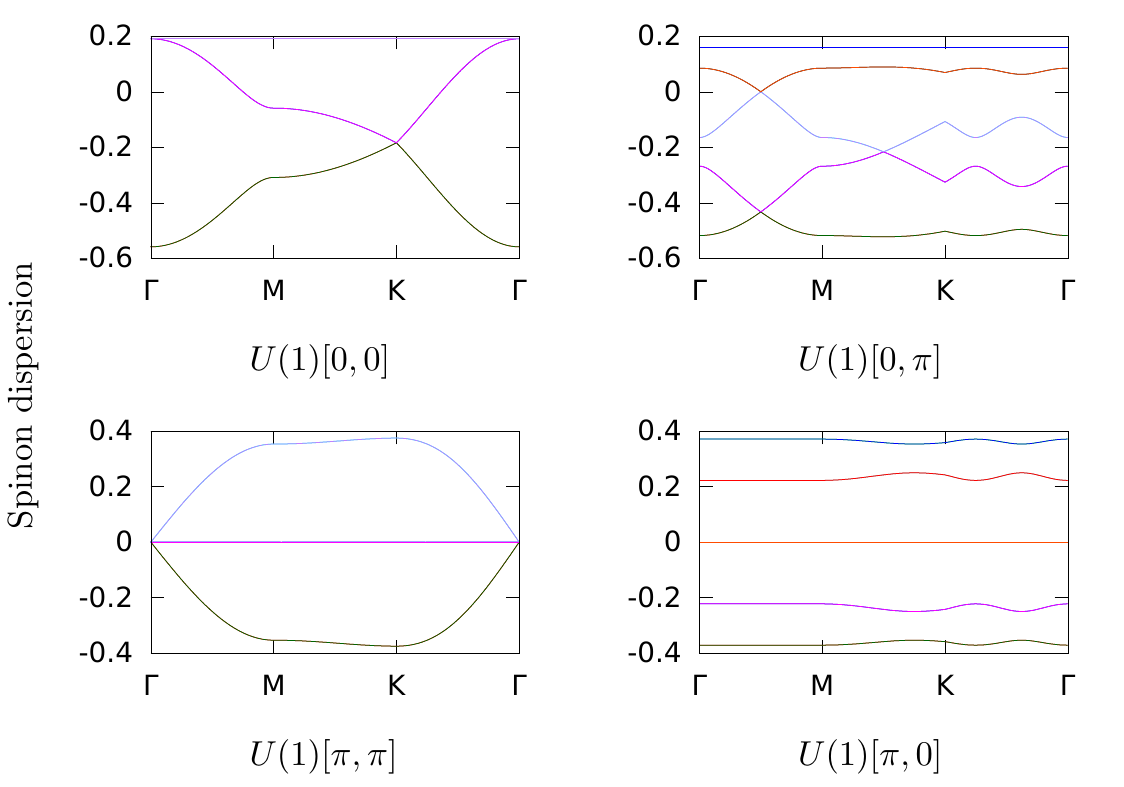}
\caption{(Color online) 
Spinon dispersion for
$U(1)$ singlet spin liquid ans\"atze.
These dispersions go from the center of the Brillouin zone
of the Kagome lattice, $\Gamma$, to the edge $M$, to
the corner $K$, and back to the center $\Gamma$.
The Fermi level is at zero energy.
}
\label{fig:dispersions}
\end{figure}

\paragraph*{$U(1)[0,0]$:} The uniform $U(1)[0,0]$ state has a Fermi surface of spinons. Since this state has uniform hopping, its dispersion is simply the band structure for the Kagome lattice, and is seen in Fig. \ref{fig:dispersions} (a). It has a flat band at the maximum of the dispersion.

\paragraph*{$U(1)[0,\pi]$:} This ansatz breaks translational symmetry of the Kagome lattice, and has Dirac points in its dispersion, as seen in Fig. \ref{fig:dispersions} (b). It has a doubly-degenerate flat band at the maximum of the dispersion, similarly to the $U(1)[0,0]$ case.

\paragraph*{$U(1)[\pi,\pi]$:} This state has a Dirac point at the center of the
Brillouin zone, but also has double-degenerate flat bands at the  Fermi level,
as seen in Fig. \ref{fig:dispersions} (c).

\paragraph*{$U(1)[\pi,0]$: } We find that there are at least two distinct
$U(1)$ states with a $[\pi,0]$ hopping flux pattern. Only one of them can stabilize
a $Z_2$ spin liquid ($Z_2[\pi,0]B$) on addition of the appropriate next-nearest-neighbor hopping
and pairing terms; therefore, this is the state that we will consider.
Like the $U(1)[\pi,\pi]$
state, there are also flat bands at zero energy, as seen in Fig.
\ref{fig:dispersions} (d). However, there is a gap to the subsequent
single-spinon excitations throughout the entire Brillouin zone.

\begin{table}
\centering
\begin{tabular}{|c||c|c|}
\hline
SL Label &  $U_{NN}$ & $U_{NNN}$
\\
\hline
\hline
$Z_2[0,0]A$ &  $\gv\tau^2,\tau^3$ & ${\gv\tau^3}$
\\
$Z_2[0,\pi]\beta$ &  $\gv\tau^2,\tau^3$ & ${\gv\tau^3}$
\\
$Z_2[0,0]B$ &  $\gv\tau^2,\tau^3$ & $\tau^3$
\\
$Z_2[0,\pi]\alpha$ &  $\gv\tau^2,\tau^3$ & $\tau^3$
\\
$Z_2[0,0]D$ &  $\tau^3$ & $\gv\tau^2,\tau^3$
\\
$Z_2[0,\pi]\gamma$ &  $\tau^3$ & $\gv\tau^2,\tau^3$
\\
$Z_2[\pi,\pi]B$ &  $\tau^2$ & $\gv\tau^3$
\\
$Z_2[\pi,0]B$ &  $\tau^2$ & $\gv\tau^3$
\\
\hline
\end{tabular}
\caption{The general structure of the singlet QSL ansatz for the $Z_2$ spin liquids. The particular bonds chosen for $U_{NN}$ and $U_{NNN}$ are
shown in Fig. \ref{fig:Kagome-bonds}. The bold terms shown in $U_{NN}$ and $U_{NNN}$ give the spinon
pairing and hopping terms whose existence is required to break the $U(1)$ parent spin liquid's gauge group down to $Z_2$. We note that we have chosen a gauge in which
$\tau^2$ terms for $Z_2[0,0]A$ and $Z_2[0,\pi]\beta$ ans\"atze
are zero.\cite{PhysRevB.83.224413}}
\label{tab:ansatzparameters}
\end{table}

On adding appropriate spinon hopping and pairing terms up to second nearest
neighbor, the IGG of the above spin liquids is broken down from $U(1)$ to
$Z_2$.\cite{PhysRevB.83.224413} There are eight such $Z_2$ spin liquids as
given in Table \ref{tab:ansatzparameters}. The addition of these terms changes
the structure of the low-energy quasiparticle excitations, with parameters
as discussed in Sec. \ref{sec:meanfielddecoupling}. In some cases, a gap
is opened ($Z_2[0,\pi]\beta$). In others, the line degeneracy of zero-energy
excitations from the $U(1)$ state's Fermi surface remains ($Z_2[0,0]B$ and
$Z_2[0,0]D$). For some other states, only band touching points
remain ($Z_2[0,0]A$, $Z_2[0,\pi]\alpha$ and $Z_2[0,\pi]\gamma$). The flat bands at zero
energy of the $U(1)[\pi,\pi]$ and $U(1)[\pi,0]$ states either remain
($Z_2[\pi,\pi]B$) or are lifted to a line degeneracy ($Z_2[\pi,0]B$). Table \ref{tab:ansatzelowenergyspinons} summarizes the structure of
the low-energy quasiparticles for the above $U(1)$ and $Z_2$ spin liquids. The
full form of $\{G_S(i)\}$ is given in Appendix \ref{sec:psgdetails}.

\subsection{``Singlet + Triplet'' Mean-Field Ans\"atze}

All of the above spin liquids on the Kagome lattice are also realized even in
absence of spin-rotational symmetry,  if time-reversal and the symmetries of
the Kagome lattice are preserved. The allowed $U_{ij}$
now has both real singlet terms and imaginary $z$-component triplet
terms for the nearest-neighbor hopping and pairing, and real singlet terms for
next-nearest-neighbor as well.

\begin{figure}
	\centering
\includegraphics[width = 0.9 \linewidth]{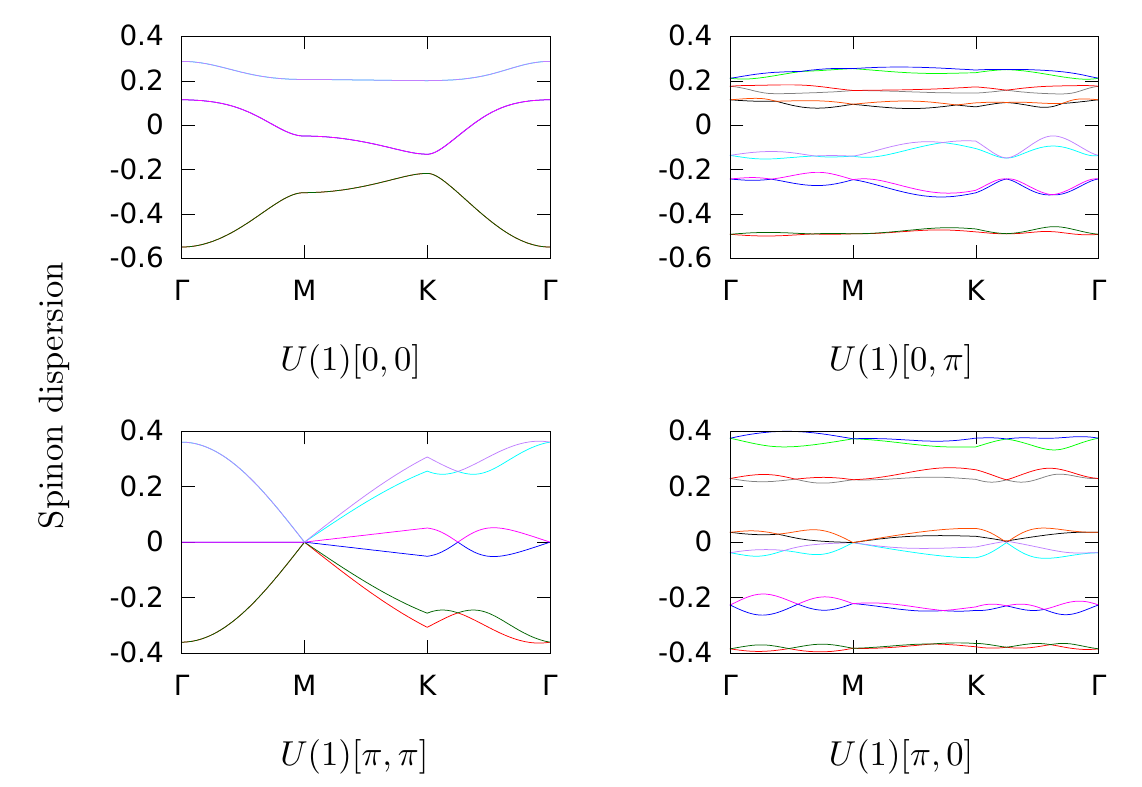}
\caption{(Color online) 
Spinon dispersion for 
U(1) spin liquid ans\"atze
with ``singlet + triplet'' channels for NN bonds.
These dispersion go from the center of the Brillouin zone
of the Kagome lattice, $\Gamma$, to the edge $M$, to
the corner $K$, and back to the center $\Gamma$. The Fermi level always lies at the zero of the energy scale.
We note that since the ``singlet + triplet'' $U(1)[0,\pi]$ spin liquid state is gapped, it is unstable to confinement due to instanton tunneling events.\cite{polyakov}
}
\label{fig:dispersions-triplet}
\end{figure}

Again, we first consider the effect of the triplet terms on the four $U(1)$ spin liquids. In all cases, the spectra show additional features in their dispersions, coming from the triplet terms that removes the flat bands previously present throughout the Brillouin zone. The $U(1)[0,0]$ state has an altered Fermi surface,  which does not split the $f_{\uparrow}$ and $f_{\downarrow}$ spinons [Fig. \ref{fig:dispersions-triplet} (a)]. The Dirac cone in the $U(1)[0,\pi]$ state [Fig. \ref{fig:dispersions} (b)]. however, is gapped out [Fig. \ref{fig:dispersions-triplet} (b)]. Further, the bands are also spin split throughout the Brillouin zone. 
We note that such a two-dimensional gapped $U(1)$ spin liquid is unstable toward a confinement transition.\cite{PhysRevB.77.224413,polyakov} So we infer that triplet decoupling may indirectly render the $U(1)[0,\pi]$ state unstable. The $U(1)[\pi,\pi]$ and $U(1)[\pi,0]$ states both have their zero-energy flat bands mostly gapped out, leaving a filled band with a line degeneracy at zero energy. The Dirac cone in the $U(1)[\pi,\pi]$ state remains intact [Fig. \ref{fig:dispersions-triplet} (c)], while newer Dirac nodes are created for the $U(1)[\pi,0]$ state on including the triplet terms [Fig. \ref{fig:dispersions-triplet} (d)].

Allowing for small triplet terms in the singlet $Z_2$ ans\"atze also changes
the low-energy structure of the quasiparticles. 
The $Z_2[0,\pi]\beta$ state has its gap altered. For states with line degeneracies, we
find that only band touching points remain ($Z_2[0,0]B$, $Z_2[0,0]D$ and $Z_2[\pi,0]B$).
Also, the band touching points are found to be robust to spin-rotation symmetry breaking perturbations ($Z_2[0,0]A$, $Z_2[0,\pi]\alpha$) and $Z_2[0,\pi]\gamma$).
In the state $Z_2[\pi,\pi]B$, the flat bands at zero energy are lifted to band touching 
points, while the Dirac node persists. These changes are summarized in Table \ref{tab:ansatzelowenergyspinons}.

With this, we have completed the description of the candidate QSL states. We shall now calculate the dynamical spin-structure factor and study their broad features that can help us identifying the nature of the spin liquid, possibly realized in Herbertsmithite, from inelastic neutron scattering and ESR experiments.

\section{Spin Correlations in $U(1)$ and $Z_2$ Spin Liquids}
\label{sec:spincorrelations}

\subsection{Structure Factor and Experimental Probes}
\label{sec:structurefactorandprobes}

To determine the spin correlations, we will calculate the dynamical structure factor; the matrix is  given by 
\begin{align}
		S^{\alpha\beta}(\v{q},\omega) = \int_{-\infty}^{\infty} \frac{dt}{2\pi} e^{i\omega t} 
		\sum_{ij} e^{i\v{q}\cdot (\v{r}_i-\v{r}_j)}
		\avg{\v{S}_i^{\alpha}(t)\v{S}_j^{\beta}(0)},
		\label{eqn:dynamicalstructurefactor}
\end{align}
where $\alpha,\beta \in \{1,2,3\}$. It characterizes spin-1 magnetic excitations of energy $\omega$ and wavevector $\v q$. Up to a magnetic form factor, $S^{\alpha\beta}(\v{q},\omega)$ can be measured directly by inelastic neutron scattering.\cite{Nature.492.406} We note that Eq. \eref{eqn:dynamicalstructurefactor} is not periodic in the reciprocal lattice vectors of the Kagome lattice. Since $r_i - r_j$ can be half of the primitive lattice vectors $\v a$ or $\v b$ in Fig. \ref{fig:Kagome-transforms}, we must extend the Brillouin zone to double its size. We will plot the dynamical structure factor along the path $\Gamma \to M' \to K' \to \Gamma$ in the extended Brillouin zone, where $\Gamma$ is the center, $M'$ is the midpoint of an edge, and $K'$ is the corner of the edge of the extended Brillouin zone (EBZ) (Fig. \ref{fig:bzcut}). We found that the equal-time ($\omega$-integrated) structure factor shows little qualitative differences among our ans\"atze, so we concentrate on the $\omega$-resolved features for different QSL states. Further, for the singlet ans\"atze, since the quadratic Hamiltonian $\HM_Q$ commutes with the total spin operator and the ground state is an eigenstate of the total spin with eigenvalue $\v 0$. Thus, the dynamical structure factor of singlet ans\"atze is zero at $\v q = \v 0$.

The dynamical structure factor can also be probed by ESR, under an external Zeeman field $H_Z \sum_i {\v S}_i \cdot \hat{z}$. Absorption of a transverse microwave field (along axis $\alpha$) of frequency $\omega$ probes the long-wavelength $\v q \approx \v 0$ dynamical structure factor $S^{\alpha\alpha}(\v 0,\omega)$.\cite{PhysRevB.67.224424}
The absorption intensity is given by
\begin{align}
		I(\omega) = \frac{H_m^2\omega}{2}\left(1-e^{-\beta\omega}\right)S^{\alpha\alpha}(\v{0},\omega),
	\label{eqn:esrabsorbtion}
\end{align}
where the field is applied in the $\alpha$ direction, with amplitude $H_m$ and frequency $\omega$.\cite{JPSJ.75.104711} The $SU(2)$-invariant terms affect the intensity trivially;
that is, $I(\omega) \propto \delta(\omega-H_Z)$, where $H_Z$ is the Zeeman field strength. Hence, the ESR line shape is sensitive to spin anisotropy, and so the triplet terms.  Thus, ESR experiments can reveal important information about the effect of the spin-rotation symmetry-breaking perturbations.

\subsection{Dynamical Structure Factor of Singlet $U(1)$ Spin Liquid Ans\"atze}

We will begin by considering the singlet $U(1)$ spinon ans\"atze. Our choice of color scale is made to accentuate the low-intensity scattering compared to the zero-intensity background.
\begin{figure*}
	\centering
		\begin{minipage}[]{ \columnwidth}
		\includegraphics[width=\columnwidth]{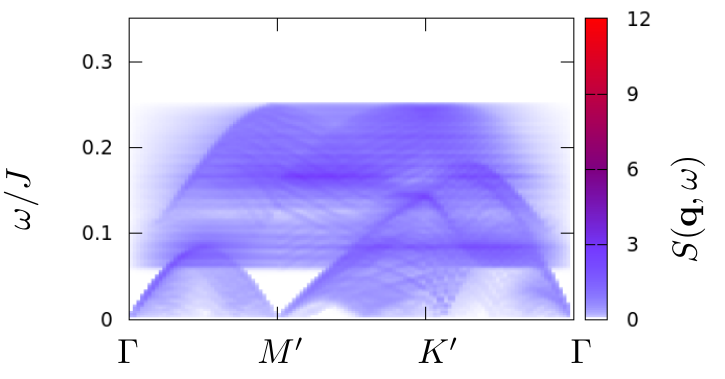}
		\caption{(Color online)
			Dynamical structure factor of the NN  
		singlet $U(1)[0,0]$ spin liquid along the high symmetry direction of the extended Brillouin zone. The characteristic low-energy domelike structure results from the spinon excitations near the Fermi surface [Fig. \ref{fig:dispersions} (a)].}\label{fig:singletdsfU1-00}
		\end{minipage}
		\hspace{0.5 \columnsep}
		\begin{minipage}[]{ \columnwidth}
		\includegraphics[width=\linewidth]{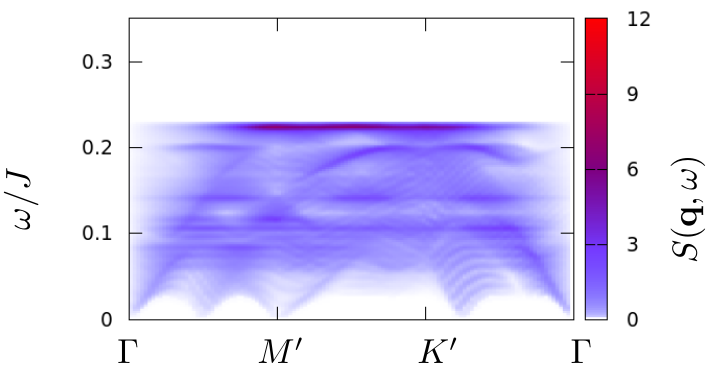}
		\caption{(Color online)
		Dynamical structure factor of the nearest-neighbor singlet
		$U(1)[0,\pi]$ spin liquid along high symmetry directions of EBZ. Compared to $U(1)[0,0]$ state, the low-energy continuum near $\Gamma$ and $M'$ points are replaced with cones of scattering which is a consequence of the Dirac node at the Fermi level for the spinons [Fig. \ref{fig:dispersions} (b)].}
	\label{fig:singletdsfU1-0pi}
		\end{minipage}
\end{figure*}
\begin{figure*}
	\centering
		\begin{minipage}[]{ \columnwidth}
		\includegraphics[width=\linewidth]{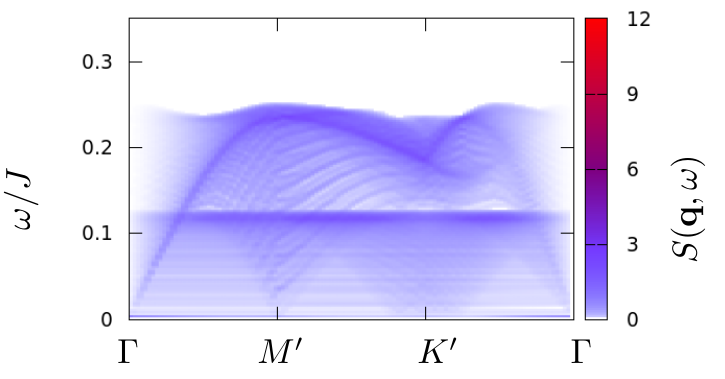}
		\caption{(Color online)
		Dynamical structure factor of the nearest-neighbor singlet
		$U(1)[\pi,\pi]$ spin liquid along the high symmetry direction of the EBZ. The extensive continuum of low energy scattering is a contribution of the flat bands at the Fermi level [Fig. \ref{fig:dispersions} (c)].}
	\label{fig:singletdsfU1-pipi}
		\end{minipage}
		\hspace{0.5 \columnsep}
		\begin{minipage}[]{ \columnwidth}
		\includegraphics[width=\linewidth]{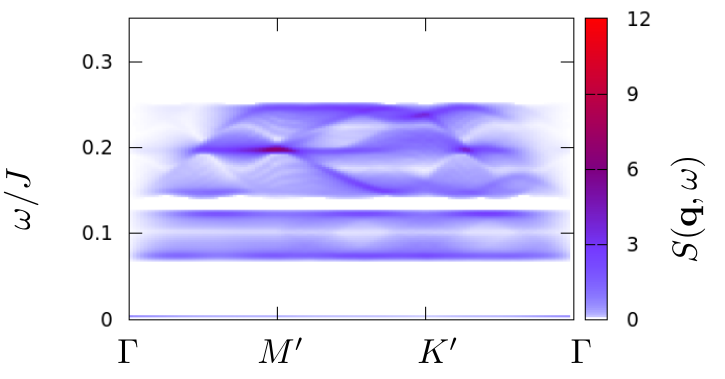}
		\caption{(Color online)
		Dynamical structure factor of the nearest-neighbor singlet
		$U(1)[\pi,0]$ spin liquid along high symmetry directions of the EBZ. There is finite scattering intensity exactly at zero energy because of the flat bands in the spinon dispersion [Fig. \ref{fig:dispersions} (d)], which is followed by a lack of scattering due a to gap in the spinon spectrum.}
	\label{fig:singletdsfU1-pi0}
		\end{minipage}
\end{figure*}

Figure \ref{fig:singletdsfU1-00} shows the dynamical spin-structure factor for the $U(1)[0,0]$ state. The low-energy domes of scattering anchored around the $\Gamma$ and $M'$ points are contributed by the excitations near the Fermi surface. Above these domes, we see additional flat scattering coming from contribution of the the higher energy, flat bands [refer to Fig. \ref{fig:dispersions} (a)].

In Fig. \ref{fig:singletdsfU1-0pi} we plot the dynamical spin-structure factor for the $U(1)[0,\pi]$ state.  In contrast with the $U(1)[0,0]$ case, low-energy cones of scattering
are seen at the $\Gamma$, $M'$, and intermediate points in the extended Brillouin zone, a consequence of the linearly-dispersing low-energy single-spinon excitations of the Dirac spin liquid. The flat, intense band of scattering seen at the highest energies are manifestations of the flat spinon bands [Fig. \ref{fig:dispersions} (b)].

Figure \ref{fig:singletdsfU1-pipi} shows the structure factor for the $U(1)[\pi,\pi]$ state. In this case, we can see intensity all the way down to zero energy across the extended Brillouin zone. Within this diffuse scattering, there are two major variations of intensity. The first is dispersive, rising from the $\Gamma$ point, due to the Dirac point in the spinon dispersion. The second is the flat band of intensity in the middle of the energy range, again due to zero-energy flat bands of the dispersion.

Finally, Fig. \ref{fig:singletdsfU1-pi0} shows the structure factor for the $U(1)[\pi,0]$ state.  While the flat bands in the dispersion contributes to the scattering {\it{exactly}} at zero-energy, the gap to subsequent excitations is seen in the absence of scattering. The highest intensity is seen at a particular point of scattering at $M'$.

We see that the low-energy spin correlations are an effective way to distinguish between these $U(1)$ spin liquids. Other dispersive scattering features characteristic to the states also show up at comparatively higher energies. 

Next, we will look at the effect of the additional pairing terms of
the $Z_2$ spin liquid ans\"atze on the dynamical structure factor, particularly
as a means to distinguish between $Z_2$ states with the same parent $U(1)$ spin liquid,
in spite of their similarities.

\subsection{Dynamical Structure Factor of Singlet $Z_2$ Spin Liquid Ans\"atze}

\begin{figure*}
	\includegraphics[width=0.9\linewidth]{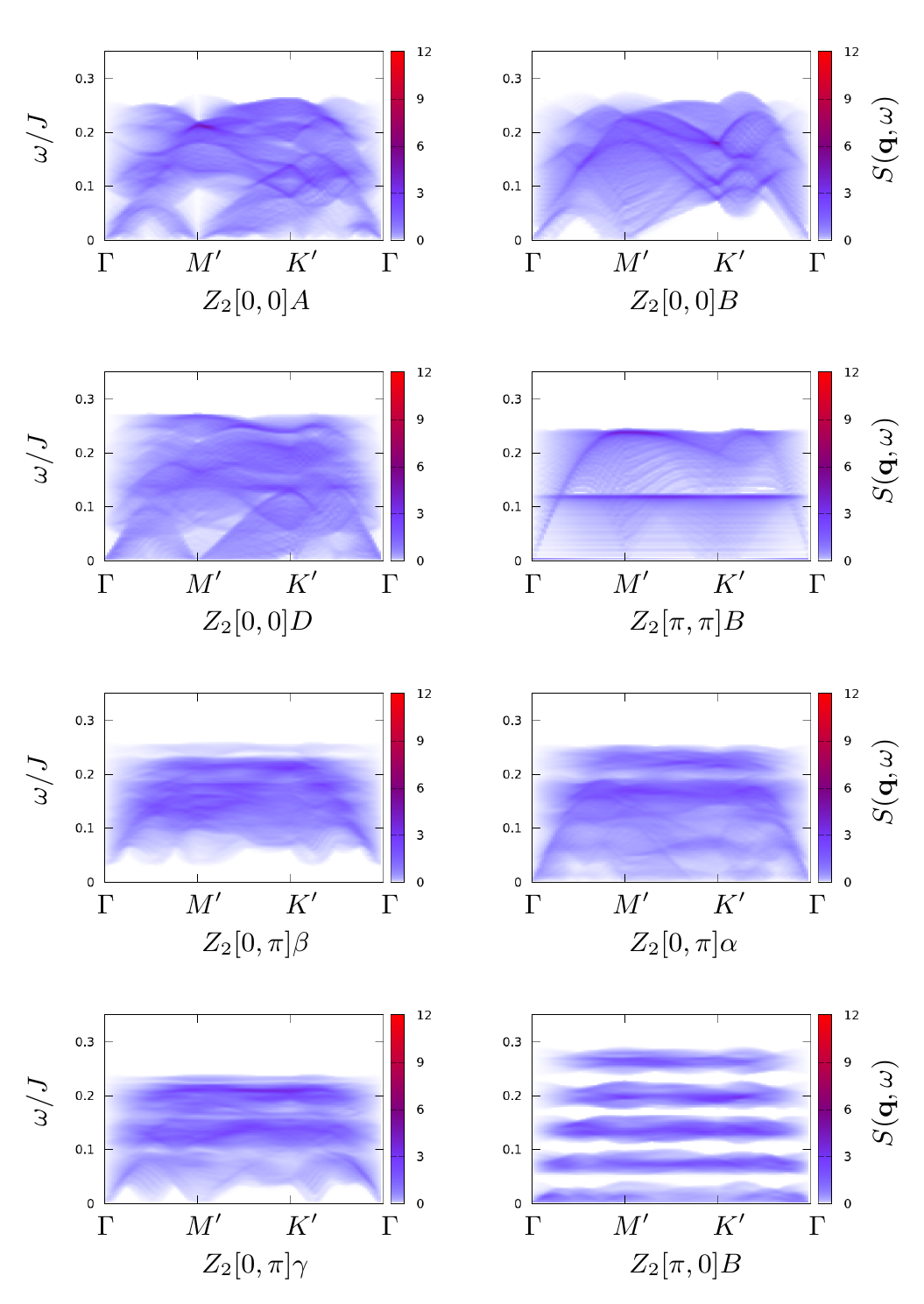}
\caption{(Color online)
Dynamical structure factor for 
$Z_2$
singlet spin liquid ans\"atze.
These structure factor are plotted from the center of the extended Brillouin zone
of the Kagome lattice, $\Gamma$, to the edge $M'$, to
the corner $K'$, and back to the center $\Gamma$.
}
\label{fig:dsf-z2-singlet-part-2}
\end{figure*}

We see that the dynamical structure factor of $Z_2$ states are similar in overall shape to that of their parent $U(1)$ states. To illustrate this point, we will  consider the $Z_2[0,0]A,$ $B$m and $D$ states, in Figs. \ref{fig:dsf-z2-singlet-part-2} (a), \ref{fig:dsf-z2-singlet-part-2} (b), and \ref{fig:dsf-z2-singlet-part-2} (c).  The $Z_2[0,0]D$ state has a structure factor that is most similar to the parent $U(1)[0,0]$ case (Fig. \ref{fig:singletdsfU1-00}), with zero-energy scattering along much of the $M'-K'$ line. Above that, the scattering is diffuse, with dispersive features that are only slightly stronger than the diffuse background. For the $Z_2[0,0]A$ state, the low-energy domes of scattering persist over small energies, and the flat features have become much more dispersive. There is a large intensity at the $M'$ point at higher energies. In the $Z_2[0,0]B$ case, there is low-energy intensity only near the $\Gamma$ and $M'$ points. Many dispersive features are seen, particularly near the $K'$ point.

The $Z_2[\pi,\pi]B$ state has a dynamical structure factor [Fig. \ref{fig:dsf-z2-singlet-part-2} (d)] which is strikingly similar to its parent $U(1)[\pi,\pi]$ state (Fig. \ref{fig:singletdsfU1-pipi}). Only the intensity of the dispersive bands increases, and scattering close to $\Gamma$ is found up to all energies, but there are no obvious qualitative differences between the two.

We also see similar resemblance in case of the the $Z_2[0,\pi]$ QSL states with their parent $U(1)[0,\pi]$ state [Figs. \ref{fig:dsf-z2-singlet-part-2} (e), \ref{fig:dsf-z2-singlet-part-2} (f), and \ref{fig:dsf-z2-singlet-part-2} (g)]. The $Z_2[0,\pi]$ states retain flat intensity at the highest energy, and relatively constant diffuse scattering across their energy range from their parent state. The $Z_2[0,\pi]\beta$ state has a fairly significant spin-gap, where the low-energy cones of scattering of the parent $U(1)$ state (Fig. \ref{fig:singletdsfU1-0pi}) have been rounded off. This spin-gap is found to be rather momentum independent. Also the flat features of the $U(1)[0,\pi]$ state are found to be broadened, and the scattering is fairly diffuse,  although some dispersive intensity can be seen coming from the $\Gamma$ point, due to the  remnant of the Dirac cone. At low energies, the intensity for the $Z_2[0,\pi]\alpha$ state has many broad domes of scattering, and zero-energy excitations exist almost throughout. At larger energies, the intensity is mostly diffuse, where the $U(1)[0,\pi]$ state's high-energy intensity has been split into two neighboring broad, diffuse, yet still almost flat bands. The low energy features of the $Z_2[0,\pi]\gamma$ state is similar to the $U(1)$ case. Above this, the intensity is also separated into two broad, diffuse bands, more prominently than the $Z_2[0,\pi]\alpha$ state.

The $Z_2[\pi,0]B$ state [Fig. \ref{fig:dsf-z2-singlet-part-2} (h)] shows many separated broad and diffuse bands, including the band near zero energy, which is the remnant of the strictly zero energy scattering coming from the flat bands of the $U(1)[\pi,0]$ state (see Fig. \ref{fig:singletdsfU1-pi0}). Furthermore, the intensity is relatively constant across the cut, only diminishing as the zone center ($\Gamma$ point) is approached.

With the exception of the $[\pi,\pi]$ states, the low-energy intensity and dispersive features allow the twelve spin-liquid ans\"atze in Table \ref{tab:ansatzparameters} to be distinguished from each other qualitatively. In the next subsection, we will see the effect of adding triplet terms to the QSL ans\"atze to the dynamical structure factor. 

\subsection{Dynamical Structure Factor of ``Singlet + Triplet'' $U(1)$ and $Z_2$ Spin Liquid Ans\"atze}

The introduction of the triplet terms $E_{ij}^z$ and $Y_{ij}^z$ breaks the
spin-rotational symmetry of the spin-polarized dynamical structure factor
$S^{\alpha\beta}(\v q,\omega)$, where $S^{xx} = S^{yy} \neq S^{zz}$. While the
singlet structure factors showed a vanishing intensity approaching 
$\v q = \v 0$, these triplet terms generate a non-zero intensity in $S_{xx}$
and $S_{yy}$. Also, the scattering intensity at low energies changes, due to
changes in the spinon dispersion (summarized in Table.
\ref{tab:ansatzelowenergyspinons}). Since the spinon bands are, in general,
split by triplet terms, the scattering becomes increasingly diffuse.

We will showcase these changes by calculating the trace of the dynamical structure factor matrix, {\it{i.e.}}, $(S^{xx} + S^{yy} + S^{zz})/3$, and then plotting the {\it difference} from the singlet structure factor. In Appendix \ref{sec:appendixdsftriplet}, we show the actual structure factor for all these ``singlet + triplet'' states.
\begin{figure*}
	\includegraphics[width=\linewidth]{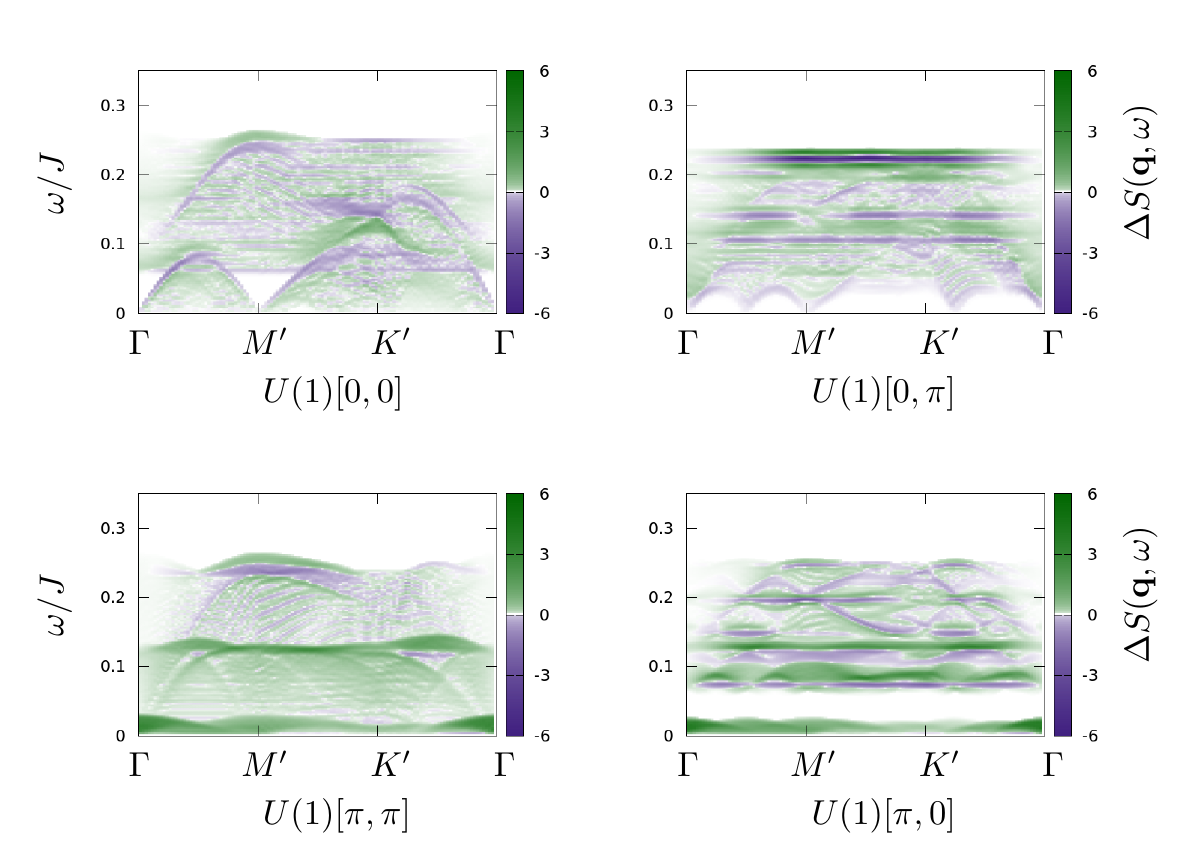}
\caption{(Color online)
Difference between the dynamical structure factor for 
$U(1)$
``singlet + triplet'' spin liquid ans\"atze.
These structure factor are plotted from the center of the extended Brillouin zone
of the Kagome lattice, $\Gamma$, to the edge $M'$, to
the corner $K'$, and back to the center $\Gamma$.
We note that since the ``singlet + triplet''  $U(1)[0,\pi]$ spin liquid state is gapped, it is unstable to instanton tunneling events.\cite{polyakov}
}
\label{fig:dsf-u1-triplet}
\end{figure*}

These differences for the ``singlet + triplet'' $U(1)$ structure factors are shown in Fig. \ref{fig:dsf-u1-triplet}.
The ``singlet + triplet'' $U(1)[0,0]$ structure factor looks very similar to the singlet
case,  particularly in the low-energy scattering. However, the diffuse
scattering at moderate energies is no longer flat. While this is also true for
the ``singlet + triplet'' $U(1)[0,\pi]$ structure factor, it also has two qualitative
differences. The first is that the cones of scattering at low energy are gapped
out; this can be seen more clearly in the full ``singlet + triplet'' structure
factor in Appendix \ref{sec:appendixdsftriplet}, Fig. \ref{fig:dsf-all-triplet-full}.
The second is that the flat band at the highest energy is split into two.
For the $U(1)[0,\pi]$ state,
though we plot the structure factor for completeness, we note that in this state the spinon spectrum is gapped and hence the state is unstable to
confinement transition due to instanton events.\cite{polyakov} 
The ``singlet + triplet'' $U(1)[\pi,\pi]$ structure factor is very similar to the
singlet case, where the low-energy scattering sees a band of intensity between
$\Gamma$ and $M'$,  as well as $\Gamma$ and $K'$. 
This is absent in the singlet case, which can distinguish between 
these states. Such a band is also seen in the ``singlet + triplet'' $U(1)[\pi,0]$ state,
which otherwise is again similar to the corresponding singlet case.

\begin{figure*}
	\includegraphics[width=0.9\linewidth]{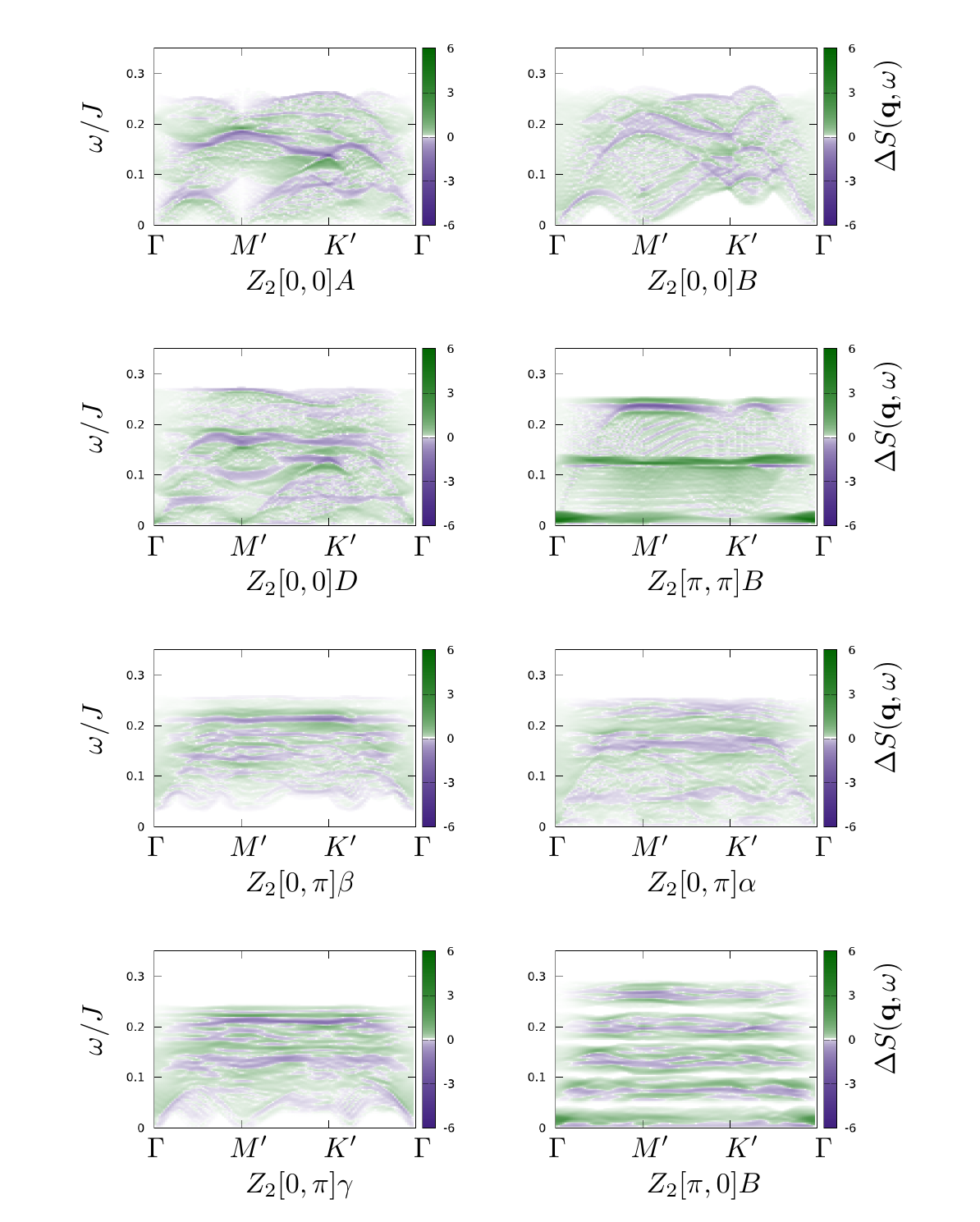}
\caption{(Color online)
Difference between the dynamical structure factor for 
$Z_2$
``singlet + triplet'' spin liquid ans\"atze.
These structure factor are plotted from the center of the extended Brillouin zone
of the Kagome lattice, $\Gamma$, to the edge $M'$, to
the corner $K'$, and back to the center $\Gamma$.
}
\label{fig:dsf-z2-triplet-part-2}
\end{figure*}

The $Z_2$ (S+T) spin liquids are shown in Fig.
\ref{fig:dsf-z2-triplet-part-2}. Most of the structure factors closely resemble
their singlet counterparts {\it{with important general differences}}. Here, we
shall limit ourselves to pointing out these differences only. Generally, due to
the triplet decoupling channels, the spinon bands are spin-split. Hence, the sharp
dispersing structures are somewhat lost and the scattering intensity becomes
more diffuse compared to the singlet only ans\"atze, with increasing strength
of the triplet terms. 
As with the $U(1)$ states,
the intensity of low energy scattering does not go to
zero at $\v q = \v 0$. Hence there is a slight enhancement of
scattering at the EBZ center compared to the singlet states. Also, for most of
the states where there is a spin-gap in the structure factor, the gap magnitude
becomes increasingly independent of the momentum.

Finally, we end this section by noting that the spin-liquid state $Z_2[0,\pi]\beta$ seems to be the best candidate in the context of present inelastic neutron scattering experiments and DMRG calculations. We shall discuss this in some detail, along with the signatures of the other spin liquids, in the next two subsections in the context of inelastic neutron scattering and ESR experiments.
\subsection{Implications for Inelastic Neutron Scattering Experiments}
\label{sec:inelasticneutronscattering}

It is useful to compare the spin-structure factor obtained above with the
results of the recent inelastic neutron scattering experiments on single
crystals of Herbertsmithite. \cite{Nature.492.406} These experiments show
almost uniform diffuse scattering over a large energy window.
Furthermore, no spin-gap is observed down to $1.5$ meV. It
also shows no obvious signature of broken lattice symmetries.

With parameters from the mean-field solution, (as discussed
later in Appendix \ref{sec:mfresults}) excitations are found for $\omega \lesssim 0.35 J$. 
In contrast, the inelastic neutron scattering shows strong diffuse intensity up to the largest measured energy
11 meV $\sim 0.65 J$. However, the numerical values from our mean-field calculation can, at best, serve as a consistency check.

Comparing with our calculated dynamical spin-structure factor with the above features of the experiment, we may infer that the $U(1)[\pi,\pi]$, $U(1)[\pi,0]$, $Z_2[\pi,\pi]B$, and $Z_2[\pi,0]B$ states clearly appear to be inconsistent with the neutron scattering data.
Further comparison of their ESR line shapes and peak distribution (see next section) may confirm/ invalidate this conclusion. 

In contrast, several of the $[0,0]$ and $[0,\pi]$ [both $U(1)$ and $Z_2$] QSLs
do have mostly-featureless diffuse scattering within some energy window
$[\omega_{\rm min},\omega_{\rm max}]$. However, in all but a few cases this
window is narrow, and these states have well-defined features at lower or
higher energies and, hence, are inconsistent with experiments.

The four states, namely $Z_2[0,0]D$, $Z_2[0,\pi]\alpha$, $Z_2[0,\pi]\gamma$ and $Z_2[0,\pi]\beta$, have dispersive features with a very weak intensity, leading to an  almost diffuse structure factor.  The dispersive features are broadened and are in relatively poor contrast with the generally diffuse background.
However, we see some modulation in intensity for the $Z_2[0,\pi]\gamma$ state
and slightly better-defined dispersive features in the $Z_2[0,0]D$ state. 
However, only the $Z_2[0,\pi]\beta$ state shows evidence of a spin gap--$Z_2[0,\pi]\alpha$ is gapless. 
As already noted, the inclusion of the spin-rotation symmetry-breaking perturbations makes the spin-gap increasingly momentum independent in $Z_2[0,\pi]\beta$.
Further these perturbations also enhances the scattering near the Brillouin
zone center somewhat in almost all cases (refer to Fig. \ref{fig:dsf-z2-triplet-part-2}).
Noting that the DMRG calculations indeed stabilize a gapped QSL state (most likely $Z_2$), in light of the inelastic neutron scattering measurements,
it is tempting to suggest the $Z_2[0,\pi]\beta$ state as a consistent choice for a candidate
ground state for Herbertsmithite. However, we should note that the issue of existence of spin-gap is still not clear in Herbertsmithite experiments, and except for the lack of spin-gap, $Z_2[0,\pi]\alpha$ is also a possible candidate.

Next, we describe the ESR absorption spectra for the QSLs that break spin-rotation symmetry.

\subsection{ESR Absorption Intensity for different spin-rotation symmetry breaking QSLs}

To observe the ESR spectra we couple the system with a Zeeman field via  
$H_{Z} \sum_i \hat z \cdot \v S_i$. 
The ESR absorption intensity is given in \eref{eqn:esrabsorbtion}. We note that
$I(\omega) \propto \omega$ at low temperatures, and peaks occur around the
value of the Zeeman field, ($\omega=H_Z$) so absorption is most easily seen at the large
values of $\omega$ and $H_Z$. We align the Zeeman field along the $z$-axis, the
axis along which the spin-rotation is broken. For 
energy scales $\omega$ near $H_Z$,
the Zeeman term breaks $SU(2)$ symmetry, with $U_{ij}$ along with the triplet terms. For this axis of the magnetic field,  $I^{xx}(\omega) = I^{yy}(\omega)$ due to the remaining $U(1)$ spin-rotational symmetry around the $z$-axis.

\begin{figure*}
	\centering
	\includegraphics[width=\linewidth]{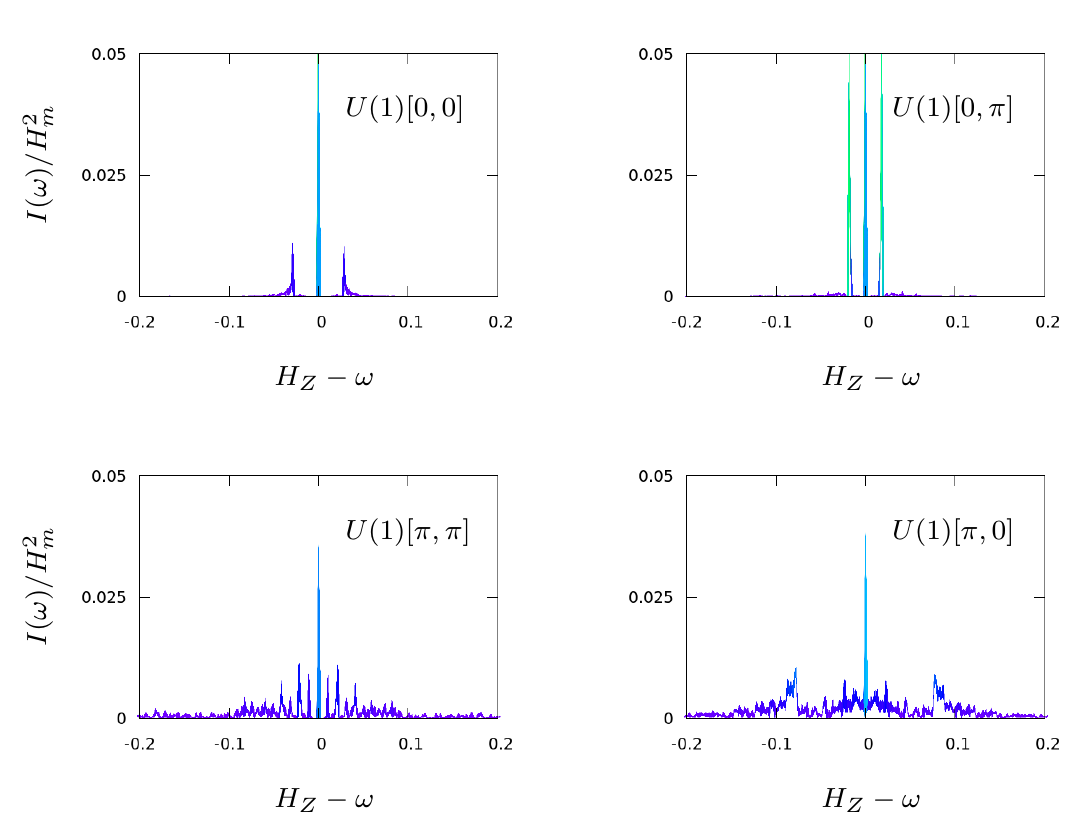}
\caption{(Color online)
Electron spin resonance absorption for
$U(1)$
``singlet + triplet'' spin liquid ans\"atze,
as a function of the Zeeman field strength $H_Z$
and microwave field frequency $\omega$.
The viewpoint is along the $H_Z = \omega$ axis.
We note that since the ``singlet + triplet'' $U(1)[0,\pi]$ spin liquid state is gapped, it is unstable to instanton tunneling events.\cite{polyakov}
}
\label{fig:esr-u1-triplet}
\end{figure*}

We now discuss the ESR absorption intensity for the four $U(1)$  and eight $Z_2$ QSL in presence of the triplet channels. We focus on the small-intensity region to show contributions from the triplet terms; these occur as satellite peaks with smaller intensities, while the peak at $\omega=H_Z$ has a much larger intensity. 
These spectra are shown in Fig. \ref{fig:esr-u1-triplet}.

Both the $U(1)[0,0]$ and $U(1)[0,\pi]$ spin liquids have an additional
`satellite' peak on either side of the main peak at $\omega=H_Z$.
The peaks of the $U(1)[0,\pi]$ state have both higher intensity and are spaced more
closely compared to the $U(1)[0,0]$ state.

However, the $U(1)[\pi,\pi]$ and $U(1)[\pi,0]$ states have wildly different
ESR absorption spectra.
There is absorption over a broad range of $H_Z - \omega$, with many 
subsequent satellite peaks almost forming a continuum. The intensity of the main peak is significantly 
diminished compared to the $U(1)[0,0]$ and $U(1)[0,\pi]$ spin liquids.
As $H_Z - \omega$ increases from zero, in the $U(1)[\pi,\pi]$ state, the primary
satellite peaks have a monotonically decreasing intensity, while the $U(1)[\pi,0]$ state features a single pair of prominent satellite peaks at a finite distance away from zero.

\begin{figure*}
	\centering
	\includegraphics[width=0.9\linewidth]{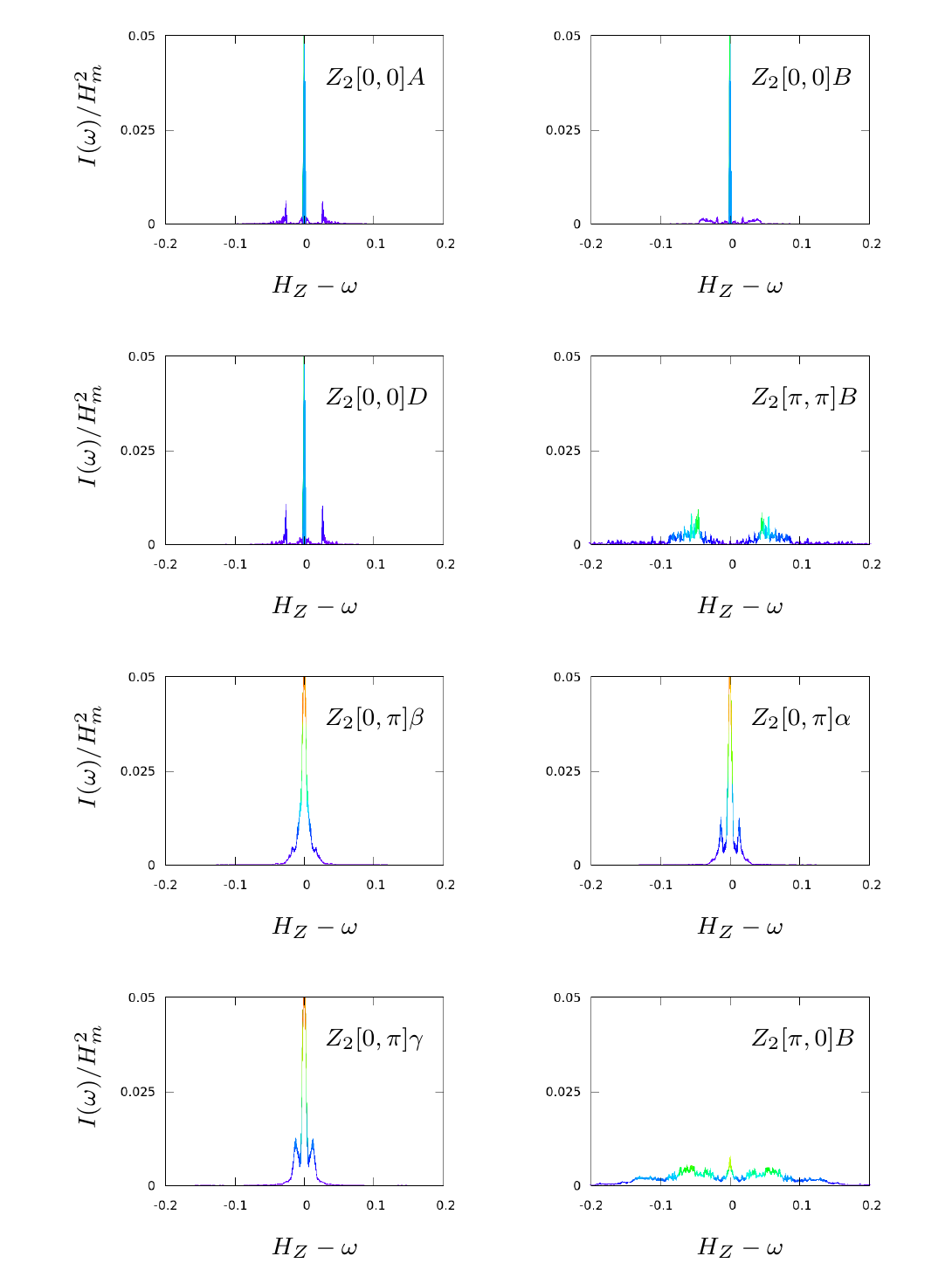}
\caption{(Color online)
Electron spin resonance absorption for
$Z_2$
``singlet + triplet'' spin liquid ans\"atze,
as a function of the Zeeman field strength $H_Z$
and microwave field frequency $\omega$.
The viewpoint is along the $H_Z = \omega$ axis.
}
\label{fig:esr-z2-triplet-part-2}
\end{figure*}

The $Z_2$ spin liquids generally lead to a broadening of the absorption peaks.
These spectra are shown in Fig. \ref{fig:esr-z2-triplet-part-2}.
The $Z_2[0,0]A$ state shows slight broadening of the all peaks
around the base, with diminished intensity compared to the the $U(1)$ case.
The $Z_2[0,0]D$ state shows a similar response, though the satellite peaks are higher
compared to the broadening of the main peak.
The $Z_2[0,0]B$ state has very broad absorption around the now significantly smaller satellite peaks. The largest satellite peaks of the $Z_2[\pi,\pi]B$ are broadened at low 
absorption, but the intensity around the main peak drops dramatically and is almost zero.

The $Z_2[0,\pi]\beta$ state shows significant broadening of the main peak
around the base, and nearly complete reduction of satellite peak intensity.
The $Z_2[0,\pi]\alpha$ and $Z_2[0,\pi]\gamma$ states have  smaller amount of broadening around the main peak, where the satellite peaks are diminished in intensity, but still visible. The intensity of the main peak of the $Z_2[\pi,0]B$ is almost zero and the satellite peaks are replaced by small, broad domes of absorption.

The shape and structure of the satellite peaks provide another qualitative clue to
distinguish between the spin liquid ans\"atze considered here, and immediate evidence of spin-rotation symmetry breaking
in the kagome planes.

Current ESR measurements down to as low as 5 K, however, find that the
absorption intensity is dominated by the impurity spins below 20 K.\cite{PhysRevLett.101.026405}
These Cu$^{2+}$ ions contribute in a nearly-paramagnetic fashion, and the
intensity displays a Curie-like response $\propto 1/T$.
The line shape remains broad.
In the regime where these impurity spins display a paramagnetic response,
isolating the kagome-layer ESR response may prove difficult. 
An increase in sample purity can curtail this effect.
However, at very low temperatures, these impurity spins may no longer 
behave in a paramagnetic fashion, and interact with the kagome planes
with a strength of $\sim 10$ K.\cite{PhysRevLett.101.026405,2013arXiv1303.1310J}
One needs to consider the effect of the coupling of these
impurities to the kagome spin liquid and the resultant ESR spectrum.
Such analysis is beyond the scope of the present paper, so we present
here the intrinsic absorption of the spin liquid layers.

\section{Discussion}
\label{sec:discussion}

We now summarize our results. In this work, we have attempted to address two
important questions to account for the unusual phenomenology of the non-magnetic
ground state of Herbertsmithite. First, we have calculated the dynamical
spin-structure factor for four $U(1)$  and eight $Z_2$ {\it{symmetric}} spin
liquids derived from the former for the NN and NNN antiferromagnetic Heisenberg
model on an isotropic Kagome lattice. These spin liquids are allowed by
spin-rotation, time-reversal, and lattice symmetries of a Kagome lattice. We
then consider the effect of small spin-rotation symmetry breaking perturbations
(DM and Ising anisotropy) on the above spin liquids and
on the corresponding spin-structure factor. Furthermore, we calculate
the ESR absorption spectra, which show nontrivial structures in the presence of
spin-rotation symmetry breaking. Since recent numerical studies suggest that
the ground state of the strictly NN antiferromagnetic Heisenberg model is very
sensitive to small perturbations of second-neighbor exchange 
and Dzyaloshinsky-Moriya interactions, we
expect that the the above perturbations may have important effects in the low
energy features of the experimentally measurable spin-structure factor.

Indeed, we find that the addition of the perturbations make the structure
factor largely diffuse over an extended energy window throughout the Brillouin
zone.  Similar scattering has been observed in the recent inelastic neutron
scattering experiment on Herbertsmithite. We particularly find {\it two} $Z_2$ spin
liquid states, the so-called $Z_2[0,\pi]\beta$ and $Z_2[0,\pi]\alpha$ states,
whose spin-structure factor features are qualitatively in
conformity with the experiments. Only the $Z_2[0,\pi]\beta$ state, however, exhibits a  
spin-gap in the structure factor. In the presence of the perturbations,
this gap becomes increasingly momentum-independent. Noting that a similar
gapped QSL was obtained as the ground-state of the NN antiferromagnetic
Heisenberg Hamiltonian in recent DMRG calculations, it is tempting to infer
that the $Z_2[0,\pi]\beta$ state may be adiabatically connected to the ground
state obtained in DMRG and also the ground state of Herbertsmithite. This
prediction can be checked further by measuring the ESR absorption spectrum. Our
present calculations suggest that the the $Z_2[0,\pi]\beta$ phase shows a
characteristic broadening of lines in ESR absorption spectra.

\begin{acknowledgments}
We thank O. Starykh, R. Schaffer and Mingxuan Fu for many insightful discussions. Y.B.K. acknowledges the support and hospitality from the Aspen Center for Physics, funded by NSF Grant No. PHY-1066293, and the KITP, funded by NSF Grant No. PHY1125915. This research was supported by the NSERC, CIFAR, and center for Quantum Materials at the University of Toronto.
\end{acknowledgments}

\appendix

\section{$U(1)$ Spin Liquid Ans\"atze}
\label{sec:u1fluxansatze}

\begin{figure}[!tbp]
	\centering
\includegraphics[width = 0.9 \columnwidth]{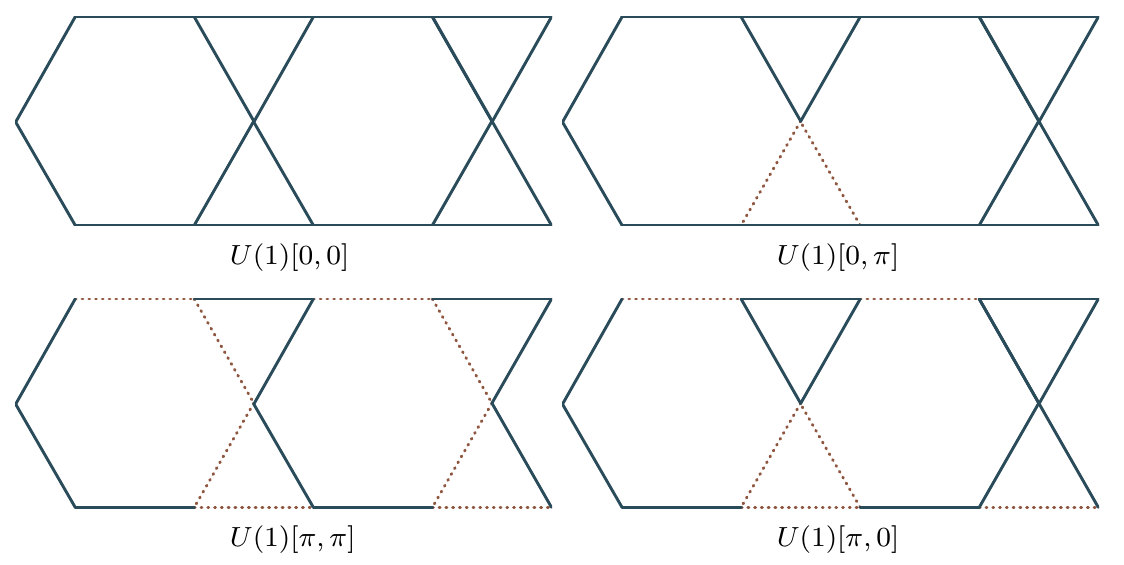}
\caption{(Color online)
Flux pattern for the real hopping terms of the
$U(1)$
singlet spin liquid ans\"atze.
Hoppings are either positive (dark blue, solid) or
negative (light brown, dashed) on nearest-neighbor bonds.
}
\label{fig:u1fluxpatterns}
\end{figure}

The flux patterns for the nearest-neighbor $U(1)$ spin liquids under
consideration are shown in Fig. \ref{fig:u1fluxpatterns}.
They consist of positive and negative real hopping terms that retain the translational
symmetry of the Kagome lattice ($U(1)[0,0]$ and $U(1)[\pi,\pi]$)
or break it ($U(1)[0,\pi]$ and $U(1)[\pi,0]$).

\section{Projective Symmetry Group}
\label{sec:psgdetails}

Determining the allowed PSGs $\{G_SS\}$, and thereafter the ans\"atze $\HM_Q$,
comes from constraints offered by the group multiplication table of 
the symmetry group SG.
With group multiplication rules like $AB=C$ that can be rewritten as $ABC^{-1} = I$,
we can place constraints on the corresponding expression $(G_AA)(G_BB)(G_CC)^{-1}$.
First, there is no net physical transformation, so the expression must reduce to
a gauge transformation. Second, each of the operations leaves $\HM_Q$ invariant, 
and so does the final expression. 
Thus, $(G_AA)(G_BB)(G_CC)^{-1} \in \mathrm{IGG}$.

As derived by Lu {\em et. al.}, 
$G_S$ are given in the singlet case by the following:
\cite{PhysRevB.83.224413}
\begin{align}
		G_{T_1}(x,y,s) &= \eta_{12}^y I
		\\
		G_{T_2}(x,y,s) &= I
		\\
		G_{\gv \sigma}(x,y,s) &= \eta_{12}^{xy}g_{\gv \sigma}(s)
		\\
		G_{C_6}(x,y,s') &= \eta_{12}^{xy+x(x+1)/2}g_{C_6}(s'), s' \in \{u,v\}
		\\
		G_{C_6}(x,y,w) &= \eta_{12}^{xy+x+y+x(x+1)/2}g_{C_6}(w)
		\\
		G_{\bf T} &= i\tau^1,
	\label{eqn:fullformpsg}
\end{align}
where 
we specify the lattice site $i$
by the position of the unit cell to which is belongs 
$\v{R} = x\v{a}+ y\v{b}$, and the sublattice index $s=u,v,w$, as indicated
in Fig. \ref{fig:Kagome-transforms}.

The PSG can also be used to generate terms on symmetry-related bonds of an ans\"atze.
The action of a lattice transformation
$S$ and an $SU(2)$ gauge transformation $W$ on the quadratic
terms of bond, given by $U_{ij}$, is
\begin{align}
		S U_{ij} \to U_{S^{-1}(i),S^{-1}(j)} \nonumber \\
		W U_{ij} \to W(i) U_{ij} W^{\dag}(j).
\end{align}
If $G_S S$ leaves the $U_{ij}$ invariant, 
as in the singlet case,
then we can combine the
above relations to generate $U_{ij}$ on symmetry-related bonds:
\begin{align}
		U_{S(i),S(j)} = G_S(S(i)) U_{ij} G_S^{\dag}(S(j)).
		\label{eqn:uijlatticetransformation}
\end{align}

We will cover the case of the ``singlet + triplet'' PSG next.

\subsection{Projective Symmetry Group for ``Singlet + Triplet''  Ans\"atze}
\label{sec:psgdetailstriplet}

Now, we discuss how time reversal and (improper) rotations
affect ``singlet + triplet'' $U_{ij}$,
as in \eref{eqn:tripletuij}. 
We will see that the PSG
$\{G_SS\}$ are the same as in the singlet case,
and that the ans\"atze allow imaginary triplet terms
as well.

We will begin by determining how the standard fermionic
time-reversal operator $\bf{T}_f$ 
changes the ansatz $U_{ij}$,
to properly capture the effect of time-reversal on the 
triplet terms.
$\bf{T}_f = \theta K$, where $K$ is the complex conjugation operator and
\begin{align}
		\theta \begin{pmatrix}
				f_{\uparrow} \\ f_{\downarrow} \end{pmatrix} = 
		\begin{pmatrix}
			-f_{\downarrow} \\ f_{\uparrow} \end{pmatrix}.
\end{align}
Consider the action on a generic $U_{ij}$:
\begin{align*}
		U = \begin{pmatrix} A & B \\ C & D \end{pmatrix}
				\xrightarrow{\bf{T}_f} 
		\begin{pmatrix} -D & C\\ B & -A \end{pmatrix}.
\end{align*}
One can instead define a modified time-reversal operator,
with an additional $i\tau^2$ gauge transformation
acting on $U_{ij}$,
${\bf T} = i \tau^2 {\bf T}_f$, giving
\begin{align*}
		U \xrightarrow{i\tau^2\bf{T}_f} 
		\begin{pmatrix} -A & -B \\ -C & -D \end{pmatrix} = -U.
	\end{align*}

Since the $C_3$ rotations are performed around the $z$-axis in both real
	space and spin space, neither the singlet nor the $E^z,Y^z$ triplet terms
in our ans\"atze are affected by them.

As all ansatz will projectively obey time-reversal symmetry,
we can use the form of $G_{\bf T}$ to simplify the ansatz before
considering the effect of reflection.
The action of ${\bf T}$ upon $U_{ij}$ is $-U_{ij}$ in both singlet and triplet
cases, thus they affect the ans\"atze in the same manner.
Thus, ${\bf T}^2 = +1$ acting on the mean-field states, as before, and we have 
the same condition $\pm G_{\bf T}(i)^2 \in IGG$.

The action in spin space of the reflection, $\gv{\sigma}$, is to flip the sign
of our triplet components, which are the $z$-components of a pseudo-vector.
Since a gauge transformation, on physical grounds,
can only mix singlet and triplet terms among themselves, this action commutes
with the $SU(2)$ gauge transformations. 
Only the lattice part of
$\gv{\sigma}$
enters in the commutation relation between $\gv{\sigma}$
and the gauge transformations $G_S$. 
Solving for the time-reversal gauge transformation $G_{\bf T}$
as in the singlet case,
the non-zero mean-field an\"atze will all have the same
$G_{\bf T} = i\tau^1$.
\cite{PhysRevB.83.224413}
 For an ansatz to be time-reversal invariant we must have
\begin{align*}
		\tau^1 U_{ij} \tau^1 = -U_{ij}.
\end{align*}
With this restriction, $U_{ij}$ can be parametrized by 
$U_{ij} = \gamma_2 \tau^2 + \gamma_3 \tau^3$, where 
$\gamma_{2,3} \in \mathbb{C}$. We note that the real components of
$\gamma_{2,3}$ are coefficients of the singlet pairing and hopping, respectively,
while the imaginary components are coefficients of triplet pairing and hopping.
Reversal of the
sign of the triplet coefficients can be performed by taking 
$U_{ij} \to U_{ij}^{\dag} = U_{ji}$. 

With this, we have the {\textit same} projective symmetry group that was
derived for singlet spin liquids with the symmetry of the Kagome
lattice.\cite{PhysRevB.83.224413} 
However, the ansatz that can be generated will differ, due to the effect of
$\gv{\sigma}$ upon the triplet terms. 

We may generate all nearest and next-nearest-neighbor bonds from just one by
using the symmetry operations of the Kagome lattice. For each, we may impose a
constraint on the allowed ansatz for these
bonds by devising a non-trivial lattice symmetry operation that takes 
$U_{ij}$ back to itself.
With the effect of lattice transformations \eref{eqn:uijlatticetransformation}
along with the transformation under $\gv{\sigma}$, we can derive the
following constraint
on the nearest-neighbor $U_{ij}$:
\begin{align}
		g_{\gv{\sigma}}(u)
		g_{C_6}(u)
		g_{C_6}(w)
		U_{\mathrm{NN}}
		g_{C_6}^{\dag}(v)
		g_{C_6}^{\dag}(w)
		g_{\gv{\sigma}}^{\dag}(v)
		= U_{\mathrm{NN}},
\end{align}
and a similar constraint on the next-nearest-neighbor $U_{ij}$:
\begin{align}
		g_{\gv{\sigma}}(u)
		g_{C_6}(u)
		U_{\mathrm{NNN}}
		g_{C_6}^{\dag}(v)
		g_{\gv{\sigma}}^{\dag}(w)
		= U_{\mathrm{NNN}}.
\end{align}
The particular bonds for $U_{\rm NN}$ and $U_{\rm NNN}$ are shown in
Fig. \ref{fig:Kagome-bonds}.
For a given $Z_2$ ansatz
in Table \ref{tab:ansatzparametersappendix}, 
these may restrict the structure
of $U_{ij}$ to disallow hopping ($\tau^3$) or pairing ($\tau^2$) terms
but do not place any conditions on whether these terms
are real or imaginary. This differs from the singlet case,
where the effect of $\gv{\sigma}$ on the triplet terms was
not considered, and the ans\"atze become restricted to only singlet terms.
Thus, spin-rotational
symmetry-breaking triplet terms are allowed within the same PSG
for the Kagome lattice.
Since $U_{ij}$ is now no longer Hermitian, $U_{ji} \neq U_{ij}$,
and the direction of the bonds matter. They are shown in 
Fig. \ref{fig:Kagome-bonds}.

\begin{table}[htp]
\centering
\begin{tabular}{|c||c|c|c|}
\hline
SL Label & $\eta_{12}$ & $g_{C_6}$ & $g_{\gv \sigma}$ 
\\
\hline
\hline
$Z_2[0,0]A$ & $+1$ & $u,v,w:I$ & $u,v,w:I$ 
\\
$Z_2[0,\pi]\beta$ & $-1$ & $u,v,w:I$ & $u,v,w:I$ 
\\
$Z_2[0,0]B$ & $+1$ & $u,v,w:i\tau^3$ & $u,v,w:I$ 
\\
$Z_2[0,\pi]\alpha$ & $-1$ & $u,v,w:i\tau^3$ & $u,v,w:I$ 
\\
$Z_2[0,0]D$ & $+1$ & $u,v,w:i\tau^3$ & $u,v,w:i\tau^3$ 
\\
$Z_2[0,\pi]\gamma$ & $-1$ & $u,v,w:i\tau^3$ & $u,v,w:i\tau^3$ \\
$Z_2[\pi,\pi]B$ & $+1$ & $u,v:I ~~ w:i\tau^1$ & $u,v,w:i\tau^3$ 
\\
$Z_2[\pi,0]B$ & $-1$ & $u,v:I ~~ w:i\tau^1$ & $u,v,w:i\tau^3$ 
\\
\hline
\end{tabular}
\caption{
Parameters $\eta_{12},g_{C_6}$ and $g_{\gv \sigma}$ characterizing the PSG 
$\{G_S S\}$ of spin liquid (SL) states, as given in \eref{eqn:fullformpsg}. 
}
\label{tab:ansatzparametersappendix}
\end{table}

\section{Mean-Field Results}
\label{sec:mfresults}

As mentioned in Sec. \ref{sec:model}, we consider the mean-field solution for each of the $Z_2$ spin liquid ans\"atze, with nearest-neighbor perturbations $D$, $\Delta$ and the next-nearest-neighbor $J_2$.  While we do not expect the self-consistent mean-field theory to give a quantitatively correct phase diagram, we can still gain some insight from the results. In particular, we would like to understand the representative mean-field  parameter values $U_{ij}$ when the states in Table \ref{tab:ansatzparameters} are stabilized.

We focus on the parameter regime $D/J, \Delta/J \in [0,0.5)$. Furthermore, we will fix $\omega_-$ as the overall nearest-neighbor energy scale, taking the regime $J_2/\omega_- \in [0,2)$. In this way, we separate the effects of $D$ and $\Delta$ from $J_2$ as much as possible. We begin by considering general trends across all ans\"atze.

\subsection{General Results}
\label{sec:mfresults-generalresults}

\paragraph*{DM interactions:}

Here, we consider the contributions from the $\hat{\chi}_+$ and $\hat{\eta}_+$
channels in Eq. \eref{eqn:singlettripletcombinedchannels}, in comparison to
contributions from $\hat{\chi}_-$ and $\hat{\eta}_-$ channels. The
self-consistent theory suggests that for $U(1)$ and $Z_2$ states labeled as
$[0,0]$ and $[0,\pi]$ states, the $\chi_-$ and $\eta_-$ channels have the
dominant contributions, and the ratio of triplet to singlet terms is a nearly linear
function of $D/J$ within our parameter window $0 < D, \Delta < J/2$.
The $[\pi,\pi]$ and $[\pi,0]$ states, however, have stable mean-field states where $\chi_+$ and $\eta_+$ are significant, and dramatically small triplet values, particularly for $D/J \lesssim 0.2$.

\paragraph*{Ising Interaction $\Delta$:} This term has a negligible effect on the $Z_2[0,0]$ and $Z_2[0,\pi]$ states, since $\hat{\chi}_+$ and $\hat{\eta}_+$ are not relevant. For $Z_2[\pi,\pi]$ and $Z_2[\pi,0]$ spin liquids, however, small values of $\Delta/J$ make $\hat{\chi}_+$ and $\hat{\eta}_+$ channels less relevant. Since $\omega_+ = (J-\Delta)/8 + O($D$)$, increasing $\Delta$ from zero  will actually decrease $\omega_+$, so these channels make little contribution to $\HM_Q$ in the mean-field theory. Within the mean-field theory, these small values of $\Delta$ yield no qualitatively new behavior.

\paragraph*{$Z_2$ Spin Liquids and Next-Nearest-neighbor Terms:} For the six (all except $Z_2[0,0]B$ and $Z_2[0,\pi]\alpha$) $Z_2$ spin liquids in Table \ref{tab:ansatzparameters} that are stabilized only in the presence of next-nearest-neighbor terms, within the mean-field theory, we must have  $0.5 \lesssim J_2/\omega_- \lesssim 1.2$. The $Z_2[0,0]$ and $Z_2[0,\pi]$ states have a next-nearest-neighbor $\chi_2$ hopping parameter that does not stabilize the $Z_2$ spin liquids from their $U(1)$ parents. Generally, $|\chi_2|$ increases with $J_2$ in a monotonic and nearly-linear fashion. However, at a large value of $J_2 \sim 1.2\omega_-$, the $Z_2[0,\pi]\gamma$ state is stabilized with small $|\chi_1|$, small $|\eta_2|$, and large $|\chi_2|$. We expect that such a state is unlikely to be realized in models where $J_2/J$ is small.
The $Z_2[\pi,\pi]B$ state is stabilized with a similar jump at $J_2 \sim 0.9\omega_-$ with
a large $|\chi_2|$ value, and a small $|\eta_1|$. 

Out of the other four, in three $Z_2$ spin liquids ($Z_2[0,0]A$, $Z_2[0,0]D$ and $Z_2[0,\pi]\beta$), $\eta$ and $\chi_2$ increases monotonically with $J_2$ beyond a critical value of $J_2/J$ which itself depends on the type of spin liquid in consideration.
While the $Z_2[\pi,0]B$ sees this behavior for smaller values of $J_2$, it also undergoes a jump around $J_2 \sim 0.9 \omega_-$,
similarly to the $Z_2[\pi,\pi]B$ state.

\subsection{Mean-Field Phase Boundaries}
\label{sec:appendixphaseboundary}

Here we present the phase boundaries
in the $J_2$-$D$ plane where these $Z_2$ spin liquid
phases can be stabilized within mean-field theory.
The $Z_2[0,0]B$ phase is stabilized throughout the phase diagram, while the
$Z_2[0,\pi]\alpha$ state is never stabilized within mean field theory of the current Hamiltonian.
The other $Z_2$ spin liquids are stabilized for sufficiently large $J_2$.

\begin{figure}[!tbp]
	\centering
	\includegraphics[width = 0.9 \linewidth]{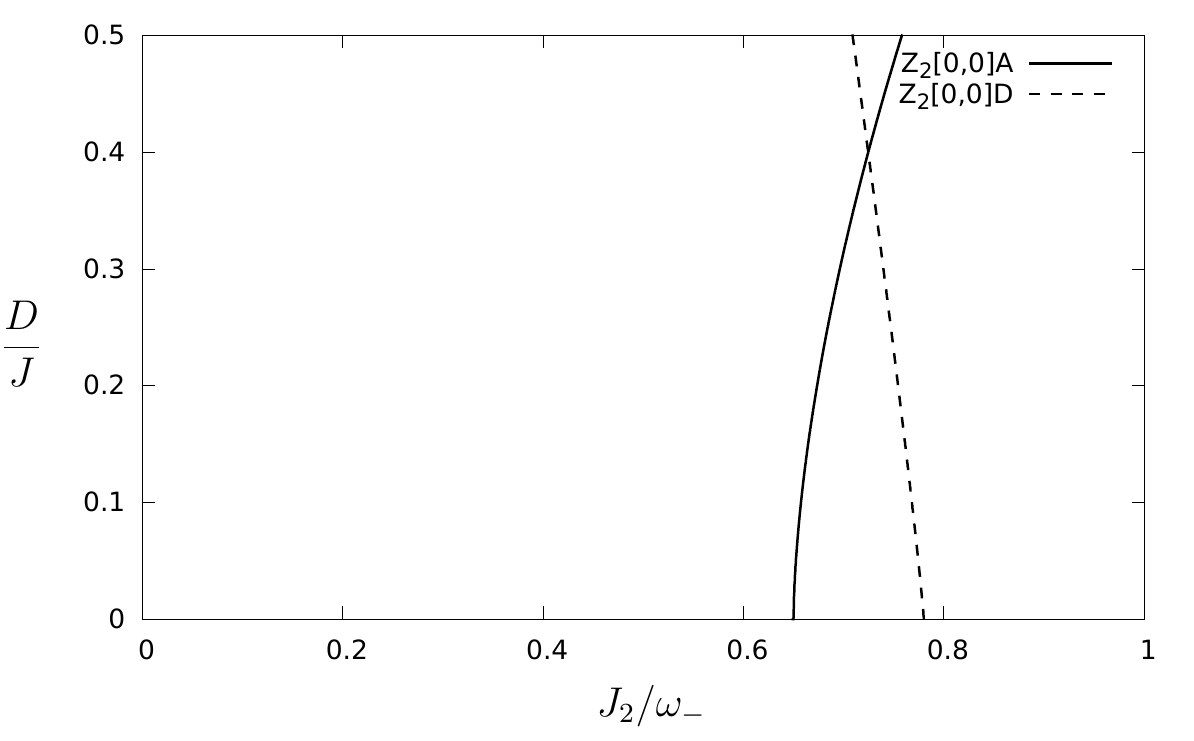}
		\caption{
				Heuristic mean-field phase boundaries in the $J_2$-$D$ plane of $Z_2[0,0]$ states,
				where $J_2$ is next-nearest-neighbor Heisenberg coupling,
				$D$ the nearest-neighbor DM interaction in
				the $\hat{z}$ direction,
				and $\omega_-$ is defined in
				\eref{eqn:singlettripletcombinedchannels}.
		}
	\label{fig:phase-00}
\end{figure}

\begin{figure}[!tbp]
	\centering
	\includegraphics[width = 0.9 \linewidth]{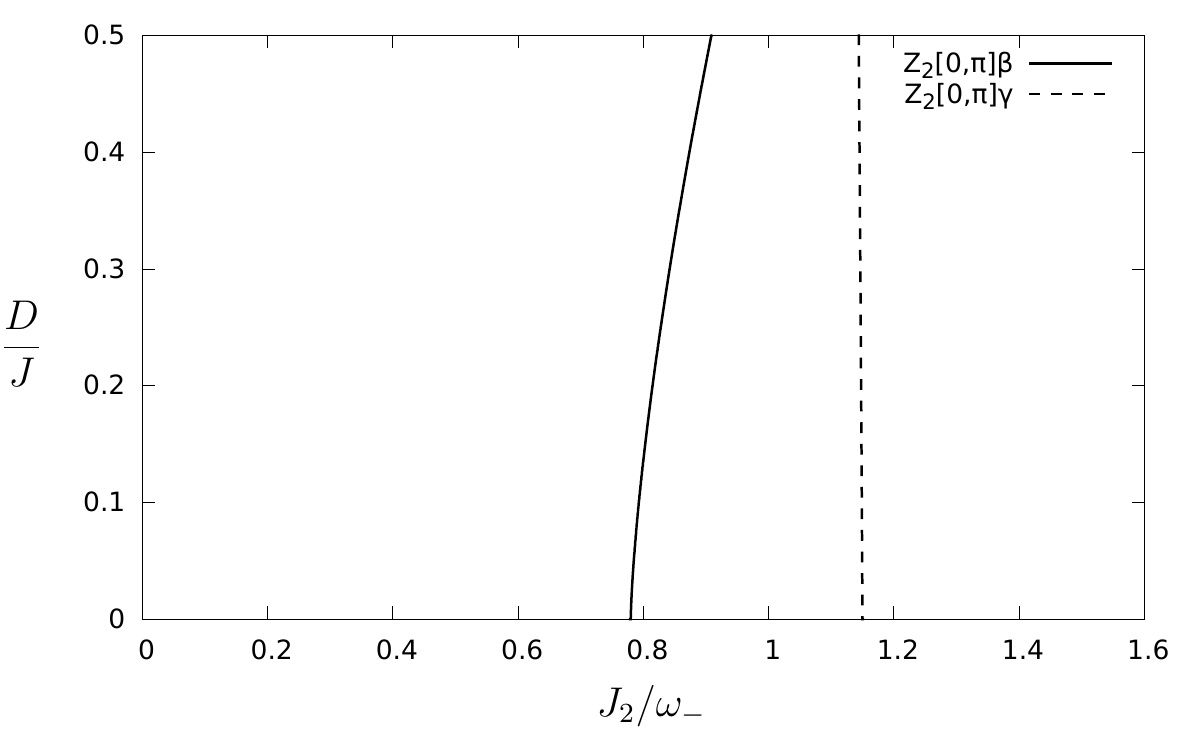}
		\caption{
				Heuristic mean-field phase boundaries in the $J_2$-$D$ plane of $Z_2[0,\pi]$ states,
				where $J_2$ is next-nearest-neighbor Heisenberg coupling,
				$D$ the nearest-neighbor DM interaction in
				the $\hat{z}$ direction,
				and $\omega_-$ is defined in
				\eref{eqn:singlettripletcombinedchannels}.
		}
	\label{fig:phase-0pi}
\end{figure}

\begin{figure}[!tbp]
	\centering
	\includegraphics[width = 0.9 \linewidth]{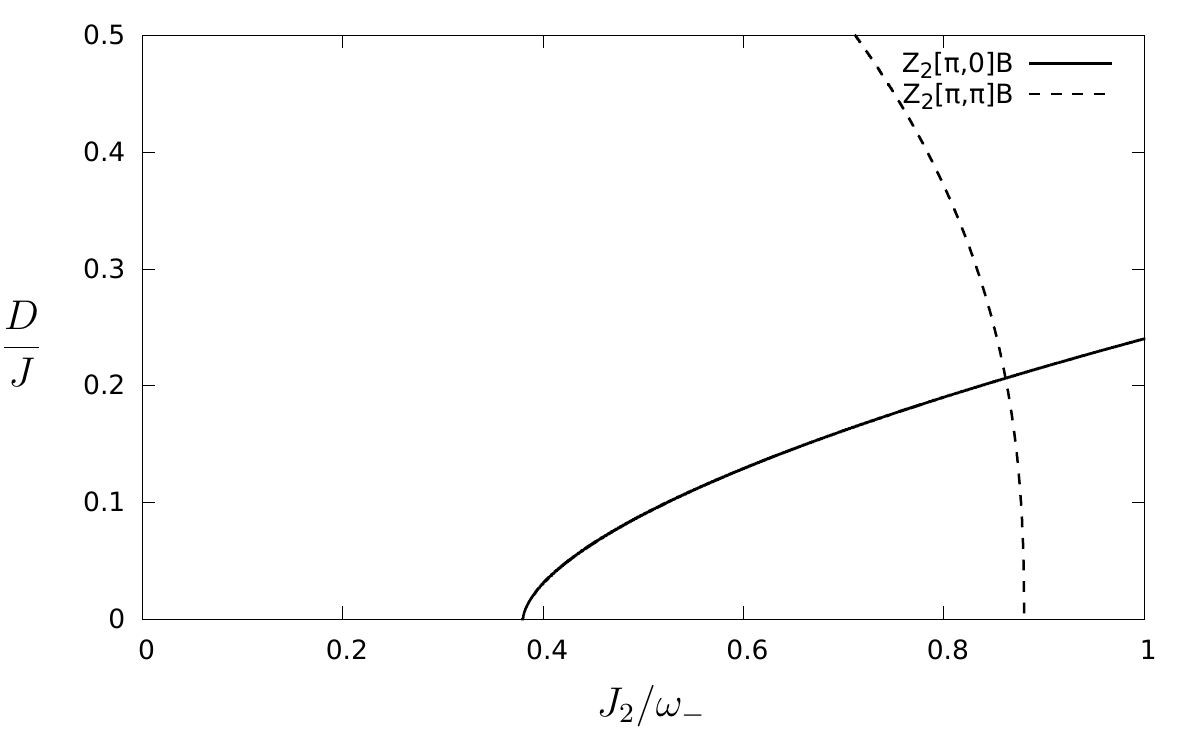}
		\caption{
				Heuristic mean-field phase boundaries in the $J_2$-$D$ plane of 
				$Z_2[\pi,\pi]B$ and $Z_2[\pi,0]B$ states,
				where $J_2$ is next-nearest-neighbor Heisenberg coupling,
				$D$ the nearest-neighbor DM interaction in
				the $\hat{z}$ direction,
				and $\omega_-$ is defined in
				\eref{eqn:singlettripletcombinedchannels}.
		}
	\label{fig:phase-pipi-pizero}
\end{figure}

Figure \ref{fig:phase-00} shows the $U(1)$-$Z_2$ phase boundary
in the $J_2$-$D$ plane for the $Z_2[0,0]$ spin liquids,
Fig. \ref{fig:phase-0pi} for the $Z_2[0,\pi]$ spin liquids,
and Fig. \ref{fig:phase-pipi-pizero} for the
$Z_2[\pi,\pi]B$ and $Z_2[\pi,0]B$ spin liquids.
The next-nearest-neighbor interaction $J_2$ stabilizes all other six $Z_2$ spin liquids.
We see that the 
$Z_2[0,0]A$, $Z_2[0,\pi]\beta$ and $Z_2[\pi,0]B$ phases 
are stabilized for smaller $J_2$ than the others, while
being destabilized by the Dzyaloshinsky-Moriya interaction $D$.
However, the $Z_2[0,0]D$, $Z_2[0,\pi]\gamma$ and $Z_2[\pi,\pi]B$ phases
are stabilized by $D$, despite requiring larger $J_2$.

\section{Dynamical Structure Factor of Singlet+Triplet Ans\"atze}
\label{sec:appendixdsftriplet}

\begin{figure*}
	\centering
	\includegraphics[width=0.9\linewidth]{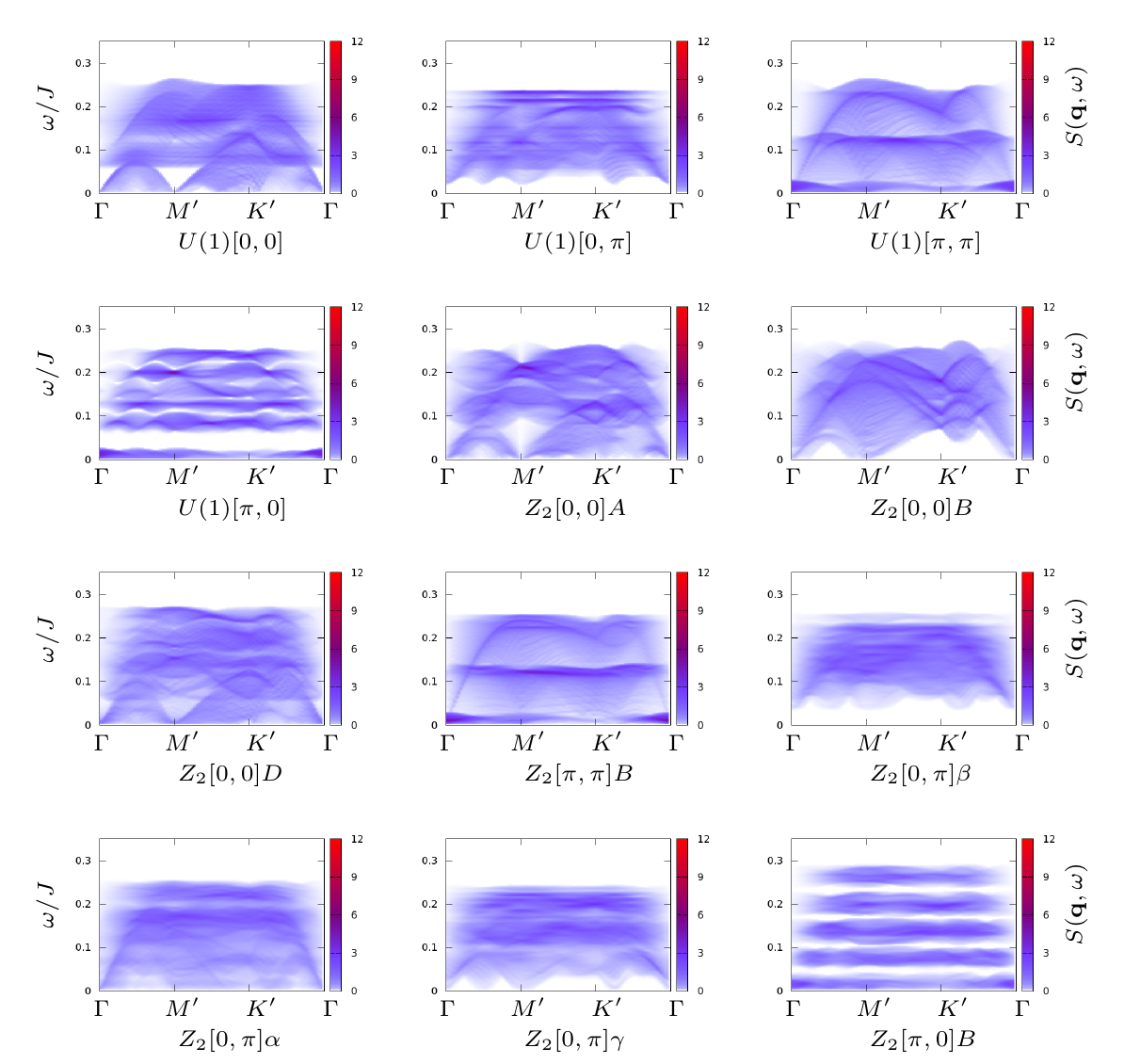}
\caption{(Color online)
Dynamical structure factor for 
$U(1)$ and $Z_2$
``singlet + triplet'' spin liquid ans\"atze.
These structure factor are plotted from the center of the extended Brillouin zone
of the Kagome lattice, $\Gamma$, to the edge $M'$, to
the corner $K'$, and back to the center $\Gamma$.
We note that since the ``singlet + triplet'' $U(1)[0,\pi]$ spin liquid state is gapped, it is unstable to instanton tunneling events.\cite{polyakov}
}
\label{fig:dsf-all-triplet-full}
\end{figure*}

In the main text, we showed the changes in the dynamical structure factor from
the singlet ans\"atze to when small triplet terms are added. Here we present
the trace of the dynamical structure factor matrix, {\it{i.e.}}, $(S^{xx} + S^{yy} +
S^{zz})/3$, for these triplet ans\"atze. 
We show the dynamical structure factor for the $U(1)$ and $Z_2$ spin liquids in Fig. \ref{fig:dsf-all-triplet-full}.

\bibliography{bibliography}

\begin{thebibliography}{65}%
\makeatletter
\providecommand \@ifxundefined [1]{%
 \@ifx{#1\undefined}
}%
\providecommand \@ifnum [1]{%
 \ifnum #1\expandafter \@firstoftwo
 \else \expandafter \@secondoftwo
 \fi
}%
\providecommand \@ifx [1]{%
 \ifx #1\expandafter \@firstoftwo
 \else \expandafter \@secondoftwo
 \fi
}%
\providecommand \natexlab [1]{#1}%
\providecommand \enquote  [1]{``#1''}%
\providecommand \bibnamefont  [1]{#1}%
\providecommand \bibfnamefont [1]{#1}%
\providecommand \citenamefont [1]{#1}%
\providecommand \href@noop [0]{\@secondoftwo}%
\providecommand \href [0]{\begingroup \@sanitize@url \@href}%
\providecommand \@href[1]{\@@startlink{#1}\@@href}%
\providecommand \@@href[1]{\endgroup#1\@@endlink}%
\providecommand \@sanitize@url [0]{\catcode `\\12\catcode `\$12\catcode
  `\&12\catcode `\#12\catcode `\^12\catcode `\_12\catcode `\%12\relax}%
\providecommand \@@startlink[1]{}%
\providecommand \@@endlink[0]{}%
\providecommand \url  [0]{\begingroup\@sanitize@url \@url }%
\providecommand \@url [1]{\endgroup\@href {#1}{\urlprefix }}%
\providecommand \urlprefix  [0]{URL }%
\providecommand \Eprint [0]{\href }%
\@ifxundefined \urlstyle {%
  \providecommand \doi  [0]{\begingroup \@sanitize@url \@doi}%
  \providecommand \@doi [1]{\endgroup \@@startlink {\doibase
  #1}doi:\discretionary {}{}{}#1\@@endlink }%
}{%
  \providecommand \doi  [0]{doi:\discretionary{}{}{}\begingroup
  \urlstyle{rm}\Url }%
}%
\providecommand \doibase [0]{http://dx.doi.org/}%
\providecommand \Doi [0]{\begingroup \@sanitize@url \@Doi }%
\providecommand \@Doi  [1]{\endgroup\@@startlink{\doibase#1}\@@Doi}%
\providecommand \@@Doi [1]{#1\@@endlink}%
\providecommand \selectlanguage [0]{\@gobble}%
\providecommand \bibinfo  [0]{\@secondoftwo}%
\providecommand \bibfield  [0]{\@secondoftwo}%
\providecommand \translation [1]{[#1]}%
\providecommand \BibitemOpen [0]{}%
\providecommand \bibitemStop [0]{}%
\providecommand \bibitemNoStop [0]{.\EOS\space}%
\providecommand \EOS [0]{\spacefactor3000\relax}%
\providecommand \BibitemShut  [1]{\csname bibitem#1\endcsname}%
\bibitem [{\citenamefont {Balents}(2010)}]{Nature.464.199}%
  \BibitemOpen
  \bibfield  {author} {\bibinfo {author} {\bibfnamefont {L.}~\bibnamefont
  {Balents}},\ }\Doi {10.1038/nature08917} {\bibfield  {journal} {\bibinfo
  {journal} {Nature},\ }\textbf {\bibinfo {volume} {464}},\ \bibinfo {pages}
  {199} (\bibinfo {year} {2010})}\BibitemShut {NoStop}%
\bibitem [{\citenamefont {Lee}(2008)}]{Lee05092008}%
  \BibitemOpen
  \bibfield  {author} {\bibinfo {author} {\bibfnamefont {P.~A.}\ \bibnamefont
  {Lee}},\ }\Doi {10.1126/science.1163196} {\bibfield  {journal} {\bibinfo
  {journal} {Science},\ }\textbf {\bibinfo {volume} {321}},\ \bibinfo {pages}
  {1306} (\bibinfo {year} {2008})}\BibitemShut {NoStop}%
\bibitem [{\citenamefont {Mila}(2000)}]{0143-0807-21-6-302}%
  \BibitemOpen
  \bibfield  {author} {\bibinfo {author} {\bibfnamefont {F.}~\bibnamefont
  {Mila}},\ }\href {http://stacks.iop.org/0143-0807/21/i=6/a=302} {\bibfield
  {journal} {\bibinfo  {journal} {European Journal of Physics},\ }\textbf
  {\bibinfo {volume} {21}},\ \bibinfo {pages} {499} (\bibinfo {year}
  {2000})}\BibitemShut {NoStop}%
\bibitem [{\citenamefont {Iqbal}\ \emph {et~al.}(2013)\citenamefont {Iqbal},
  \citenamefont {Becca}, \citenamefont {Sorella},\ and\ \citenamefont
  {Poilblanc}}]{PhysRevB.87.060405}%
  \BibitemOpen
  \bibfield  {author} {\bibinfo {author} {\bibfnamefont {Y.}~\bibnamefont
  {Iqbal}}, \bibinfo {author} {\bibfnamefont {F.}~\bibnamefont {Becca}},
  \bibinfo {author} {\bibfnamefont {S.}~\bibnamefont {Sorella}}, \ and\
  \bibinfo {author} {\bibfnamefont {D.}~\bibnamefont {Poilblanc}},\ }\Doi
  {10.1103/PhysRevB.87.060405} {\bibfield  {journal} {\bibinfo  {journal}
  {Phys. Rev. B},\ }\textbf {\bibinfo {volume} {87}},\ \bibinfo {pages}
  {060405} (\bibinfo {year} {2013})}\BibitemShut {NoStop}%
\bibitem [{\citenamefont {Depenbrock}\ \emph {et~al.}(2012)\citenamefont
  {Depenbrock}, \citenamefont {McCulloch},\ and\ \citenamefont
  {Schollw\"ock}}]{PhysRevLett.109.067201}%
  \BibitemOpen
  \bibfield  {author} {\bibinfo {author} {\bibfnamefont {S.}~\bibnamefont
  {Depenbrock}}, \bibinfo {author} {\bibfnamefont {I.~P.}\ \bibnamefont
  {McCulloch}}, \ and\ \bibinfo {author} {\bibfnamefont {U.}~\bibnamefont
  {Schollw\"ock}},\ }\Doi {10.1103/PhysRevLett.109.067201} {\bibfield
  {journal} {\bibinfo  {journal} {Phys. Rev. Lett.},\ }\textbf {\bibinfo
  {volume} {109}},\ \bibinfo {pages} {067201} (\bibinfo {year}
  {2012})}\BibitemShut {NoStop}%
\bibitem [{\citenamefont {Iqbal}\ \emph {et~al.}(2012)\citenamefont {Iqbal},
  \citenamefont {Becca},\ and\ \citenamefont {Poilblanc}}]{NewJPhys.14.115031}%
  \BibitemOpen
  \bibfield  {author} {\bibinfo {author} {\bibfnamefont {Y.}~\bibnamefont
  {Iqbal}}, \bibinfo {author} {\bibfnamefont {F.}~\bibnamefont {Becca}}, \ and\
  \bibinfo {author} {\bibfnamefont {D.}~\bibnamefont {Poilblanc}},\ }\href
  {http://stacks.iop.org/1367-2630/14/i=11/a=115031} {\bibfield  {journal}
  {\bibinfo  {journal} {New Journal of Physics},\ }\textbf {\bibinfo {volume}
  {14}},\ \bibinfo {pages} {115031} (\bibinfo {year} {2012})}\BibitemShut
  {NoStop}%
\bibitem [{\citenamefont {Jiang}\ \emph {et~al.}(2012)\citenamefont {Jiang},
  \citenamefont {Wang},\ and\ \citenamefont {Balents}}]{NatPhys.8.902}%
  \BibitemOpen
  \bibfield  {author} {\bibinfo {author} {\bibfnamefont {H.-C.}\ \bibnamefont
  {Jiang}}, \bibinfo {author} {\bibfnamefont {Z.}~\bibnamefont {Wang}}, \ and\
  \bibinfo {author} {\bibfnamefont {L.}~\bibnamefont {Balents}},\ }\Doi
  {10.1038/nphys2465} {\bibfield  {journal} {\bibinfo  {journal} {Nat Phys},\
  }\textbf {\bibinfo {volume} {8}},\ \bibinfo {pages} {902} (\bibinfo {year}
  {2012})}\BibitemShut {NoStop}%
\bibitem [{\citenamefont {Han}\ \emph {et~al.}(2012){\natexlab{a}}\citenamefont
  {Han}, \citenamefont {Chu},\ and\ \citenamefont
  {Lee}}]{PhysRevLett.108.157202}%
  \BibitemOpen
  \bibfield  {author} {\bibinfo {author} {\bibfnamefont {T.}~\bibnamefont
  {Han}}, \bibinfo {author} {\bibfnamefont {S.}~\bibnamefont {Chu}}, \ and\
  \bibinfo {author} {\bibfnamefont {Y.~S.}\ \bibnamefont {Lee}},\ }\Doi
  {10.1103/PhysRevLett.108.157202} {\bibfield  {journal} {\bibinfo  {journal}
  {Phys. Rev. Lett.},\ }\textbf {\bibinfo {volume} {108}},\ \bibinfo {pages}
  {157202} (\bibinfo {year} {2012}{\natexlab{a}})}\BibitemShut {NoStop}%
\bibitem [{\citenamefont {Han}\ \emph {et~al.}(2012){\natexlab{b}}\citenamefont
  {Han}, \citenamefont {Helton}, \citenamefont {Chu}, \citenamefont {Nocera},
  \citenamefont {Rodriguez-Rivera}, \citenamefont {Broholm},\ and\
  \citenamefont {Lee}}]{Nature.492.406}%
  \BibitemOpen
  \bibfield  {author} {\bibinfo {author} {\bibfnamefont {T.-H.}\ \bibnamefont
  {Han}}, \bibinfo {author} {\bibfnamefont {J.~S.}\ \bibnamefont {Helton}},
  \bibinfo {author} {\bibfnamefont {S.}~\bibnamefont {Chu}}, \bibinfo {author}
  {\bibfnamefont {D.~G.}\ \bibnamefont {Nocera}}, \bibinfo {author}
  {\bibfnamefont {J.~A.}\ \bibnamefont {Rodriguez-Rivera}}, \bibinfo {author}
  {\bibfnamefont {C.}~\bibnamefont {Broholm}}, \ and\ \bibinfo {author}
  {\bibfnamefont {Y.~S.}\ \bibnamefont {Lee}},\ }\Doi {10.1038/nature11659}
  {\bibfield  {journal} {\bibinfo  {journal} {Nature},\ }\textbf {\bibinfo
  {volume} {492}},\ \bibinfo {pages} {406} (\bibinfo {year}
  {2012}{\natexlab{b}})}\BibitemShut {NoStop}%
\bibitem [{\citenamefont {Lu}\ \emph {et~al.}(2011)\citenamefont {Lu},
  \citenamefont {Ran},\ and\ \citenamefont {Lee}}]{PhysRevB.83.224413}%
  \BibitemOpen
  \bibfield  {author} {\bibinfo {author} {\bibfnamefont {Y.-M.}\ \bibnamefont
  {Lu}}, \bibinfo {author} {\bibfnamefont {Y.}~\bibnamefont {Ran}}, \ and\
  \bibinfo {author} {\bibfnamefont {P.~A.}\ \bibnamefont {Lee}},\ }\href@noop
  {} {\bibfield  {journal} {\bibinfo  {journal} {Phys. Rev. B},\ }\textbf
  {\bibinfo {volume} {83}},\ \bibinfo {pages} {224413} (\bibinfo {year}
  {2011})}\BibitemShut {NoStop}%
\bibitem [{\citenamefont {Messio}\ \emph {et~al.}(2010)\citenamefont {Messio},
  \citenamefont {C\'epas},\ and\ \citenamefont
  {Lhuillier}}]{PhysRevB.81.064428}%
  \BibitemOpen
  \bibfield  {author} {\bibinfo {author} {\bibfnamefont {L.}~\bibnamefont
  {Messio}}, \bibinfo {author} {\bibfnamefont {O.}~\bibnamefont {C\'epas}}, \
  and\ \bibinfo {author} {\bibfnamefont {C.}~\bibnamefont {Lhuillier}},\ }\Doi
  {10.1103/PhysRevB.81.064428} {\bibfield  {journal} {\bibinfo  {journal}
  {Phys. Rev. B},\ }\textbf {\bibinfo {volume} {81}},\ \bibinfo {pages}
  {064428} (\bibinfo {year} {2010})}\BibitemShut {NoStop}%
\bibitem [{\citenamefont {C\'epas}\ \emph {et~al.}(2008)\citenamefont
  {C\'epas}, \citenamefont {Fong}, \citenamefont {Leung},\ and\ \citenamefont
  {Lhuillier}}]{PhysRevB.78.140405}%
  \BibitemOpen
  \bibfield  {author} {\bibinfo {author} {\bibfnamefont {O.}~\bibnamefont
  {C\'epas}}, \bibinfo {author} {\bibfnamefont {C.~M.}\ \bibnamefont {Fong}},
  \bibinfo {author} {\bibfnamefont {P.~W.}\ \bibnamefont {Leung}}, \ and\
  \bibinfo {author} {\bibfnamefont {C.}~\bibnamefont {Lhuillier}},\ }\Doi
  {10.1103/PhysRevB.78.140405} {\bibfield  {journal} {\bibinfo  {journal}
  {Phys. Rev. B},\ }\textbf {\bibinfo {volume} {78}},\ \bibinfo {pages}
  {140405} (\bibinfo {year} {2008})}\BibitemShut {NoStop}%
\bibitem [{\citenamefont {Zorko}\ \emph {et~al.}(2008)\citenamefont {Zorko},
  \citenamefont {Nellutla}, \citenamefont {van Tol}, \citenamefont {Brunel},
  \citenamefont {Bert}, \citenamefont {Duc}, \citenamefont {Trombe},
  \citenamefont {de~Vries}, \citenamefont {Harrison},\ and\ \citenamefont
  {Mendels}}]{PhysRevLett.101.026405}%
  \BibitemOpen
  \bibfield  {author} {\bibinfo {author} {\bibfnamefont {A.}~\bibnamefont
  {Zorko}}, \bibinfo {author} {\bibfnamefont {S.}~\bibnamefont {Nellutla}},
  \bibinfo {author} {\bibfnamefont {J.}~\bibnamefont {van Tol}}, \bibinfo
  {author} {\bibfnamefont {L.~C.}\ \bibnamefont {Brunel}}, \bibinfo {author}
  {\bibfnamefont {F.}~\bibnamefont {Bert}}, \bibinfo {author} {\bibfnamefont
  {F.}~\bibnamefont {Duc}}, \bibinfo {author} {\bibfnamefont {J.-C.}\
  \bibnamefont {Trombe}}, \bibinfo {author} {\bibfnamefont {M.~A.}\
  \bibnamefont {de~Vries}}, \bibinfo {author} {\bibfnamefont {A.}~\bibnamefont
  {Harrison}}, \ and\ \bibinfo {author} {\bibfnamefont {P.}~\bibnamefont
  {Mendels}},\ }\Doi {10.1103/PhysRevLett.101.026405} {\bibfield  {journal}
  {\bibinfo  {journal} {Phys. Rev. Lett.},\ }\textbf {\bibinfo {volume}
  {101}},\ \bibinfo {pages} {026405} (\bibinfo {year} {2008})}\BibitemShut
  {NoStop}%
\bibitem [{\citenamefont {Helton}\ \emph {et~al.}(2007)\citenamefont {Helton},
  \citenamefont {Matan}, \citenamefont {Shores}, \citenamefont {Nytko},
  \citenamefont {Bartlett}, \citenamefont {Yoshida}, \citenamefont {Takano},
  \citenamefont {Suslov}, \citenamefont {Qiu}, \citenamefont {Chung},
  \citenamefont {Nocera},\ and\ \citenamefont {Lee}}]{PhysRevLett.98.107204}%
  \BibitemOpen
  \bibfield  {author} {\bibinfo {author} {\bibfnamefont {J.~S.}\ \bibnamefont
  {Helton}}, \bibinfo {author} {\bibfnamefont {K.}~\bibnamefont {Matan}},
  \bibinfo {author} {\bibfnamefont {M.~P.}\ \bibnamefont {Shores}}, \bibinfo
  {author} {\bibfnamefont {E.~A.}\ \bibnamefont {Nytko}}, \bibinfo {author}
  {\bibfnamefont {B.~M.}\ \bibnamefont {Bartlett}}, \bibinfo {author}
  {\bibfnamefont {Y.}~\bibnamefont {Yoshida}}, \bibinfo {author} {\bibfnamefont
  {Y.}~\bibnamefont {Takano}}, \bibinfo {author} {\bibfnamefont
  {A.}~\bibnamefont {Suslov}}, \bibinfo {author} {\bibfnamefont
  {Y.}~\bibnamefont {Qiu}}, \bibinfo {author} {\bibfnamefont {J.-H.}\
  \bibnamefont {Chung}}, \bibinfo {author} {\bibfnamefont {D.~G.}\ \bibnamefont
  {Nocera}}, \ and\ \bibinfo {author} {\bibfnamefont {Y.~S.}\ \bibnamefont
  {Lee}},\ }\Doi {10.1103/PhysRevLett.98.107204} {\bibfield  {journal}
  {\bibinfo  {journal} {Phys. Rev. Lett.},\ }\textbf {\bibinfo {volume} {98}},\
  \bibinfo {pages} {107204} (\bibinfo {year} {2007})}\BibitemShut {NoStop}%
\bibitem [{\citenamefont {Mendels}\ \emph {et~al.}(2007)\citenamefont
  {Mendels}, \citenamefont {Bert}, \citenamefont {de~Vries}, \citenamefont
  {Olariu}, \citenamefont {Harrison}, \citenamefont {Duc}, \citenamefont
  {Trombe}, \citenamefont {Lord}, \citenamefont {Amato},\ and\ \citenamefont
  {Baines}}]{PhysRevLett.98.077204}%
  \BibitemOpen
  \bibfield  {author} {\bibinfo {author} {\bibfnamefont {P.}~\bibnamefont
  {Mendels}}, \bibinfo {author} {\bibfnamefont {F.}~\bibnamefont {Bert}},
  \bibinfo {author} {\bibfnamefont {M.~A.}\ \bibnamefont {de~Vries}}, \bibinfo
  {author} {\bibfnamefont {A.}~\bibnamefont {Olariu}}, \bibinfo {author}
  {\bibfnamefont {A.}~\bibnamefont {Harrison}}, \bibinfo {author}
  {\bibfnamefont {F.}~\bibnamefont {Duc}}, \bibinfo {author} {\bibfnamefont
  {J.~C.}\ \bibnamefont {Trombe}}, \bibinfo {author} {\bibfnamefont {J.~S.}\
  \bibnamefont {Lord}}, \bibinfo {author} {\bibfnamefont {A.}~\bibnamefont
  {Amato}}, \ and\ \bibinfo {author} {\bibfnamefont {C.}~\bibnamefont
  {Baines}},\ }\Doi {10.1103/PhysRevLett.98.077204} {\bibfield  {journal}
  {\bibinfo  {journal} {Phys. Rev. Lett.},\ }\textbf {\bibinfo {volume} {98}},\
  \bibinfo {pages} {077204} (\bibinfo {year} {2007})}\BibitemShut {NoStop}%
\bibitem [{\citenamefont {Elhajal}\ \emph {et~al.}(2002)\citenamefont
  {Elhajal}, \citenamefont {Canals},\ and\ \citenamefont
  {Lacroix}}]{PhysRevB.66.014422}%
  \BibitemOpen
  \bibfield  {author} {\bibinfo {author} {\bibfnamefont {M.}~\bibnamefont
  {Elhajal}}, \bibinfo {author} {\bibfnamefont {B.}~\bibnamefont {Canals}}, \
  and\ \bibinfo {author} {\bibfnamefont {C.}~\bibnamefont {Lacroix}},\ }\Doi
  {10.1103/PhysRevB.66.014422} {\bibfield  {journal} {\bibinfo  {journal}
  {Phys. Rev. B},\ }\textbf {\bibinfo {volume} {66}},\ \bibinfo {pages}
  {014422} (\bibinfo {year} {2002})}\BibitemShut {NoStop}%
\bibitem [{\citenamefont {Nikolic}\ and\ \citenamefont
  {Senthil}(2003)}]{PhysRevB.68.214415}%
  \BibitemOpen
  \bibfield  {author} {\bibinfo {author} {\bibfnamefont {P.}~\bibnamefont
  {Nikolic}}\ and\ \bibinfo {author} {\bibfnamefont {T.}~\bibnamefont
  {Senthil}},\ }\Doi {10.1103/PhysRevB.68.214415} {\bibfield  {journal}
  {\bibinfo  {journal} {Phys. Rev. B},\ }\textbf {\bibinfo {volume} {68}},\
  \bibinfo {pages} {214415} (\bibinfo {year} {2003})}\BibitemShut {NoStop}%
\bibitem [{\citenamefont {Singh}\ and\ \citenamefont
  {Huse}(2008)}]{PhysRevB.77.144415}%
  \BibitemOpen
  \bibfield  {author} {\bibinfo {author} {\bibfnamefont {R.~R.~P.}\
  \bibnamefont {Singh}}\ and\ \bibinfo {author} {\bibfnamefont {D.~A.}\
  \bibnamefont {Huse}},\ }\Doi {10.1103/PhysRevB.77.144415} {\bibfield
  {journal} {\bibinfo  {journal} {Phys. Rev. B},\ }\textbf {\bibinfo {volume}
  {77}},\ \bibinfo {pages} {144415} (\bibinfo {year} {2008})}\BibitemShut
  {NoStop}%
\bibitem [{\citenamefont {Marston}\ and\ \citenamefont
  {Zeng}(1991)}]{JApplPhys.69.5962}%
  \BibitemOpen
  \bibfield  {author} {\bibinfo {author} {\bibfnamefont {J.~B.}\ \bibnamefont
  {Marston}}\ and\ \bibinfo {author} {\bibfnamefont {C.}~\bibnamefont {Zeng}},\
  }\Doi {10.1063/1.347830} {\bibfield  {journal} {\bibinfo  {journal} {Journal
  of Applied Physics},\ }\textbf {\bibinfo {volume} {69}},\ \bibinfo {pages}
  {5962 } (\bibinfo {year} {1991})}\BibitemShut {NoStop}%
\bibitem [{\citenamefont {Bert}\ \emph {et~al.}(2005)\citenamefont {Bert},
  \citenamefont {Bono}, \citenamefont {Mendels}, \citenamefont {Ladieu},
  \citenamefont {Duc}, \citenamefont {Trombe},\ and\ \citenamefont
  {Millet}}]{PhysRevLett.95.087203}%
  \BibitemOpen
  \bibfield  {author} {\bibinfo {author} {\bibfnamefont {F.}~\bibnamefont
  {Bert}}, \bibinfo {author} {\bibfnamefont {D.}~\bibnamefont {Bono}}, \bibinfo
  {author} {\bibfnamefont {P.}~\bibnamefont {Mendels}}, \bibinfo {author}
  {\bibfnamefont {F.}~\bibnamefont {Ladieu}}, \bibinfo {author} {\bibfnamefont
  {F.}~\bibnamefont {Duc}}, \bibinfo {author} {\bibfnamefont {J.-C.}\
  \bibnamefont {Trombe}}, \ and\ \bibinfo {author} {\bibfnamefont
  {P.}~\bibnamefont {Millet}},\ }\Doi {10.1103/PhysRevLett.95.087203}
  {\bibfield  {journal} {\bibinfo  {journal} {Phys. Rev. Lett.},\ }\textbf
  {\bibinfo {volume} {95}},\ \bibinfo {pages} {087203} (\bibinfo {year}
  {2005})}\BibitemShut {NoStop}%
\bibitem [{\citenamefont {Yoshida}\ \emph {et~al.}(2009)\citenamefont
  {Yoshida}, \citenamefont {Takigawa}, \citenamefont {Yoshida}, \citenamefont
  {Okamoto},\ and\ \citenamefont {Hiroi}}]{PhysRevLett.103.077207}%
  \BibitemOpen
  \bibfield  {author} {\bibinfo {author} {\bibfnamefont {M.}~\bibnamefont
  {Yoshida}}, \bibinfo {author} {\bibfnamefont {M.}~\bibnamefont {Takigawa}},
  \bibinfo {author} {\bibfnamefont {H.}~\bibnamefont {Yoshida}}, \bibinfo
  {author} {\bibfnamefont {Y.}~\bibnamefont {Okamoto}}, \ and\ \bibinfo
  {author} {\bibfnamefont {Z.}~\bibnamefont {Hiroi}},\ }\Doi
  {10.1103/PhysRevLett.103.077207} {\bibfield  {journal} {\bibinfo  {journal}
  {Phys. Rev. Lett.},\ }\textbf {\bibinfo {volume} {103}},\ \bibinfo {pages}
  {077207} (\bibinfo {year} {2009})}\BibitemShut {NoStop}%
\bibitem [{\citenamefont {Moessner}(2001)}]{CanJPhys.79.1283}%
  \BibitemOpen
  \bibfield  {author} {\bibinfo {author} {\bibfnamefont {R.}~\bibnamefont
  {Moessner}},\ }\Doi {10.1139/p01-123} {\bibfield  {journal} {\bibinfo
  {journal} {Canadian Journal of Physics},\ }\textbf {\bibinfo {volume} {79}},\
  \bibinfo {pages} {1283} (\bibinfo {year} {2001})}\BibitemShut {NoStop}%
\bibitem [{\citenamefont {Diep}(2004)}]{diep2004frustrated}%
  \BibitemOpen
  \bibfield  {author} {\bibinfo {author} {\bibfnamefont {H.}~\bibnamefont
  {Diep}},\ }\href@noop {} {\emph {\bibinfo {title} {Frustrated spin
  systems}}}\ (\bibinfo  {publisher} {World Scientific},\ \bibinfo {year}
  {2004})\BibitemShut {NoStop}%
\bibitem [{\citenamefont {Okamoto}\ \emph {et~al.}(2009)\citenamefont
  {Okamoto}, \citenamefont {Yoshida},\ and\ \citenamefont
  {Hiroi}}]{JPSJ.78.033701}%
  \BibitemOpen
  \bibfield  {author} {\bibinfo {author} {\bibfnamefont {Y.}~\bibnamefont
  {Okamoto}}, \bibinfo {author} {\bibfnamefont {H.}~\bibnamefont {Yoshida}}, \
  and\ \bibinfo {author} {\bibfnamefont {Z.}~\bibnamefont {Hiroi}},\ }\Doi
  {10.1143/JPSJ.78.033701} {\bibfield  {journal} {\bibinfo  {journal} {Journal
  of the Physical Society of Japan},\ }\textbf {\bibinfo {volume} {78}},\
  \bibinfo {pages} {033701} (\bibinfo {year} {2009})}\BibitemShut {NoStop}%
\bibitem [{\citenamefont {Sachdev}(1992)}]{PhysRevB.45.12377}%
  \BibitemOpen
  \bibfield  {author} {\bibinfo {author} {\bibfnamefont {S.}~\bibnamefont
  {Sachdev}},\ }\Doi {10.1103/PhysRevB.45.12377} {\bibfield  {journal}
  {\bibinfo  {journal} {Phys. Rev. B},\ }\textbf {\bibinfo {volume} {45}},\
  \bibinfo {pages} {12377} (\bibinfo {year} {1992})}\BibitemShut {NoStop}%
\bibitem [{\citenamefont {Shores}\ \emph {et~al.}(2005)\citenamefont {Shores},
  \citenamefont {Nytko}, \citenamefont {Bartlett},\ and\ \citenamefont
  {Nocera}}]{JAmChemSoc.127.13462}%
  \BibitemOpen
  \bibfield  {author} {\bibinfo {author} {\bibfnamefont {M.~P.}\ \bibnamefont
  {Shores}}, \bibinfo {author} {\bibfnamefont {E.~A.}\ \bibnamefont {Nytko}},
  \bibinfo {author} {\bibfnamefont {B.~M.}\ \bibnamefont {Bartlett}}, \ and\
  \bibinfo {author} {\bibfnamefont {D.~G.}\ \bibnamefont {Nocera}},\ }\Doi
  {10.1021/ja053891p} {\bibfield  {journal} {\bibinfo  {journal} {Journal of
  the American Chemical Society},\ }\textbf {\bibinfo {volume} {127}},\
  \bibinfo {pages} {13462} (\bibinfo {year} {2005})}\BibitemShut {NoStop}%
\bibitem [{\citenamefont {Yan}\ \emph {et~al.}(2011)\citenamefont {Yan},
  \citenamefont {Huse},\ and\ \citenamefont {White}}]{Yan03062011}%
  \BibitemOpen
  \bibfield  {author} {\bibinfo {author} {\bibfnamefont {S.}~\bibnamefont
  {Yan}}, \bibinfo {author} {\bibfnamefont {D.~A.}\ \bibnamefont {Huse}}, \
  and\ \bibinfo {author} {\bibfnamefont {S.~R.}\ \bibnamefont {White}},\ }\Doi
  {10.1126/science.1201080} {\bibfield  {journal} {\bibinfo  {journal}
  {Science},\ }\textbf {\bibinfo {volume} {332}},\ \bibinfo {pages} {1173}
  (\bibinfo {year} {2011})}\BibitemShut {NoStop}%
\bibitem [{\citenamefont {Wen}(2004)}]{wen2004quantum}%
  \BibitemOpen
  \bibfield  {author} {\bibinfo {author} {\bibfnamefont {X.}~\bibnamefont
  {Wen}},\ }\href@noop {} {\emph {\bibinfo {title} {Quantum field theory of
  many-body systems}}},\ Oxford Graduate Texts\ (\bibinfo  {publisher} {Oxford
  University Press},\ \bibinfo {year} {2004})\BibitemShut {NoStop}%
\bibitem [{\citenamefont {Wen}(2002)}]{PRB.65.165113}%
  \BibitemOpen
  \bibfield  {author} {\bibinfo {author} {\bibfnamefont {X.-G.}\ \bibnamefont
  {Wen}},\ }\Doi {10.1103/PhysRevB.65.165113} {\bibfield  {journal} {\bibinfo
  {journal} {Phys. Rev. B},\ }\textbf {\bibinfo {volume} {65}},\ \bibinfo
  {pages} {165113} (\bibinfo {year} {2002})}\BibitemShut {NoStop}%
\bibitem [{\citenamefont {de~Vries}\ \emph {et~al.}(2009)\citenamefont
  {de~Vries}, \citenamefont {Stewart}, \citenamefont {Deen}, \citenamefont
  {Piatek}, \citenamefont {Nilsen}, \citenamefont {R\o{}nnow},\ and\
  \citenamefont {Harrison}}]{PhysRevLett.103.237201}%
  \BibitemOpen
  \bibfield  {author} {\bibinfo {author} {\bibfnamefont {M.~A.}\ \bibnamefont
  {de~Vries}}, \bibinfo {author} {\bibfnamefont {J.~R.}\ \bibnamefont
  {Stewart}}, \bibinfo {author} {\bibfnamefont {P.~P.}\ \bibnamefont {Deen}},
  \bibinfo {author} {\bibfnamefont {J.~O.}\ \bibnamefont {Piatek}}, \bibinfo
  {author} {\bibfnamefont {G.~J.}\ \bibnamefont {Nilsen}}, \bibinfo {author}
  {\bibfnamefont {H.~M.}\ \bibnamefont {R\o{}nnow}}, \ and\ \bibinfo {author}
  {\bibfnamefont {A.}~\bibnamefont {Harrison}},\ }\Doi
  {10.1103/PhysRevLett.103.237201} {\bibfield  {journal} {\bibinfo  {journal}
  {Phys. Rev. Lett.},\ }\textbf {\bibinfo {volume} {103}},\ \bibinfo {pages}
  {237201} (\bibinfo {year} {2009})}\BibitemShut {NoStop}%
\bibitem [{\citenamefont {Imai}\ \emph {et~al.}(2008)\citenamefont {Imai},
  \citenamefont {Nytko}, \citenamefont {Bartlett}, \citenamefont {Shores},\
  and\ \citenamefont {Nocera}}]{PhysRevLett.100.077203}%
  \BibitemOpen
  \bibfield  {author} {\bibinfo {author} {\bibfnamefont {T.}~\bibnamefont
  {Imai}}, \bibinfo {author} {\bibfnamefont {E.~A.}\ \bibnamefont {Nytko}},
  \bibinfo {author} {\bibfnamefont {B.~M.}\ \bibnamefont {Bartlett}}, \bibinfo
  {author} {\bibfnamefont {M.~P.}\ \bibnamefont {Shores}}, \ and\ \bibinfo
  {author} {\bibfnamefont {D.~G.}\ \bibnamefont {Nocera}},\ }\Doi
  {10.1103/PhysRevLett.100.077203} {\bibfield  {journal} {\bibinfo  {journal}
  {Phys. Rev. Lett.},\ }\textbf {\bibinfo {volume} {100}},\ \bibinfo {pages}
  {077203} (\bibinfo {year} {2008})}\BibitemShut {NoStop}%
\bibitem [{\citenamefont {Imai}\ \emph {et~al.}(2011)\citenamefont {Imai},
  \citenamefont {Fu}, \citenamefont {Han},\ and\ \citenamefont
  {Lee}}]{PhysRevB.84.020411}%
  \BibitemOpen
  \bibfield  {author} {\bibinfo {author} {\bibfnamefont {T.}~\bibnamefont
  {Imai}}, \bibinfo {author} {\bibfnamefont {M.}~\bibnamefont {Fu}}, \bibinfo
  {author} {\bibfnamefont {T.~H.}\ \bibnamefont {Han}}, \ and\ \bibinfo
  {author} {\bibfnamefont {Y.~S.}\ \bibnamefont {Lee}},\ }\Doi
  {10.1103/PhysRevB.84.020411} {\bibfield  {journal} {\bibinfo  {journal}
  {Phys. Rev. B},\ }\textbf {\bibinfo {volume} {84}},\ \bibinfo {pages}
  {020411} (\bibinfo {year} {2011})}\BibitemShut {NoStop}%
\bibitem [{\citenamefont {Jeong}\ \emph {et~al.}(2011)\citenamefont {Jeong},
  \citenamefont {Bert}, \citenamefont {Mendels}, \citenamefont {Duc},
  \citenamefont {Trombe}, \citenamefont {de~Vries},\ and\ \citenamefont
  {Harrison}}]{PhysRevLett.107.237201}%
  \BibitemOpen
  \bibfield  {author} {\bibinfo {author} {\bibfnamefont {M.}~\bibnamefont
  {Jeong}}, \bibinfo {author} {\bibfnamefont {F.}~\bibnamefont {Bert}},
  \bibinfo {author} {\bibfnamefont {P.}~\bibnamefont {Mendels}}, \bibinfo
  {author} {\bibfnamefont {F.}~\bibnamefont {Duc}}, \bibinfo {author}
  {\bibfnamefont {J.~C.}\ \bibnamefont {Trombe}}, \bibinfo {author}
  {\bibfnamefont {M.~A.}\ \bibnamefont {de~Vries}}, \ and\ \bibinfo {author}
  {\bibfnamefont {A.}~\bibnamefont {Harrison}},\ }\Doi
  {10.1103/PhysRevLett.107.237201} {\bibfield  {journal} {\bibinfo  {journal}
  {Phys. Rev. Lett.},\ }\textbf {\bibinfo {volume} {107}},\ \bibinfo {pages}
  {237201} (\bibinfo {year} {2011})}\BibitemShut {NoStop}%
\bibitem [{\citenamefont {Wulferding}\ \emph {et~al.}(2010)\citenamefont
  {Wulferding}, \citenamefont {Lemmens}, \citenamefont {Scheib}, \citenamefont
  {R\"oder}, \citenamefont {Mendels}, \citenamefont {Chu}, \citenamefont
  {Han},\ and\ \citenamefont {Lee}}]{PhysRevB.82.144412}%
  \BibitemOpen
  \bibfield  {author} {\bibinfo {author} {\bibfnamefont {D.}~\bibnamefont
  {Wulferding}}, \bibinfo {author} {\bibfnamefont {P.}~\bibnamefont {Lemmens}},
  \bibinfo {author} {\bibfnamefont {P.}~\bibnamefont {Scheib}}, \bibinfo
  {author} {\bibfnamefont {J.}~\bibnamefont {R\"oder}}, \bibinfo {author}
  {\bibfnamefont {P.}~\bibnamefont {Mendels}}, \bibinfo {author} {\bibfnamefont
  {S.}~\bibnamefont {Chu}}, \bibinfo {author} {\bibfnamefont {T.}~\bibnamefont
  {Han}}, \ and\ \bibinfo {author} {\bibfnamefont {Y.~S.}\ \bibnamefont
  {Lee}},\ }\Doi {10.1103/PhysRevB.82.144412} {\bibfield  {journal} {\bibinfo
  {journal} {Phys. Rev. B},\ }\textbf {\bibinfo {volume} {82}},\ \bibinfo
  {pages} {144412} (\bibinfo {year} {2010})}\BibitemShut {NoStop}%
\bibitem [{\citenamefont {de~Vries}\ \emph {et~al.}(2008)\citenamefont
  {de~Vries}, \citenamefont {Kamenev}, \citenamefont {Kockelmann},
  \citenamefont {Sanchez-Benitez},\ and\ \citenamefont
  {Harrison}}]{PhysRevLett.100.157205}%
  \BibitemOpen
  \bibfield  {author} {\bibinfo {author} {\bibfnamefont {M.~A.}\ \bibnamefont
  {de~Vries}}, \bibinfo {author} {\bibfnamefont {K.~V.}\ \bibnamefont
  {Kamenev}}, \bibinfo {author} {\bibfnamefont {W.~A.}\ \bibnamefont
  {Kockelmann}}, \bibinfo {author} {\bibfnamefont {J.}~\bibnamefont
  {Sanchez-Benitez}}, \ and\ \bibinfo {author} {\bibfnamefont {A.}~\bibnamefont
  {Harrison}},\ }\Doi {10.1103/PhysRevLett.100.157205} {\bibfield  {journal}
  {\bibinfo  {journal} {Phys. Rev. Lett.},\ }\textbf {\bibinfo {volume}
  {100}},\ \bibinfo {pages} {157205} (\bibinfo {year} {2008})}\BibitemShut
  {NoStop}%
\bibitem [{\citenamefont {Evenbly}\ and\ \citenamefont
  {Vidal}(2010)}]{PhysRevLett.104.187203}%
  \BibitemOpen
  \bibfield  {author} {\bibinfo {author} {\bibfnamefont {G.}~\bibnamefont
  {Evenbly}}\ and\ \bibinfo {author} {\bibfnamefont {G.}~\bibnamefont
  {Vidal}},\ }\Doi {10.1103/PhysRevLett.104.187203} {\bibfield  {journal}
  {\bibinfo  {journal} {Phys. Rev. Lett.},\ }\textbf {\bibinfo {volume}
  {104}},\ \bibinfo {pages} {187203} (\bibinfo {year} {2010})}\BibitemShut
  {NoStop}%
\bibitem [{\citenamefont {Poilblanc}\ \emph {et~al.}(2010)\citenamefont
  {Poilblanc}, \citenamefont {Mambrini},\ and\ \citenamefont
  {Schwandt}}]{PhysRevB.81.180402}%
  \BibitemOpen
  \bibfield  {author} {\bibinfo {author} {\bibfnamefont {D.}~\bibnamefont
  {Poilblanc}}, \bibinfo {author} {\bibfnamefont {M.}~\bibnamefont {Mambrini}},
  \ and\ \bibinfo {author} {\bibfnamefont {D.}~\bibnamefont {Schwandt}},\ }\Doi
  {10.1103/PhysRevB.81.180402} {\bibfield  {journal} {\bibinfo  {journal}
  {Phys. Rev. B},\ }\textbf {\bibinfo {volume} {81}},\ \bibinfo {pages}
  {180402} (\bibinfo {year} {2010})}\BibitemShut {NoStop}%
\bibitem [{\citenamefont {Budnik}\ and\ \citenamefont
  {Auerbach}(2004)}]{PhysRevLett.93.187205}%
  \BibitemOpen
  \bibfield  {author} {\bibinfo {author} {\bibfnamefont {R.}~\bibnamefont
  {Budnik}}\ and\ \bibinfo {author} {\bibfnamefont {A.}~\bibnamefont
  {Auerbach}},\ }\Doi {10.1103/PhysRevLett.93.187205} {\bibfield  {journal}
  {\bibinfo  {journal} {Phys. Rev. Lett.},\ }\textbf {\bibinfo {volume} {93}},\
  \bibinfo {pages} {187205} (\bibinfo {year} {2004})}\BibitemShut {NoStop}%
\bibitem [{\citenamefont {Hermele}\ \emph {et~al.}(2008)\citenamefont
  {Hermele}, \citenamefont {Ran}, \citenamefont {Lee},\ and\ \citenamefont
  {Wen}}]{PhysRevB.77.224413}%
  \BibitemOpen
  \bibfield  {author} {\bibinfo {author} {\bibfnamefont {M.}~\bibnamefont
  {Hermele}}, \bibinfo {author} {\bibfnamefont {Y.}~\bibnamefont {Ran}},
  \bibinfo {author} {\bibfnamefont {P.~A.}\ \bibnamefont {Lee}}, \ and\
  \bibinfo {author} {\bibfnamefont {X.-G.}\ \bibnamefont {Wen}},\ }\Doi
  {10.1103/PhysRevB.77.224413} {\bibfield  {journal} {\bibinfo  {journal}
  {Phys. Rev. B},\ }\textbf {\bibinfo {volume} {77}},\ \bibinfo {pages}
  {224413} (\bibinfo {year} {2008})}\BibitemShut {NoStop}%
\bibitem [{\citenamefont {Jiang}\ \emph {et~al.}(2008)\citenamefont {Jiang},
  \citenamefont {Weng},\ and\ \citenamefont {Sheng}}]{PhysRevLett.101.117203}%
  \BibitemOpen
  \bibfield  {author} {\bibinfo {author} {\bibfnamefont {H.~C.}\ \bibnamefont
  {Jiang}}, \bibinfo {author} {\bibfnamefont {Z.~Y.}\ \bibnamefont {Weng}}, \
  and\ \bibinfo {author} {\bibfnamefont {D.~N.}\ \bibnamefont {Sheng}},\ }\Doi
  {10.1103/PhysRevLett.101.117203} {\bibfield  {journal} {\bibinfo  {journal}
  {Phys. Rev. Lett.},\ }\textbf {\bibinfo {volume} {101}},\ \bibinfo {pages}
  {117203} (\bibinfo {year} {2008})}\BibitemShut {NoStop}%
\bibitem [{\citenamefont {Hastings}(2000)}]{PhysRevB.63.014413}%
  \BibitemOpen
  \bibfield  {author} {\bibinfo {author} {\bibfnamefont {M.~B.}\ \bibnamefont
  {Hastings}},\ }\Doi {10.1103/PhysRevB.63.014413} {\bibfield  {journal}
  {\bibinfo  {journal} {Phys. Rev. B},\ }\textbf {\bibinfo {volume} {63}},\
  \bibinfo {pages} {014413} (\bibinfo {year} {2000})}\BibitemShut {NoStop}%
\bibitem [{\citenamefont {Mila}(1998)}]{PhysRevLett.81.2356}%
  \BibitemOpen
  \bibfield  {author} {\bibinfo {author} {\bibfnamefont {F.}~\bibnamefont
  {Mila}},\ }\Doi {10.1103/PhysRevLett.81.2356} {\bibfield  {journal} {\bibinfo
   {journal} {Phys. Rev. Lett.},\ }\textbf {\bibinfo {volume} {81}},\ \bibinfo
  {pages} {2356} (\bibinfo {year} {1998})}\BibitemShut {NoStop}%
\bibitem [{\citenamefont {Wang}\ and\ \citenamefont
  {Vishwanath}(2006)}]{PhysRevB.74.174423}%
  \BibitemOpen
  \bibfield  {author} {\bibinfo {author} {\bibfnamefont {F.}~\bibnamefont
  {Wang}}\ and\ \bibinfo {author} {\bibfnamefont {A.}~\bibnamefont
  {Vishwanath}},\ }\Doi {10.1103/PhysRevB.74.174423} {\bibfield  {journal}
  {\bibinfo  {journal} {Phys. Rev. B},\ }\textbf {\bibinfo {volume} {74}},\
  \bibinfo {pages} {174423} (\bibinfo {year} {2006})}\BibitemShut {NoStop}%
\bibitem [{\citenamefont {White}(2012)}]{BAPS.2012.MAR.L19.1}%
  \BibitemOpen
  \bibfield  {author} {\bibinfo {author} {\bibfnamefont {S.~R.}\ \bibnamefont
  {White}},\ }\href {http://meetings.aps.org/link/BAPS.2012.MAR.L19.1}
  {\bibfield  {journal} {\bibinfo  {journal} {Bull. Am. Phys. Soc.},\ }\textbf
  {\bibinfo {volume} {57}},\ \bibinfo {pages} {MAR.L19.1} (\bibinfo {year}
  {2012})}\BibitemShut {NoStop}%
\bibitem [{\citenamefont {{Sindzingre, P.}}\ and\ \citenamefont {{Lhuillier,
  C.}}(2009)}]{epl.2009.27009}%
  \BibitemOpen
  \bibfield  {author} {\bibinfo {author} {\bibnamefont {{Sindzingre, P.}}}\
  and\ \bibinfo {author} {\bibnamefont {{Lhuillier, C.}}},\ }\Doi
  {10.1209/0295-5075/88/27009} {\bibfield  {journal} {\bibinfo  {journal}
  {EPL},\ }\textbf {\bibinfo {volume} {88}},\ \bibinfo {pages} {27009}
  (\bibinfo {year} {2009})}\BibitemShut {NoStop}%
\bibitem [{\citenamefont {Huh}\ \emph {et~al.}(2010)\citenamefont {Huh},
  \citenamefont {Fritz},\ and\ \citenamefont {Sachdev}}]{PhysRevB.81.144432}%
  \BibitemOpen
  \bibfield  {author} {\bibinfo {author} {\bibfnamefont {Y.}~\bibnamefont
  {Huh}}, \bibinfo {author} {\bibfnamefont {L.}~\bibnamefont {Fritz}}, \ and\
  \bibinfo {author} {\bibfnamefont {S.}~\bibnamefont {Sachdev}},\ }\Doi
  {10.1103/PhysRevB.81.144432} {\bibfield  {journal} {\bibinfo  {journal}
  {Phys. Rev. B},\ }\textbf {\bibinfo {volume} {81}},\ \bibinfo {pages}
  {144432} (\bibinfo {year} {2010})}\BibitemShut {NoStop}%
\bibitem [{\citenamefont {Ofer}\ \emph {et~al.}(2011)\citenamefont {Ofer},
  \citenamefont {Keren}, \citenamefont {Brewer}, \citenamefont {Han},\ and\
  \citenamefont {Lee}}]{JPhysCondMatt.23.164207}%
  \BibitemOpen
  \bibfield  {author} {\bibinfo {author} {\bibfnamefont {O.}~\bibnamefont
  {Ofer}}, \bibinfo {author} {\bibfnamefont {A.}~\bibnamefont {Keren}},
  \bibinfo {author} {\bibfnamefont {J.~H.}\ \bibnamefont {Brewer}}, \bibinfo
  {author} {\bibfnamefont {T.~H.}\ \bibnamefont {Han}}, \ and\ \bibinfo
  {author} {\bibfnamefont {Y.~S.}\ \bibnamefont {Lee}},\ }\href
  {http://stacks.iop.org/0953-8984/23/i=16/a=164207} {\bibfield  {journal}
  {\bibinfo  {journal} {Journal of Physics: Condensed Matter},\ }\textbf
  {\bibinfo {volume} {23}},\ \bibinfo {pages} {164207} (\bibinfo {year}
  {2011})}\BibitemShut {NoStop}%
\bibitem [{\citenamefont {Ofer}\ and\ \citenamefont
  {Keren}(2009)}]{PhysRevB.79.134424}%
  \BibitemOpen
  \bibfield  {author} {\bibinfo {author} {\bibfnamefont {O.}~\bibnamefont
  {Ofer}}\ and\ \bibinfo {author} {\bibfnamefont {A.}~\bibnamefont {Keren}},\
  }\Doi {10.1103/PhysRevB.79.134424} {\bibfield  {journal} {\bibinfo  {journal}
  {Phys. Rev. B},\ }\textbf {\bibinfo {volume} {79}},\ \bibinfo {pages}
  {134424} (\bibinfo {year} {2009})}\BibitemShut {NoStop}%
\bibitem [{\citenamefont {Maeda}\ and\ \citenamefont
  {Oshikawa}(2003)}]{PhysRevB.67.224424}%
  \BibitemOpen
  \bibfield  {author} {\bibinfo {author} {\bibfnamefont {Y.}~\bibnamefont
  {Maeda}}\ and\ \bibinfo {author} {\bibfnamefont {M.}~\bibnamefont
  {Oshikawa}},\ }\Doi {10.1103/PhysRevB.67.224424} {\bibfield  {journal}
  {\bibinfo  {journal} {Phys. Rev. B},\ }\textbf {\bibinfo {volume} {67}},\
  \bibinfo {pages} {224424} (\bibinfo {year} {2003})}\BibitemShut {NoStop}%
\bibitem [{\citenamefont {Derzhko}\ and\ \citenamefont
  {Verkholyak}(2006)}]{JPSJ.75.104711}%
  \BibitemOpen
  \bibfield  {author} {\bibinfo {author} {\bibfnamefont {O.}~\bibnamefont
  {Derzhko}}\ and\ \bibinfo {author} {\bibfnamefont {T.}~\bibnamefont
  {Verkholyak}},\ }\Doi {10.1143/JPSJ.75.104711} {\bibfield  {journal}
  {\bibinfo  {journal} {Journal of the Physical Society of Japan},\ }\textbf
  {\bibinfo {volume} {75}},\ \bibinfo {pages} {104711} (\bibinfo {year}
  {2006})}\BibitemShut {NoStop}%
\bibitem [{\citenamefont {Povarov}\ \emph {et~al.}(2011)\citenamefont
  {Povarov}, \citenamefont {Smirnov}, \citenamefont {Starykh}, \citenamefont
  {Petrov},\ and\ \citenamefont {Shapiro}}]{PhysRevLett.107.037204}%
  \BibitemOpen
  \bibfield  {author} {\bibinfo {author} {\bibfnamefont {K.~Y.}\ \bibnamefont
  {Povarov}}, \bibinfo {author} {\bibfnamefont {A.~I.}\ \bibnamefont
  {Smirnov}}, \bibinfo {author} {\bibfnamefont {O.~A.}\ \bibnamefont
  {Starykh}}, \bibinfo {author} {\bibfnamefont {S.~V.}\ \bibnamefont {Petrov}},
  \ and\ \bibinfo {author} {\bibfnamefont {A.~Y.}\ \bibnamefont {Shapiro}},\
  }\Doi {10.1103/PhysRevLett.107.037204} {\bibfield  {journal} {\bibinfo
  {journal} {Phys. Rev. Lett.},\ }\textbf {\bibinfo {volume} {107}},\ \bibinfo
  {pages} {037204} (\bibinfo {year} {2011})}\BibitemShut {NoStop}%
\bibitem [{\citenamefont {Oshikawa}\ and\ \citenamefont
  {Affleck}(2002)}]{PhysRevB.65.134410}%
  \BibitemOpen
  \bibfield  {author} {\bibinfo {author} {\bibfnamefont {M.}~\bibnamefont
  {Oshikawa}}\ and\ \bibinfo {author} {\bibfnamefont {I.}~\bibnamefont
  {Affleck}},\ }\Doi {10.1103/PhysRevB.65.134410} {\bibfield  {journal}
  {\bibinfo  {journal} {Phys. Rev. B},\ }\textbf {\bibinfo {volume} {65}},\
  \bibinfo {pages} {134410} (\bibinfo {year} {2002})}\BibitemShut {NoStop}%
\bibitem [{\citenamefont {Asano}\ \emph {et~al.}(2000)\citenamefont {Asano},
  \citenamefont {Nojiri}, \citenamefont {Inagaki}, \citenamefont {Boucher},
  \citenamefont {Sakon}, \citenamefont {Ajiro},\ and\ \citenamefont
  {Motokawa}}]{PhysRevLett.84.5880}%
  \BibitemOpen
  \bibfield  {author} {\bibinfo {author} {\bibfnamefont {T.}~\bibnamefont
  {Asano}}, \bibinfo {author} {\bibfnamefont {H.}~\bibnamefont {Nojiri}},
  \bibinfo {author} {\bibfnamefont {Y.}~\bibnamefont {Inagaki}}, \bibinfo
  {author} {\bibfnamefont {J.~P.}\ \bibnamefont {Boucher}}, \bibinfo {author}
  {\bibfnamefont {T.}~\bibnamefont {Sakon}}, \bibinfo {author} {\bibfnamefont
  {Y.}~\bibnamefont {Ajiro}}, \ and\ \bibinfo {author} {\bibfnamefont
  {M.}~\bibnamefont {Motokawa}},\ }\Doi {10.1103/PhysRevLett.84.5880}
  {\bibfield  {journal} {\bibinfo  {journal} {Phys. Rev. Lett.},\ }\textbf
  {\bibinfo {volume} {84}},\ \bibinfo {pages} {5880} (\bibinfo {year}
  {2000})}\BibitemShut {NoStop}%
\bibitem [{\citenamefont {Baskaran}\ and\ \citenamefont
  {Anderson}(1988)}]{PhysRevB.37.580}%
  \BibitemOpen
  \bibfield  {author} {\bibinfo {author} {\bibfnamefont {G.}~\bibnamefont
  {Baskaran}}\ and\ \bibinfo {author} {\bibfnamefont {P.~W.}\ \bibnamefont
  {Anderson}},\ }\Doi {10.1103/PhysRevB.37.580} {\bibfield  {journal} {\bibinfo
   {journal} {Phys. Rev. B},\ }\textbf {\bibinfo {volume} {37}},\ \bibinfo
  {pages} {580} (\bibinfo {year} {1988})}\BibitemShut {NoStop}%
\bibitem [{\citenamefont {Baskaran}\ \emph {et~al.}(1987)\citenamefont
  {Baskaran}, \citenamefont {Zou},\ and\ \citenamefont
  {Anderson}}]{Baskaran1987973}%
  \BibitemOpen
  \bibfield  {author} {\bibinfo {author} {\bibfnamefont {G.}~\bibnamefont
  {Baskaran}}, \bibinfo {author} {\bibfnamefont {Z.}~\bibnamefont {Zou}}, \
  and\ \bibinfo {author} {\bibfnamefont {P.~W.}\ \bibnamefont {Anderson}},\
  }\Doi {10.1016/0038-1098(87)90642-9} {\bibfield  {journal} {\bibinfo
  {journal} {Solid State Communications},\ }\textbf {\bibinfo {volume} {63}},\
  \bibinfo {pages} {973} (\bibinfo {year} {1987})}\BibitemShut {NoStop}%
\bibitem [{\citenamefont {Affleck}\ and\ \citenamefont
  {Marston}(1988)}]{PhysRevB.37.3774}%
  \BibitemOpen
  \bibfield  {author} {\bibinfo {author} {\bibfnamefont {I.}~\bibnamefont
  {Affleck}}\ and\ \bibinfo {author} {\bibfnamefont {J.~B.}\ \bibnamefont
  {Marston}},\ }\Doi {10.1103/PhysRevB.37.3774} {\bibfield  {journal} {\bibinfo
   {journal} {Phys. Rev. B},\ }\textbf {\bibinfo {volume} {37}},\ \bibinfo
  {pages} {3774} (\bibinfo {year} {1988})}\BibitemShut {NoStop}%
\bibitem [{\citenamefont {Kotliar}\ and\ \citenamefont
  {Liu}(1988)}]{PhysRevB.38.5142}%
  \BibitemOpen
  \bibfield  {author} {\bibinfo {author} {\bibfnamefont {G.}~\bibnamefont
  {Kotliar}}\ and\ \bibinfo {author} {\bibfnamefont {J.}~\bibnamefont {Liu}},\
  }\Doi {10.1103/PhysRevB.38.5142} {\bibfield  {journal} {\bibinfo  {journal}
  {Phys. Rev. B},\ }\textbf {\bibinfo {volume} {38}},\ \bibinfo {pages} {5142}
  (\bibinfo {year} {1988})}\BibitemShut {NoStop}%
\bibitem [{\citenamefont {Lee}\ \emph {et~al.}(2006)\citenamefont {Lee},
  \citenamefont {Nagaosa},\ and\ \citenamefont {Wen}}]{RevModPhys.78.17}%
  \BibitemOpen
  \bibfield  {author} {\bibinfo {author} {\bibfnamefont {P.~A.}\ \bibnamefont
  {Lee}}, \bibinfo {author} {\bibfnamefont {N.}~\bibnamefont {Nagaosa}}, \ and\
  \bibinfo {author} {\bibfnamefont {X.-G.}\ \bibnamefont {Wen}},\ }\Doi
  {10.1103/RevModPhys.78.17} {\bibfield  {journal} {\bibinfo  {journal} {Rev.
  Mod. Phys.},\ }\textbf {\bibinfo {volume} {78}},\ \bibinfo {pages} {17}
  (\bibinfo {year} {2006})}\BibitemShut {NoStop}%
\bibitem [{\citenamefont {Shindou}\ and\ \citenamefont
  {Momoi}(2009)}]{PhysRevB.80.064410}%
  \BibitemOpen
  \bibfield  {author} {\bibinfo {author} {\bibfnamefont {R.}~\bibnamefont
  {Shindou}}\ and\ \bibinfo {author} {\bibfnamefont {T.}~\bibnamefont
  {Momoi}},\ }\Doi {10.1103/PhysRevB.80.064410} {\bibfield  {journal} {\bibinfo
   {journal} {Phys. Rev. B},\ }\textbf {\bibinfo {volume} {80}},\ \bibinfo
  {pages} {064410} (\bibinfo {year} {2009})}\BibitemShut {NoStop}%
\bibitem [{\citenamefont {Bhattacharjee}\ \emph {et~al.}(2012)\citenamefont
  {Bhattacharjee}, \citenamefont {Kim}, \citenamefont {Lee},\ and\
  \citenamefont {Lee}}]{PhysRevB.85.224428}%
  \BibitemOpen
  \bibfield  {author} {\bibinfo {author} {\bibfnamefont {S.}~\bibnamefont
  {Bhattacharjee}}, \bibinfo {author} {\bibfnamefont {Y.~B.}\ \bibnamefont
  {Kim}}, \bibinfo {author} {\bibfnamefont {S.-S.}\ \bibnamefont {Lee}}, \ and\
  \bibinfo {author} {\bibfnamefont {D.-H.}\ \bibnamefont {Lee}},\ }\Doi
  {10.1103/PhysRevB.85.224428} {\bibfield  {journal} {\bibinfo  {journal}
  {Phys. Rev. B},\ }\textbf {\bibinfo {volume} {85}},\ \bibinfo {pages}
  {224428} (\bibinfo {year} {2012})}\BibitemShut {NoStop}%
\bibitem [{\citenamefont {Lawler}\ \emph {et~al.}(2008)\citenamefont {Lawler},
  \citenamefont {Kee}, \citenamefont {Kim},\ and\ \citenamefont
  {Vishwanath}}]{PhysRevLett.100.227201}%
  \BibitemOpen
  \bibfield  {author} {\bibinfo {author} {\bibfnamefont {M.~J.}\ \bibnamefont
  {Lawler}}, \bibinfo {author} {\bibfnamefont {H.-Y.}\ \bibnamefont {Kee}},
  \bibinfo {author} {\bibfnamefont {Y.~B.}\ \bibnamefont {Kim}}, \ and\
  \bibinfo {author} {\bibfnamefont {A.}~\bibnamefont {Vishwanath}},\ }\Doi
  {10.1103/PhysRevLett.100.227201} {\bibfield  {journal} {\bibinfo  {journal}
  {Phys. Rev. Lett.},\ }\textbf {\bibinfo {volume} {100}},\ \bibinfo {pages}
  {227201} (\bibinfo {year} {2008})}\BibitemShut {NoStop}%
\bibitem [{\citenamefont {Schaffer}\ \emph {et~al.}(2012)\citenamefont
  {Schaffer}, \citenamefont {Bhattacharjee},\ and\ \citenamefont
  {Kim}}]{PhysRevB.86.224417}%
  \BibitemOpen
  \bibfield  {author} {\bibinfo {author} {\bibfnamefont {R.}~\bibnamefont
  {Schaffer}}, \bibinfo {author} {\bibfnamefont {S.}~\bibnamefont
  {Bhattacharjee}}, \ and\ \bibinfo {author} {\bibfnamefont {Y.~B.}\
  \bibnamefont {Kim}},\ }\Doi {10.1103/PhysRevB.86.224417} {\bibfield
  {journal} {\bibinfo  {journal} {Phys. Rev. B},\ }\textbf {\bibinfo {volume}
  {86}},\ \bibinfo {pages} {224417} (\bibinfo {year} {2012})}\BibitemShut
  {NoStop}%
\bibitem [{\citenamefont {Senthil}\ and\ \citenamefont
  {Fisher}(2000)}]{PhysRevB.62.7850}%
  \BibitemOpen
  \bibfield  {author} {\bibinfo {author} {\bibfnamefont {T.}~\bibnamefont
  {Senthil}}\ and\ \bibinfo {author} {\bibfnamefont {M.~P.~A.}\ \bibnamefont
  {Fisher}},\ }\Doi {10.1103/PhysRevB.62.7850} {\bibfield  {journal} {\bibinfo
  {journal} {Phys. Rev. B},\ }\textbf {\bibinfo {volume} {62}},\ \bibinfo
  {pages} {7850} (\bibinfo {year} {2000})}\BibitemShut {NoStop}%
\bibitem [{\citenamefont {Polyakov}(1987)}]{polyakov}%
  \BibitemOpen
  \bibfield  {author} {\bibinfo {author} {\bibfnamefont {A.~M.}\ \bibnamefont
  {Polyakov}},\ }\href@noop {} {\emph {\bibinfo {title} {{Gauge Fields and
  Strings}}}}\ (\bibinfo  {publisher} {Harwood},\ \bibinfo {address} {New
  York},\ \bibinfo {year} {1987})\BibitemShut {NoStop}%
\bibitem [{\citenamefont {{Jeschke}}\ \emph {et~al.}(2013)\citenamefont
  {{Jeschke}}, \citenamefont {{Salvat-Pujol}},\ and\ \citenamefont
  {{Valenti}}}]{2013arXiv1303.1310J}%
  \BibitemOpen
  \bibfield  {author} {\bibinfo {author} {\bibfnamefont {H.~O.}\ \bibnamefont
  {{Jeschke}}}, \bibinfo {author} {\bibfnamefont {F.}~\bibnamefont
  {{Salvat-Pujol}}}, \ and\ \bibinfo {author} {\bibfnamefont {R.}~\bibnamefont
  {{Valenti}}},\ }\href@noop {} {\bibfield  {journal} {\bibinfo  {journal}
  {ArXiv e-prints}} (\bibinfo {year} {2013})},\ \Eprint
  {http://arxiv.org/abs/1303.1310} {arXiv:1303.1310 [cond-mat.str-el]}
  \BibitemShut {NoStop}%
\end{thebibliography}%
\end{document}